\documentclass{aa}  
\usepackage{graphicx}
\usepackage[varg]{txfonts}
\usepackage{longtable}
\usepackage{natbib}
\usepackage{url}
\usepackage{color}
\usepackage{multirow,bigdelim}
\usepackage{cases}
\usepackage{rotating}
\usepackage[colorinlistoftodos]{todonotes}
\bibpunct{(}{)}{;}{a}{}{,} 
\newcommand\kms{{\rm\,km\,s^{-1}}}

\usepackage{hyperref}
\hypersetup{
     breaklinks,
     colorlinks   = true,
     citecolor    = blue,
     urlcolor     = blue,
     linkcolor    = blue
}

\begin{document} 

\title{Kinematic structures of the Solar neighbourhood\\ 
revealed by \textit{\bfseries Gaia} DR1/TGAS and RAVE}

\titlerunning{Kinematic structures of the Solar neighbourhood revealed by \textit{Gaia} DR1/TGAS and RAVE} 

\author{
I.~Kushniruk\inst{1}
\and
T.~Schirmer\inst{1,2}
\and 
T.~Bensby\inst{1}
}

\institute{
Lund Observatory, Department of Astronomy and Theoretical Physics, Box 43, SE-221\,00 Lund, Sweden\\
\email{iryna@astro.lu.se}
\and
Ecole Normale Sup\'erieure Paris-Saclay, D\'epartement de physique, 61 avenue du Pr\'esident Wilson
94 235 Cachan Cedex, France
}

\date{Received 10 May 2017 / Accepted 19 September 2017}

 
\abstract
{The velocity distribution of stars in the Solar neighbourhood is inhomogeneous and rich with stellar streams and kinematic structures. These may retain important clues of the formation and dynamical history of the Milky Way. However, the nature and origin of many of the streams and structures is unclear, hindering our understanding of how the Milky Way formed and evolved.
}
{We aim to study the velocity distribution of stars of the Solar neighbourhood and investigate the properties of individual kinematic structures in order to improve our understanding of their origins.    
}
{Using the astrometric data provided by {\it Gaia}~DR1/TGAS and radial velocities from RAVE~DR5 we perform a wavelet analysis with the {\it \`a trous} algorithm to 55\,831 stars that have $U$ and $V$ velocity uncertainties less than $4\,\kms$. An auto-convolution histogram method is used to filter the output data, and we then run Monte Carlo simulations to verify that the detected structures are real due to velocity uncertainties. Additionally we analysed our stellar sample by splitting all stars into a nearby sample ($<300$\,pc) and a distant sample ($>300$\,pc), and two chemically defined samples that to a first degree represent the thin and the thick disks.}
{We detect 19 kinematic structures in the Solar neighbourhood between scales $3-16\,\kms$ at the $3\sigma$ confidence level. Among them we identified well-known groups (such as Hercules, Sirius, Coma Berenices, Pleiades, and Wolf 630), confirmed recently detected groups (such as Antoja12 and Bobylev16), and detected a new structure at $(U,V)\approx(37, 8)\,\kms$. Another three new groups are tentatively detected, but require further confirmation. Some of the detected groups show clear dependence on distance in the sense that they are only present in the nearby sample ($<300$\,pc), and others appear to be correlated with chemistry as they are only present in either of the chemically defined thin and thick disk samples.
}
{With the much enlarged stellar sample and much increased precision in distances, proper motions, provided by {\it Gaia}~DR1 TGAS we have shown that the velocity distribution of stars in the Solar neighbourhood contains more structures than previously known. A new feature is discovered and three recently detected groups are confirmed at high confidence level. Dividing the sample based on distance and/or metallicity shows that there are variety of structures which are as large-scale and small-scale groups, some of them have clear trends on metallicities, others are a mixture of both disk stars and based on that we discuss possible origin of each group.
}


\maketitle
%

\section{Introduction}

Studies of the velocity distribution of stars in the Solar neighbourhood have shown that it contains a plethora of kinematic structures, with stars that have similar space velocities $(U,V,W)$. There are several possibilities to why different stars have similar kinematic properties: they could be from evaporated open clusters; they could be due to dynamical resonances within the Milky Way; or they could even be remnants of accreted satellite galaxies that merged with the Milky Way billions of years ago. This means that stellar streams retain a lot of information of various dynamical processes of the Milky Way's past and provide clues to our understanding of the formation of the Galaxy. Mapping the structure and properties of the Milky Way, that is a benchmark galaxy, will also aid our attempts to understand the evolution and formation of large spiral galaxies in general. A detailed characterisation of the kinematic properties together with chemical composition of the stars of such structures are crucial when trying to trace their origins \citep[e.g.][]{_freeman02}. 

The release of Hipparcos Catalogue twenty years ago \citep{esa1997} boosted the study of kinematic properties of the Solar neighbourhood. For example, \citet{_dehnen98} studied the distribution of 14\,369 kinematically selected stars using a maximum likelihood estimate method and detected 14 features in the $U-V$ plane of Galactic space velocities. The $W$ direction did not appear very rich in structures with only four moving groups detected. The sample was then split based on ($B-V$) colour index to study the behaviour of young and old stars separately. They found that there are moving groups composed of red stars (supposed to be older), while younger structures were composed of stars with different colours. This was an argument against the theory previously proposed by \cite{_eggen96}, that kinematic structures could be remnants of disrupted open clusters, in which stars are chemically homogeneous. Instead, \citet{_dehnen98} propose that moving groups that follow eccentric orbits could be formed through resonances with the Galactic bar.

\citet{_skuljan99} studied a sample of 4\,597 Hipparcos stars and, unlike \citet{_dehnen98}, used radial velocities provided in Hipparcos Input Catalogue \citep{_turon92}. \citet{_skuljan99} applied a wavelet analysis method for kinematic group detection, identified the most significant structures in the $U-V$ plane, and found that the velocity distribution has a more complicated structure than just moving groups and has a larger, branch-like structure. 

Later, using a combination of CORAVEL radial velocities \citep{_baranne79} and ages, together with Tycho-2 astrometry, \citet{_famaey05} investigated a stellar sample of 5\,952 stars and found well-known streams like Hercules, Sirius, Hyades and Pleiades. They suggest that stellar groups are of dynamical origin as isochrones of associated stars with the moving groups show a big dispersion in ages. A deeper study of the origin of moving groups provided by \citet{_famaey08} involved wavelet transform applied to the same data as in \citet{_famaey05} and checked the theory of kinematic groups being remnants of open clusters. After fitting isochrones inherent for open clusters to stars associated with the Sirius, Hyades and Pleiades streams, \citet{_famaey08} claimed dynamical origins for these groups, as they did not match. 

\citet{_antoja08} investigated a larger sample of 24\,910 stars using wavelet techniques and analysed the age-metallicity distribution of the kinematic branches of Sirius, Hercules, Hyades-Pleiades and Coma Berenices. Each branch showed a wide spread of metallicities, especially Hercules. Sirius group stars were older and peaked at about 400 Myr, compared to Hyades-Pleiades which consist of mainly younger stars.

\citet{_zhao09} later detected 22 structures by applying a wavelet analysis to a sample of 14\,000 dwarf stars from \citet{_nordstrom04} and 6\,000 giant stars from \citet{_famaey05}. That study provided a comprehensive comparison of the positions of all kinematic structures detected by different authors. Eleven of 22 groups had previously been found in the literature, while the remaining 11 were discovered for the first time. 

\citet{_antoja12} identified 19 structures in the Solar neighbourhood by analysing a sample of over 200\,000 stars with available RAVE radial velocities and compared their results with those obtained by using the Geneva-Copenhagen survey \citep{_nordstrom04}. They found 19 structures in the Solar neighbourhood from a sample of over 110\,000 stars and support the dynamical origin of stellar branches based on age-metallicity distribution from \citet{_antoja08} and the fact that the main groups are large-scale structures that are detectable even beyond the Solar neighbourhood.

An alternative approach (than analysis in the $U-V$ velocity plane) is to search for streams in the plane defined by the integrals of motions. This way of searching for kinematic over-densities is important as one may discover stellar streams of possible resonant or even extra-galactic origin. Stars in associated kinematic over-densities keep the same angular momenta and in the Solar neighbourhood behave the same way as moving groups of the cluster disruption origin. Together with the approximation of Keplerian orbits \citep[see][]{_dekker76}, \citet{_arifyanto06}, \citet{_klement08} studied the distribution of stars in $\sqrt{U^2+2V^2}$ and $V$ plane as a consequence of integral of motion approach, first discussed in \citet{_helmi99}. \citet{_arifyanto06} applied wavelet transform to the thin and thick disk samples that consist of nearby subdwarfs with metallicities [Fe/H]\,$>-$0.6 and [Fe/H]\,$\le-$0.6. They found Pleiades, Hyades and Hercules in the thin disk and Arcturus stream in the thick disk. \citet{_klement08} were the first to use RAVE DR1 archive. Their sample consisted of 7\,015 stars that allowed them to detect 8 groups in the $\sqrt{U^2+2V^2}$ and $V$ plane. Later, \citet{_zhao14} focused on the search for kinematic structures for the thick disk population based on LAMOST survey \citep[see][]{_zhao12}. Their stellar sample consisted of 7\,993 stars. Thus, only a few kinematic structures were detected. 

Usually the origin of kinematic structures is studied with help from our knowledge about other components of the Galaxy, but \citet{_antoja14} did the opposite: assuming that the Hercules stream was caused by resonances of Galactic bar, they used the Hercules to constrain the Galactic bar's pattern speed and the local circular frequency. This paper demonstrated further the importance of the study of kinematic structures, as they could be a key to better understanding of the Milky Way formation. 

Cross-matching the first astrometric data release of {\it Gaia}~DR1 \citep{_lindegren16} and the radial velocities from the RAVE~DR5 catalogue \citep{_kunder17}, we now have an access to the most up-to-date and precise astrometric measurements for more than 200\,000 stars. This is a substantially larger sample than most sample that has been previously available, and the precision in the data is also much better than before. Recently, using TGAS and RAVE, the kinematics of halo stars was investigated by \citet{_helmi17}, who studied the velocity correlation function and the energy versus angular momentum space of about 1000 stars with metallicities $\rm [M/H] \le - 1.5$. They found that the distribution of stars in the space defined by integrals of motion has complex kinematic structure and more that a half of them follow retrograde orbits. Halo substructure with TGAS and RAVE was also studied in \citet{_myeong17}. The clump of 14 co-moving metal-poor giants was discovered. Based on small spreads of the metallicity within the group, authors explain its origin as being a dissolving remnant of a globular cluster. \citet{_liang17} applied a wavelet analysis technique to a sample that is a combination of the LAMOST DR3 \citep{_zhao12} and the {\it Gaia} TGAS \citep{_michalik15} catalogues. They detected 16 kinematic structures were found and four of them are potential new stream candidates.

The list of works on kinematic groups could be extended and all of them prove that the velocity distribution in the Solar neighbourhood is inhomogeneous and has a complex, branch-like structure. The question on how did the stellar streams formed is still actual. Some of the mentioned above papers attempts to resolve this question, and as a result exists a variety of theories that could explain the origin of stellar streams, but there is no exact agreement among them even for the well-studied groups, which further demonstrates the importance of the study of kinematic structures. 

Using the recent TGAS and RAVE data releases, this study focus on the velocity distribution of stars in the $U-V$ plane to reveal the structures and to further analyse some properties of each structure in terms of distance and chemistry. The paper is organised in the following way: in Sect.~\ref{sec_data} we discuss the stellar sample and its properties; Sect.~\ref{sec:numerical_methods} contains an explanation of numerical methods (wavelet analysis) used in this work; Sect.~\ref{sec:maps} covers the description of input and output maps for the the wavelet analysis; all the results including tables and figures of kinematic structures we present in Sect.~\ref{sec:results}; finally, the summary and discussion of this work are in Sect.~\ref{sec:discussion}.

\section{Stellar sample}
\label{sec_data}

To detect stellar structures in velocity space we will perform a wavelet analysis applied for  a data sample in the $U-V$ plane (see Sect.~\ref{sec:numerical_methods}), where $U, V, W$ are the space velocities of the stars in a right-handed Cartesian coordinate system $(X, Y, Z)$. $X$ axis points towards Galactic centre, $Y$ axis defines the direction of Galactic rotation, and the $Z$ axis points towards Galactic North Pole. The sample should be as large as possible and contain precise measurements of distances, proper motions, and radial velocities, for the calculation of accurate space velocities.

\subsection{Distances and radial velocities}\label{sec_catalogues}

Since the {\it Gaia} satellite was launched in 2013 we are expecting the most precise astrometric measurements for billions of stars in the Milky Way. The first {\it Gaia} data release (DR1) \citep{_lindegren16} due to the shortage of observations is still far from declared precision: for a star with a magnitude $V=15$ it is expected to get positions, proper motions and parallaxes with the precision up to 5-25 $\mu$as \citep[see][]{_michalik15}. However, adding astrometry from the Hipparcos catalogue, TGAS gives astrometric solutions for 2.5 million stars with precise measurements of all required astrometric data \citep{_michalik15}. According to \citet{_brown16} it is recommended that a systematic error of 0.3 mas has to be accounted. Later, \citet{_lindegren17} states that parallax uncertainties already represent the total uncertainty and additional account of systematic error could lead to overestimation. So, in this work we used original TGAS data. In order to calculate the three-dimensional movements of the stars, i.e. the $U$, $V$, and $W$ space velocities, the TGAS data needs to be complemented with radial velocities.

The Radial Velocity Experiment (RAVE) is a medium-resolution spectroscopic survey with the primary goal of determining radial velocities and stellar parameters for hundreds of thousands of stars in the Southern Hemisphere using spectra of resolving power R$=$7\,500 \citep{steinmetz2003}. The final release of RAVE (DR5) gives precise measurements of radial velocities of 457\,588 unique stars \citep{_kunder17}. Cross-matching RAVE~DR5 with TGAS provide us a sample of 159\,299 stars with known coordinates ($\alpha$, $\delta$), proper motions ($\mu_{\alpha}$, $\mu_{\delta}$), positive parallaxes ($\pi$), radial velocities ($v_{rad}$), abundances of Mg and Fe and their associated uncertainties for all stars. The RAVE catalogue contains multiple observations for some stars. In those cases, the median value of every parameter were used in this work. 

To further expand our sample we will also explore the option to include the data releases from The Large sky Area Multi-Object Fibre Spectroscopic Telescope \citep[LAMOST DR2,][]{_luo15}. This is a Northern hemisphere survey which contains spectra of almost 2 million stars with the resolution of R$=$2\,000. The cross-matching of A, F, G and K type stars in the LAMOST DR2 catalogue with TGAS  \footnote{Using the gaia\_tools Python package developed by Jo Bovy that is available at \url{https://github.com/jobovy/gaia\_tools}} leaves us a sample of 107\,501 stars with positive parallax.

\subsection{Space velocities and their uncertainties}
\label{sec:vel_unc}

Space velocities and their uncertainties, which are dependent on the accuracy of the proper motions, the parallaxes, and the radial velocities, were computed according to the equations in \citep{_johnson87}. 

\begin{figure}
   \centering
   \resizebox{\hsize}{!}{
   \includegraphics{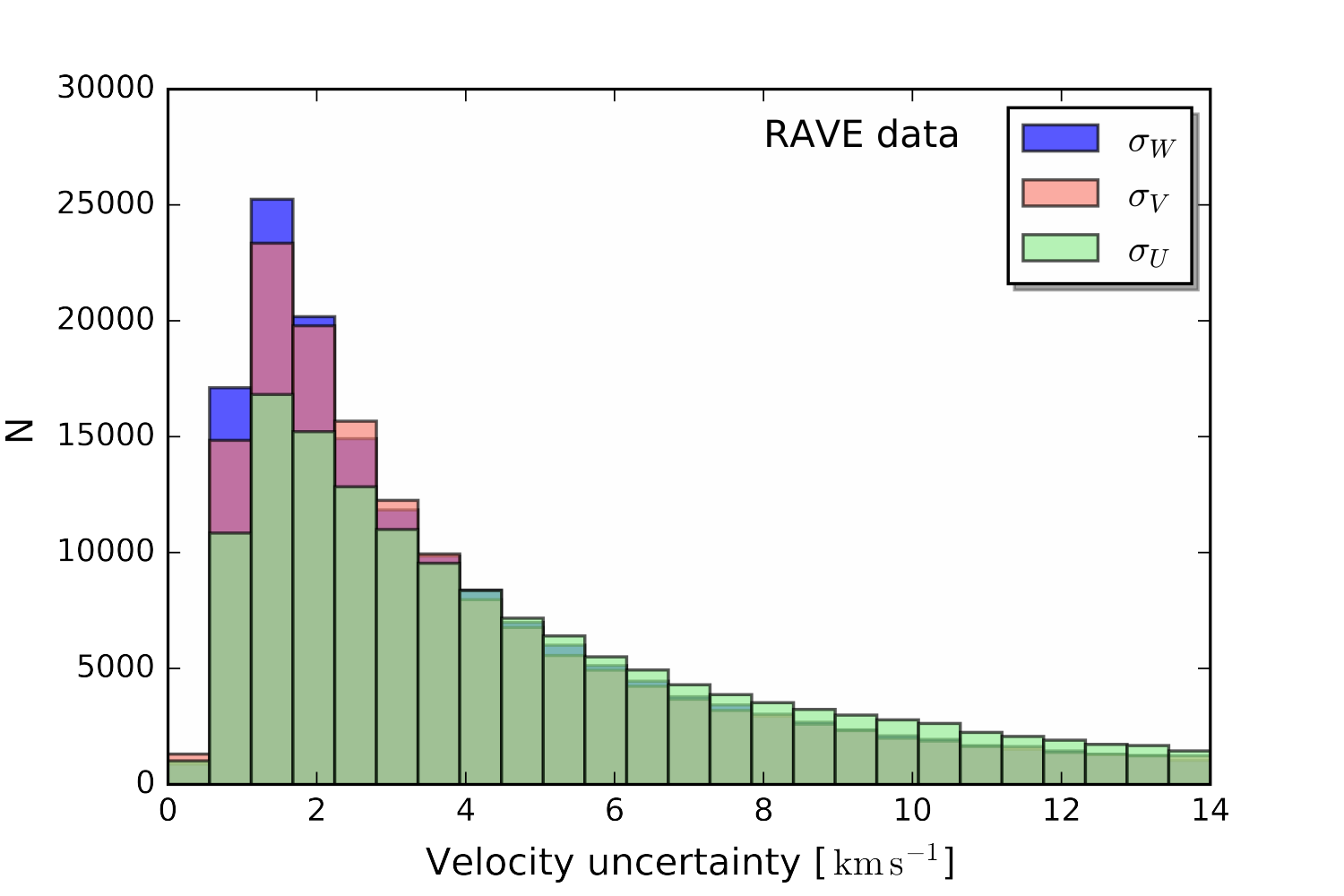}}
   \resizebox{\hsize}{!}{
    \includegraphics{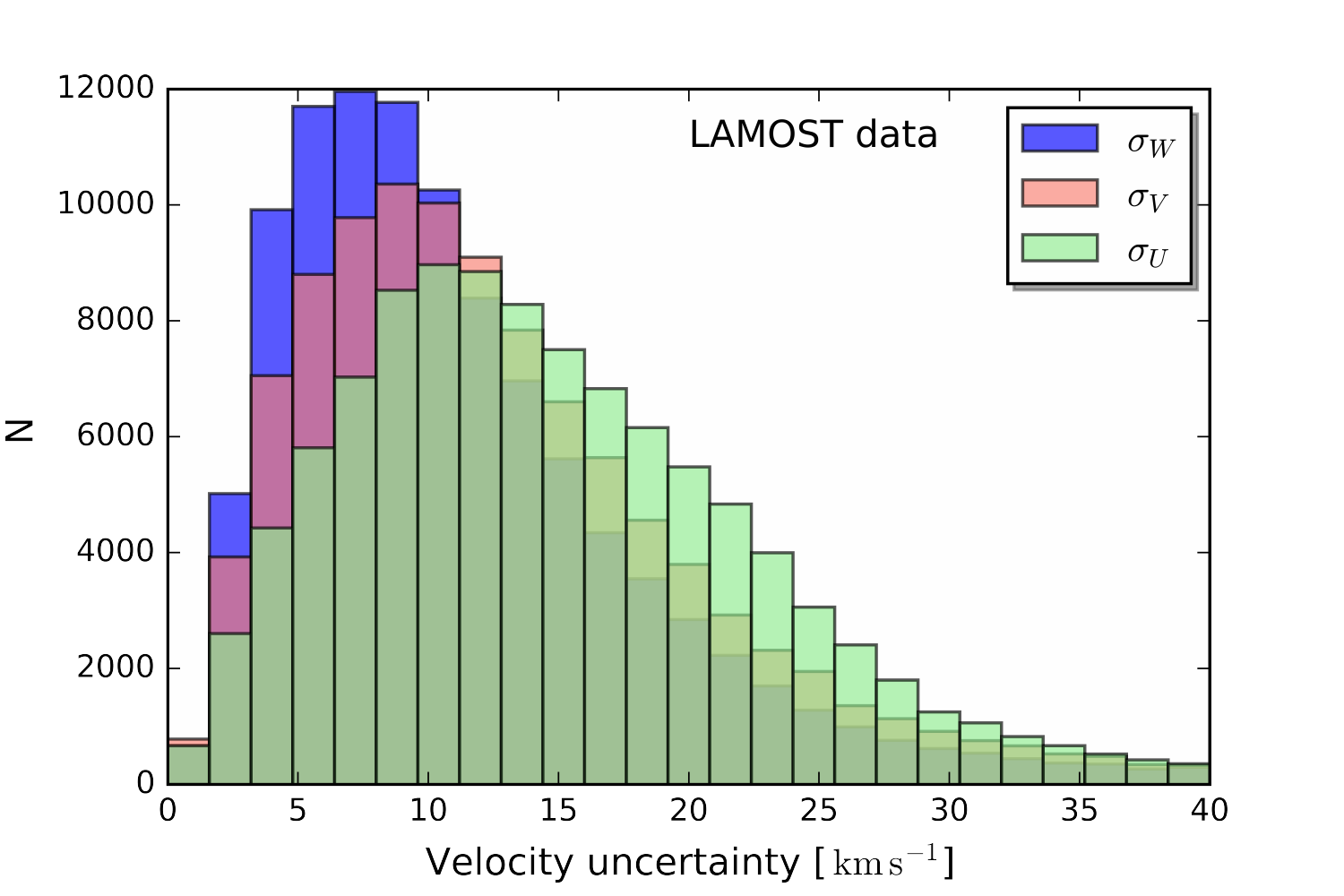}}
   \caption{Distribution of the velocity uncertainties, $\sigma_{U}$ is shown with green colour, $\sigma_{V}$ -- pink, $\sigma_{W}$ is presented in blue. Top: 159\,299 RAVE stars. Bottom: 107\,501 LAMOST stars.}
   \label{_guvhist}
\end{figure}

Figure~\ref{_guvhist} shows the distributions of the uncertainties in the $U$, $V$, and $W$ velocities for 159\,299 RAVE stars (top) and 107\,501 LAMOST stars (bottom). Each velocity component is indicated with a different colour. About 35\% (55\,831) of the RAVE stars have velocities with uncertainties smaller than 4$\kms$, while only 0.8\% (905) LAMOST stars belong to the same region. Such a comparably low accuracy of LAMOST velocities can be explained with high uncertainties of radial velocities, which are one of the main components when computing $\sigma_U$, $\sigma_V$ and $\sigma_W$. \cite{_tian15} cross-matched LAMOST DR1 with APOGEE and discovered an offset of $\sim$5.7$\kms$ of LAMOST radial velocities. \citet{_schonrich17} report that LAMOST line-of-site velocities are underestimated and have to be corrected by 5$\kms$. The accuracy of space velocities is crucial for detection of kinematic groups which will be shown later in the Sect.~\ref{sec:maps}. Taking into account high uncertainties for the LAMOST stars we decided to focus our analysis on the RAVE sample only, which gives us a sample of 159\,299 stars. 

\begin{figure}
   \centering
   \resizebox{\hsize}{!}{
   \includegraphics[viewport = 0 0 410 280,clip]{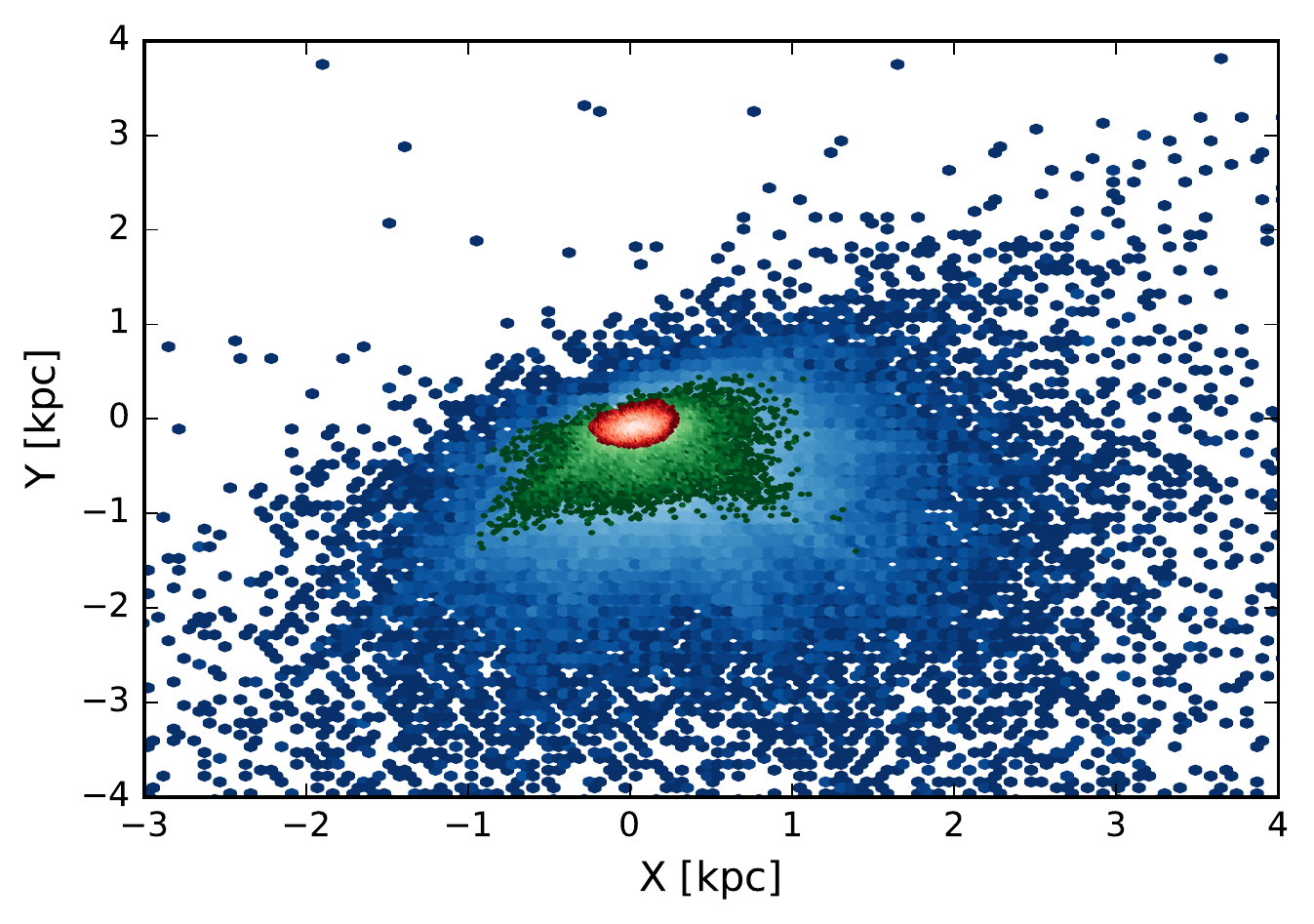}}
    \resizebox{\hsize}{!}{
   \includegraphics[viewport = 0 0 410 280,clip]{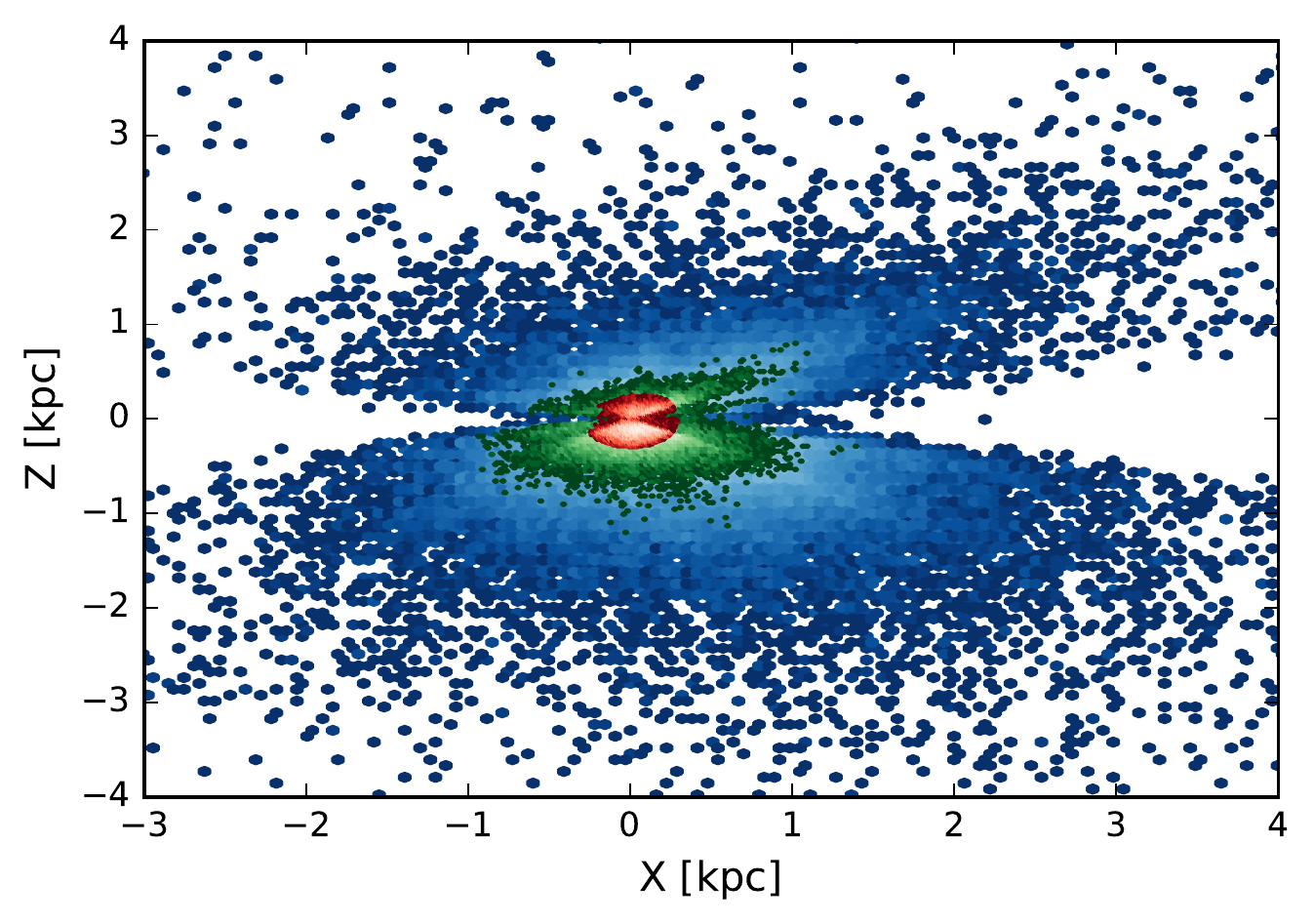}}
   \caption{Spatial distribution of RAVE stars in the $X-Y$ (top) and $X-Z$ (bottom) planes shown as a log-scale density plot. Blue colour shows a sample of 159\,299 stars, green colour is for a sample of 55\,831 stars with $\sigma_{U}$ and $\sigma_{V}$ < 4 $\kms$, red colour describes the distribution of 31\,533 stars associated with the solar neighbourhood ($d<300$ pc). The lighter shades of each colour show higher number of stars in distributions.
   \label{fig:xyz}
   }
\end{figure}

\begin{figure}
   \centering
   \resizebox{\hsize}{!}{
   \includegraphics[viewport = 0 0 410 280,clip]{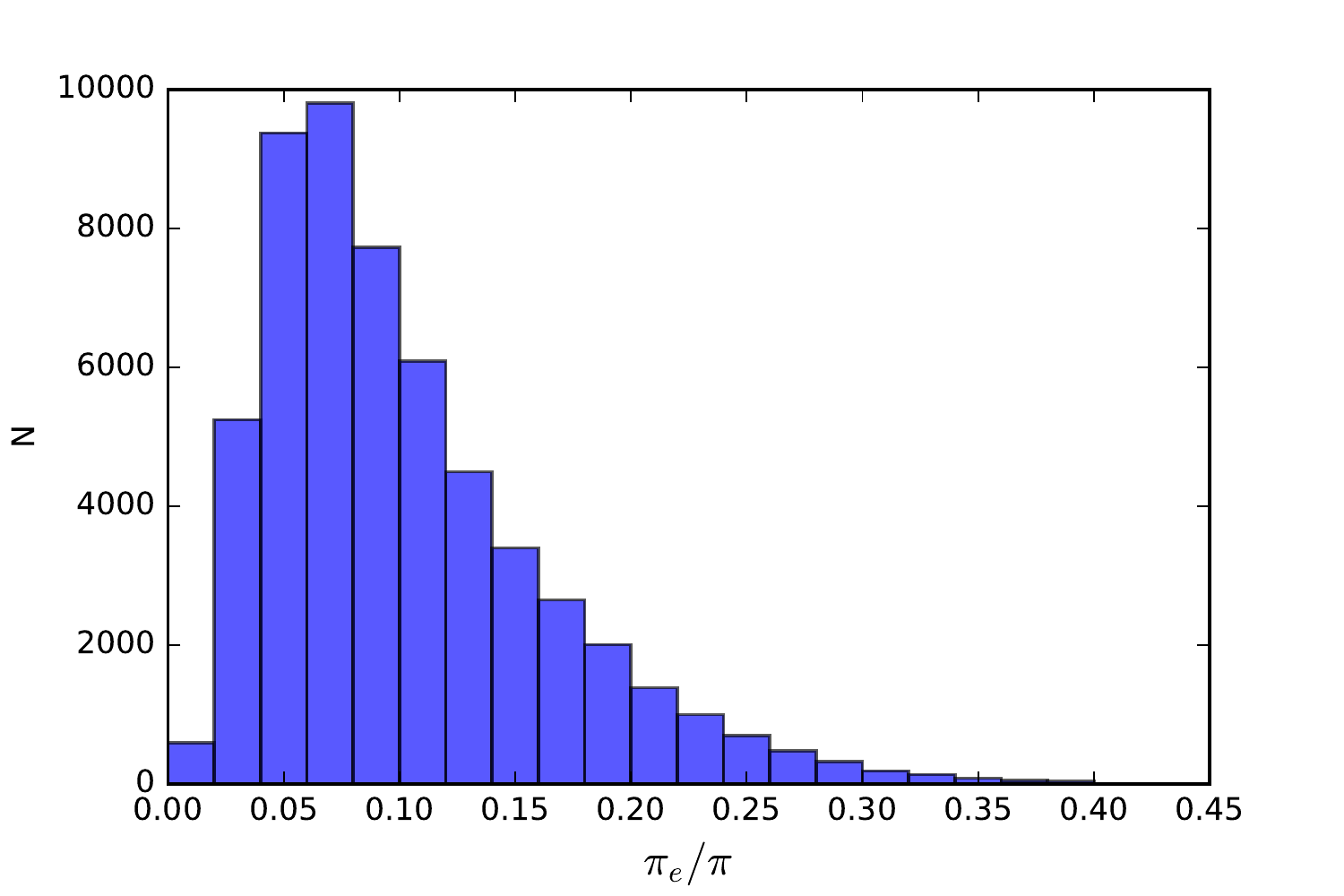}}
   \caption{The distribution of parallax relative uncertainties for 55\,831 stars with $\sigma_{U}$ and $\sigma_{V}<4.0\kms$.
   }
   \label{_p_e}
\end{figure}

The spatial distribution of our RAVE-TGAS star sample in the $X-Y$ and $X-Z$ planes is shown in Fig.~\ref{fig:xyz}. In this plot we show three distributions: blue colour is for the sample of all 159\,299 stars, green colour shows 55\,831 stars with $\sigma_U$, $\sigma_V < 4 \kms$, and the red colour indicates the same stars as the green but with distance limit $< 300$\,pc. As will be shown later in Sect.~\ref{sec:maps} we will focus on the analysis of the last two sub-samples. The precision of the parallax distances provided by TGAS is high enough and additionally the cut on velocity uncertainties ($\sigma_U$ and $\sigma_V$) less than $4\,\kms$ already cut stars by parallax too. In Fig.~\ref{_p_e} the distribution of parallax relative uncertainties $\pi_e/\pi$ for the total sample is shown, where $\pi$ is the parallax and $\pi_e$ is its uncertainty. Most of stars have uncertainties less than 30\,\%. This cut is necessary to get robust positions of kinematic structures.

The question if the Local Standard of Rest (LSR) should be included in the space velocities, or not, in the detection analysis for kinematic groups is debatable. In several works the space velocities were not adjusted for the peculiar Solar motion \citep[e.g.][]{_skuljan99,_antoja08,_zhao09,_antoja12}, while in some papers it was \citep[e.g.][]{_klement08,_zhao14}. We chose to not correct our space velocities to the LSR as the adopted solar motion relative to the LSR may differ between studies \citep[e.g][]{schonrich2012} and if so, would make direct comparisons of the detected structures more difficult.

\section{Numerical methods}
\label{sec:numerical_methods}

Different statistical methods have been used to highlight kinematic over-densities: wavelet analysis \citep[e.g.][]{_skuljan99,_antoja08,_zhao09,_antoja12}, maximum likelihood algorithm \citep[e.g.][]{_dehnen98, _famaey05}, and adaptive kernel estimate \citep[e.g.][]{_skuljan99}. We chose the most efficient technique for our purposes: the wavelet analysis with the {\it \`a trous} algorithm \citep{_starck98} as it is a powerful tool which gives signal characteristics in terms of location, both in position and scale (size of the structure) simultaneously. The utility of this analysis method applied to the detection of moving groups in the Solar neighbourhood has already been demonstrated in several studies \citep[e.g.][]{_cherelul98, _skuljan99, _famaey08, _antoja08,_antoja12}. 

The analysis consists of applying a set of filters at different scales to the original data in order to determine wavelet coefficients. Detected structures, which correspond to local maxima in the wavelet space, can be either physical (kinematic groups) or `artefacts'. The latter can have two origins: (1) The wavelet coefficients contain Poisson noise due to that the stellar sample is finite. Details on the filtering of the Poisson noise can be found in Sect.~\ref{sec:sec:filtering}; (2) The space velocities of the stars contain significant uncertainties. Details on the verification of the robustness of results are given in Sect.~\ref{sec:sec:MC}.

\subsection{The {\it \`a trous} algorithm}
\label{sec:sec:algorithm}
We focused on the wavelet analysis with the {\it \`a trous} algorithm because it has an advantage compared to other statistical methods: it does not require any assumptions on the initial stellar distribution. So, the input data correspond only to the original star count map in the $U-V$ plane. The algorithm implies applying a set of filters at different scales $s_{j}=2^{j}\times\Delta$ in order to decompose the 2-D signal $c_{0}(i_x,i_y)$ into a set of wavelet coefficients $(w_1,\,...,\,w_n)$ that contain the information about kinematic structures. Here, ($i_x,i_y$) is a position at the input grid, $j$ is the scale index ($j\in\left[1,\,p\right]$), $p$ is the maximum scale and $\Delta$ is the bin size of the input pixel grid which is used to detect structures which have sizes between $s_{j-1}$ and $s_{j}$ $\kms$ (for details on the algorithm see \citealt{starck_astronomical_2002}). 

For one position ($i_x,i_y$), a positive wavelet coefficient corresponds to an over-density in the velocity space. We followed the documentation provided with the MR software and we used a maximum scale $p$ equal to $\mathrm{log}_2(N-1)$, where $N$, assuming that the input star count map has a size $N \times M$, is the number of pixels in the smaller direction.

\subsection{Image filtering and detection of significant structures}
\label{sec:sec:filtering}

Given that the data sample is finite, wavelet coefficients at each scale except the information about the structures contain also noise which follows Poisson statistics. The procedure to determine if a wavelet coefficient is significant or not depends on the kind of data. First, we determined the multi-resolution support of the resulting image, which is a logical\footnote{if $w_{j}(i_x,i_y)$ is significant for a given scale $j$ and position $(i_x,i_y)$, then $M_{j}(i_x,i_y)=1$, otherwise, $M_{j}(i_x,i_y)=0$} way to store information about the significance of a wavelet coefficient at a given scale $j$ and a position $(i_x,i_y)$. Our data contains a large number of pixels with less than 30 star counts, which is called the case of ``few events''. In order to remove the Poisson noise in the case of ``few events'' we used the auto-convolution histogram method~\cite[][]{1993ApJ...409..517S} which has already been successfully used to detect structures in data with few events such as low intensity X-ray images~\cite[][]{1998A&AS..128..397S}.
    
With the final set of wavelet coefficients we used an algorithm provided with the MR software that groups coefficients into structures that are characterised by the level of confidence $\epsilon$. A structure detected with a 3$\sigma$ confidence level corresponds to a 99.86\,\% probability that the structure was not produced by the Poisson noise. Then, the algorithm approximates the shape of the structure by an ellipse, characterised by its centre, its semi-minor axis, its semi-major axis, and the angle between the major axis and the horizontal axis of the input map. These parameters are useful for the estimation of the number of stars that belongs to the structure, assuming that all the stars inside the ellipse can be associated with the structure.

   \begin{figure}
   \centering
   \resizebox{\hsize}{!}{
   \includegraphics[viewport = 70 0 480 410,clip]{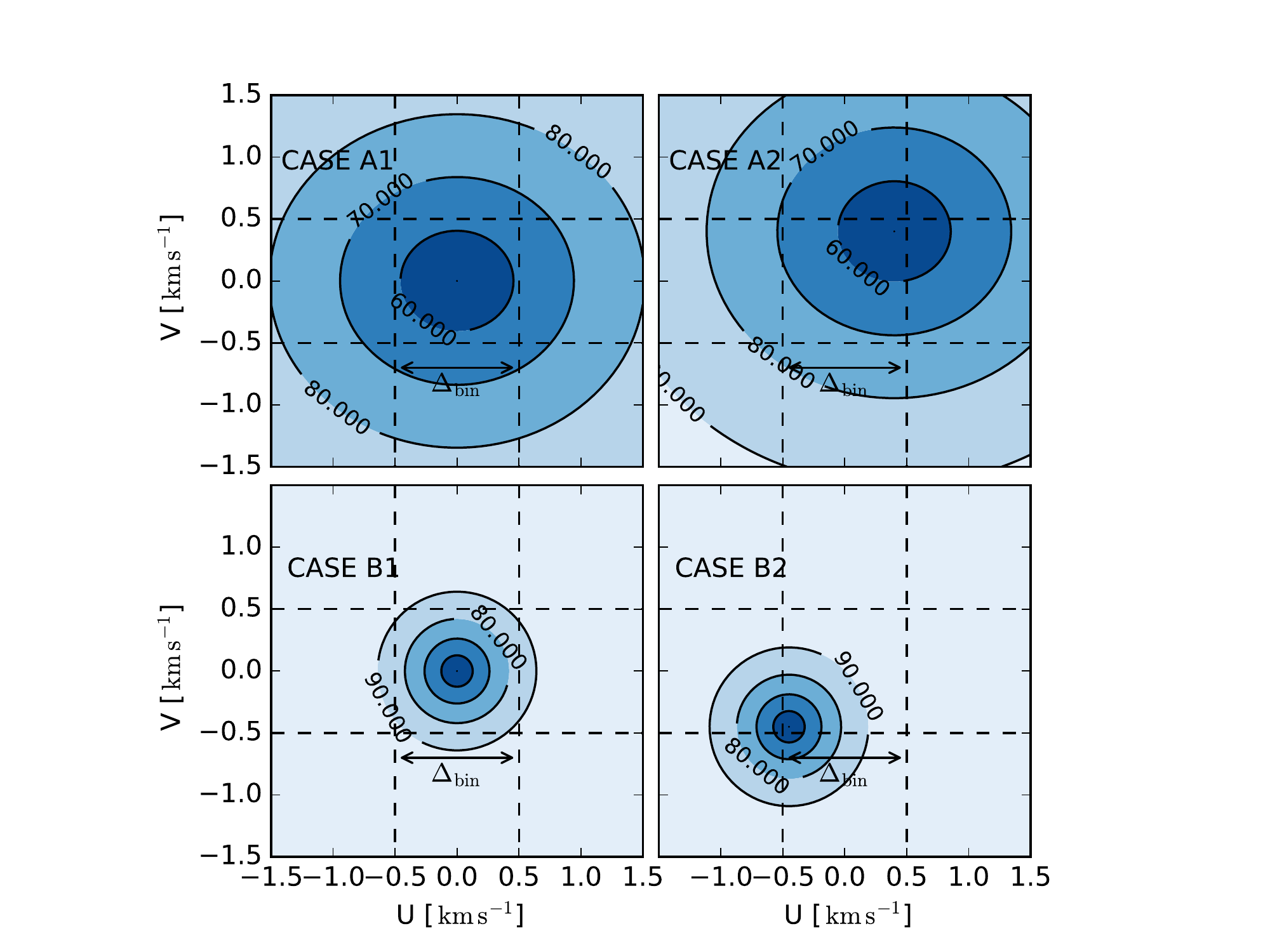}
   } 
   \caption{
   Probability density functions of the location of stars in the $U-V$ plane for different cases and $\Delta_{\rm{bin}}$=1.0 $\kms$. Case A1: the star is located in the centre of a bin and the uncertainty is equal to the average uncertainty of the stars in our sample ($\sigma_U=\langle\sigma_{U}\rangle=1.80$ $\kms$ and $\sigma_V=\langle\sigma_{V}\rangle=1.60$ $\kms$). Case A2: the star is located close to the edge of a bin and the uncertainty is equal to the average uncertainty of the stars in our sample ($\sigma_U=\langle\sigma_{U}\rangle=1.80$ $\kms$ and $\sigma_V=\langle\sigma_{V}\rangle=1.60$ $\kms$). Case B1: the stars is located in the centre of a bin and the uncertainty is lower than the average uncertainty of the stars in our sample ($\sigma_U=\langle\sigma_{U}\rangle=0.90\,\kms$ and
   $\sigma_V=\langle\sigma_{V}\rangle=0.80$ $\kms$). Case B2: the star is located close to the edge of a bin and the uncertainty is lower than the average uncertainty of the stars in our sample ($\sigma_U=\langle\sigma_{U}\rangle=0.90$ $\kms$ and $\sigma_V=\langle\sigma_{V}\rangle=0.80$ $\kms$). Numbers at each concentric circle show the probability in percents for the star to fall inside the circle.
   }
   \label{_bins}
   \end{figure}

\section{Analysis}
\label{sec:maps}


\subsection{Input data}
\label{sec:sec:input}

The constraints on velocity uncertainties and the choice of the bin size of the input star count map are linked. First, the uncertainties have to be at the same time large enough in order to provide us with as a large sample as possible, and at the same time small enough to take advantage of the high-precision data provided by {\it Gaia} DR1/TGAS and RAVE. Second, the bin size of the star count map has to be consistent with the space velocity uncertainty of the stars in order to get robust positions of the structures.

This means that the bin size needs to be roughly equal to the average velocity uncertainty of the sample, otherwise the probability that a star falls into the particular bin will be reduced and therefore the precision of the positions of kinematic structures will also decrease. If the bin size is higher than $\sim$5\,$\kms$, the first scale ($J=1$) would be $10-20\,\kms$, but from previous studies it has been shown that the typical size of structures is of the order $10\,\kms$ \citep[e.g.][]{_antoja12}. Thus, a bin size larger than about 5\,$\kms$ should not be used as too many structures would be lost. With a restriction on $\sigma_U$ and $\sigma_V$ equal to $4\,\kms$ it should possible to get robust measurements of positions of structures, and that leaves us with a sample of 55\,831 stars that have average velocity uncertainties of $\langle\sigma_{U}\rangle=1.8\,\kms$ and $\langle\sigma_{V}\rangle=1.6\,\kms$. We then chose a bin size of $\Delta_{\rm{bin}}=1 \kms$. With this value the scales of the output images from the wavelet transform will be: $J=1$ (2-4\,$\kms$), $J=2$ (4-8\,$\kms$), $J=3$ (8-16\,$\kms$), $J=4$ (16-32\,$\kms$), $J=5$ (32-64\,$\kms$). 

\subsection{Monte-Carlo simulations}
\label{sec:sec:MC}

The space velocities of the stars have uncertainties that will influence the ability to detect kinematic structures and how robust these detections will be. Figure~\ref{_bins} shows the probability density function of one star to be located in the centre of one bin in $U-V$ plane (plots on the left-hand side) or at the edge of the bin (plots on the right-hand side), given that the velocity uncertainties are equal to the average uncertainties of the sample (upper plots) or half of the average uncertainties (lower plots). The probability (see numbers at each concentric circle) that a star can fall into different bins is non-zero and consequently, can lead either to that structures being `fake detections', or miss the detection of real physical structures. Hence, we perform Monte-Carlo (MC) simulations in order to estimate the robustness of the detected structures.

$N_{MC}$ synthetic samples are generated from the original one by randomly drawing 55\,831 new couples $(U,V)$ assuming that the stars have Gaussian velocity distributions, where mean values are positions $(U_i,V_i)$ and the standard deviations are uncertainties $(\sigma_{U_i},\,\sigma_{V_i})$, where $i\in[1,\,N_{\rm{stars}}]$ refers to the $i^{\rm{th}}$ star in the original sample. Then, the wavelet analysis and the structure detection algorithm are applied to the $N_{MC}$ synthetic stellar samples and the positions of all structures at all scales are stored for each simulated sample.

\subsection{Output data}
\label{sec:sec:output}
    
Following the computations described in Sect.~\ref{sec:numerical_methods} and MC simulations as in Sect.~\ref{sec:sec:MC}, the wavelet analysis was applied to these $N_{MC}$ samples giving:
    (1) $N_{MC}$ sets of wavelet coefficients $\left[\left(w_{1}^{1},\,w_{2}^{1},\,...,\,w_{J}^{1}\right),\,...,\left(w_{1}^{N_{MC}},\,w_{2}^{N_{MC}},\,...,\,w_{J}^{N_{MC}}\right)\right]$;
    (2) the multi-resolution support for $J$ scales and $N_{MC}$ simulations, which gives: $\left[\left(M_{1}^{1},\,M_{2}^{1},\,...\,, M_{J}^{1}\right),\,..., \left(M_{1}^{N_{MC}},\,M_{2}^{N_{MC}},\,...,\,M_{J}^{N_{MC}}\right)\right]$;  
    (3) positions of detected structures for $J$ scales and $N_{MC}$ simulations.

\begin{figure}
   \centering
   \resizebox{\hsize}{!}{
   \includegraphics[viewport= 20 30 550 350,clip]{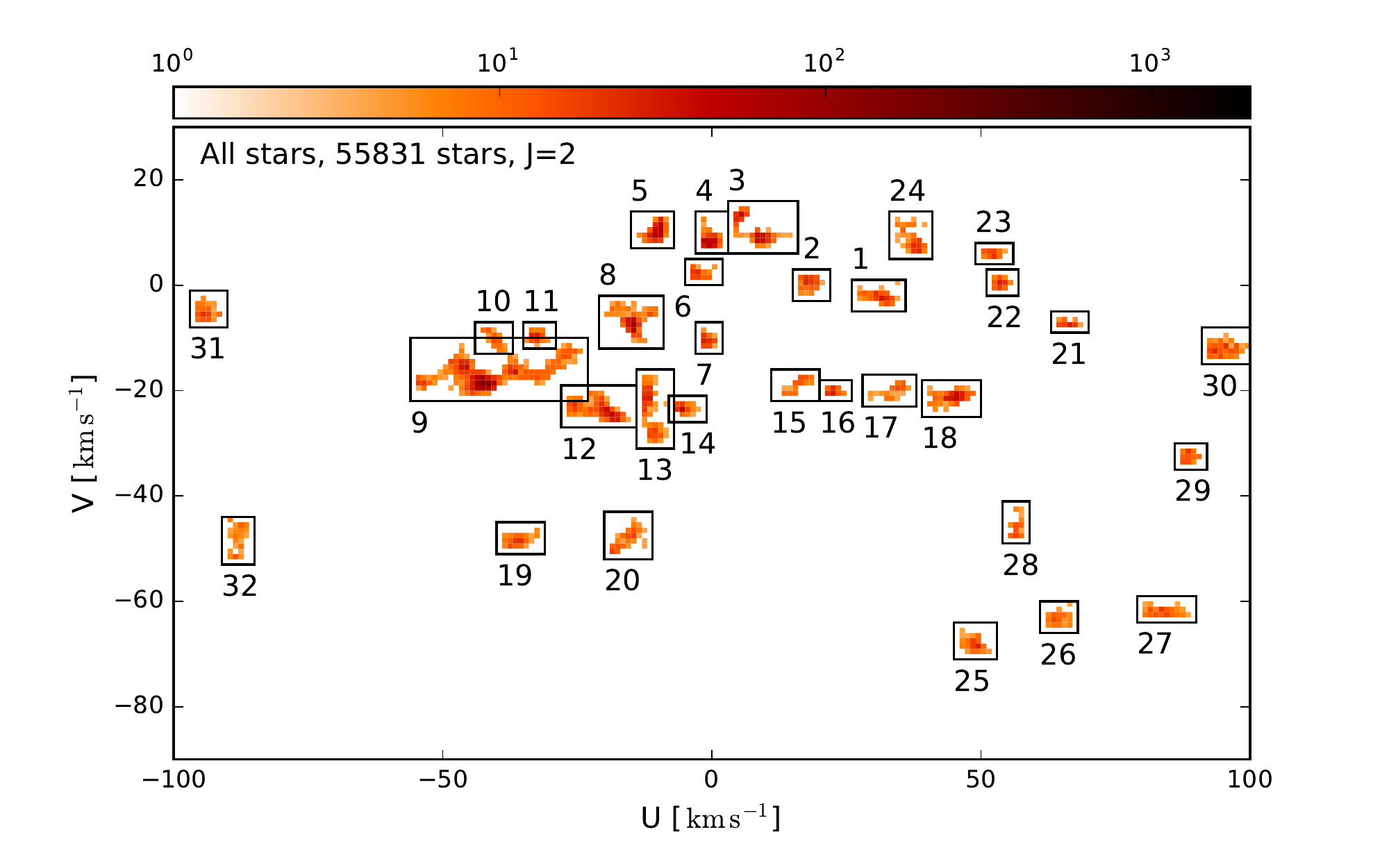}}
   \resizebox{\hsize}{!}{
   \includegraphics[viewport= 20 30 550 360,clip]{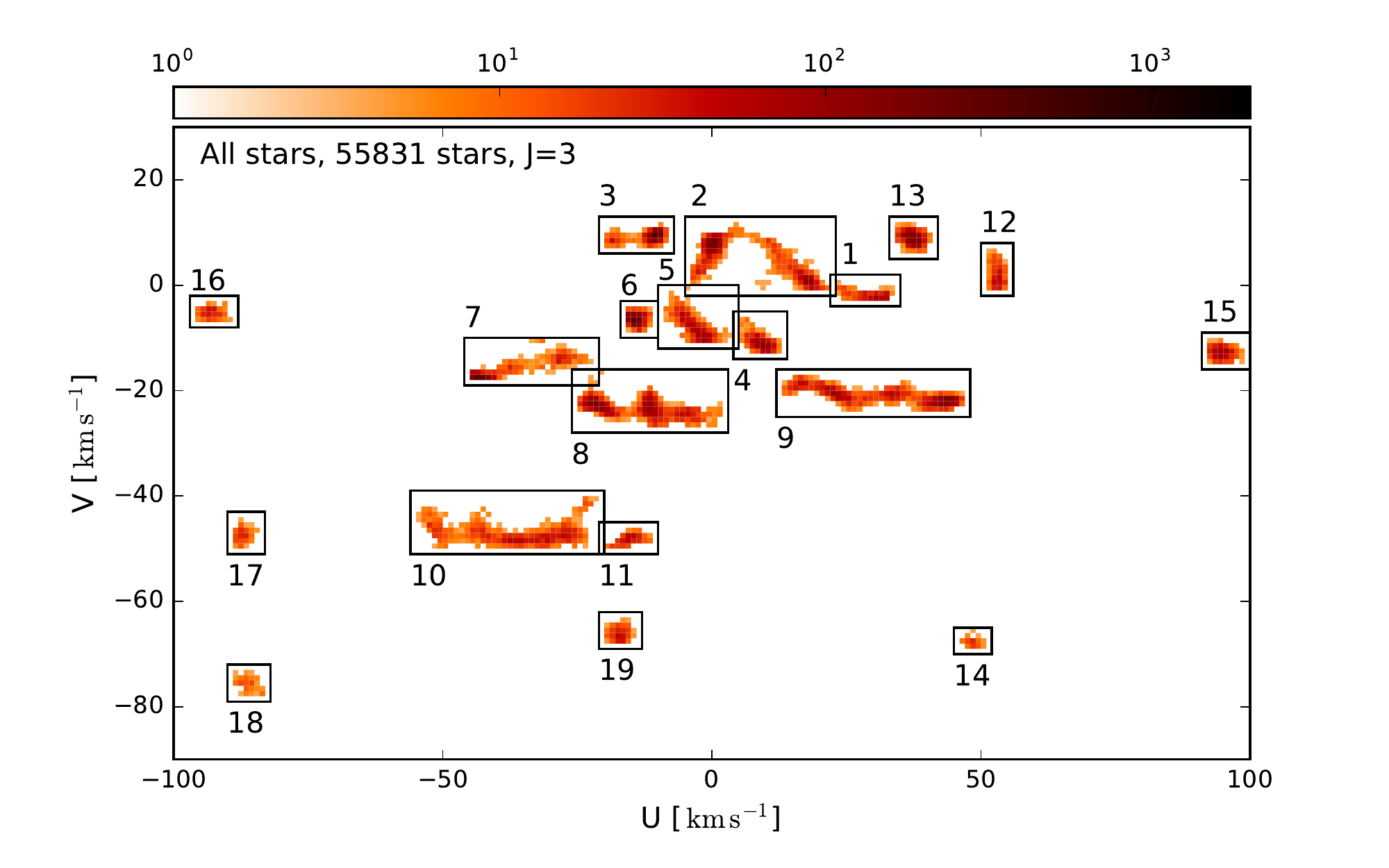}}
   \resizebox{\hsize}{!}{
   \includegraphics[viewport= 20 10 550 360,clip]{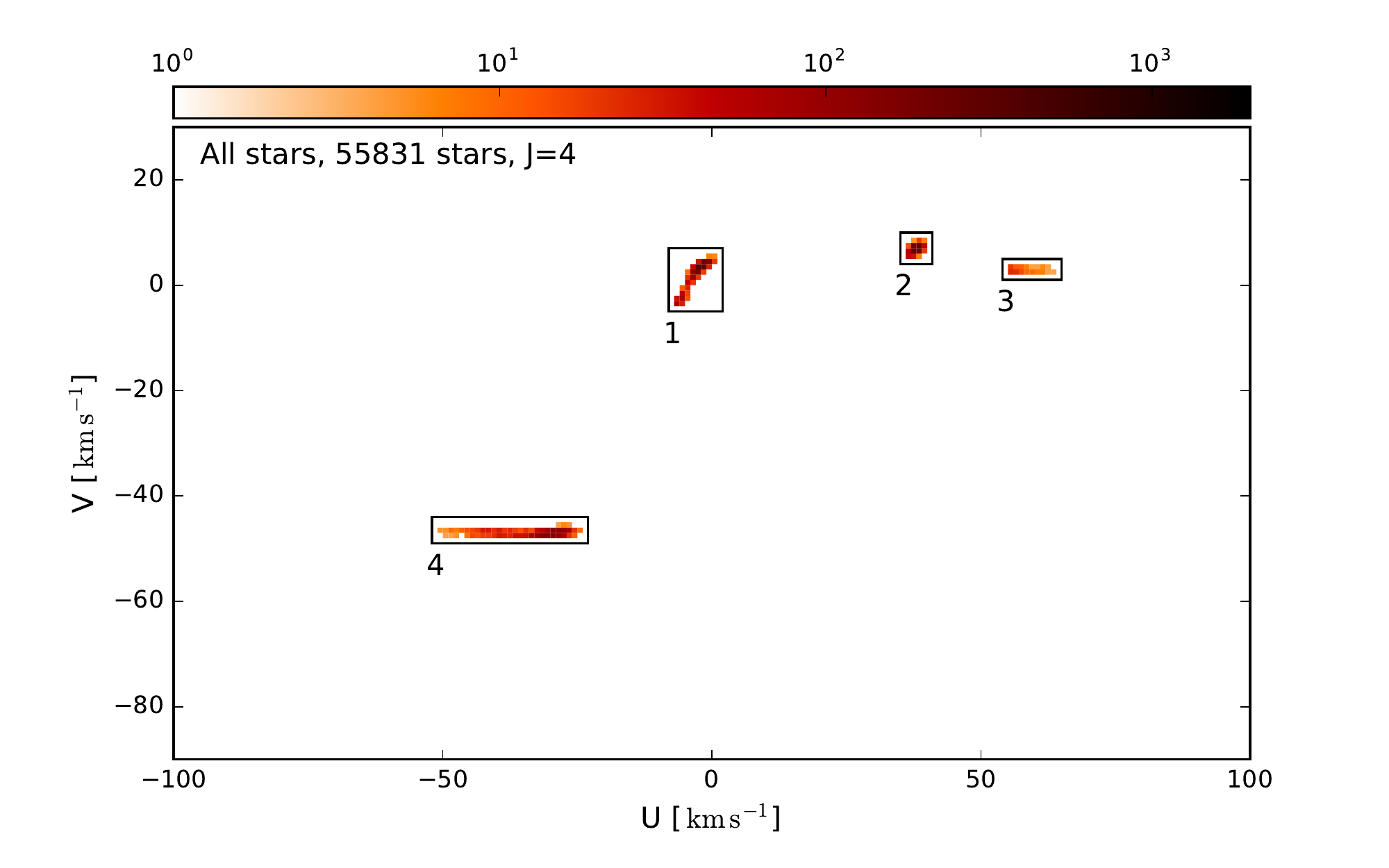}}
   \caption{Positions of kinematic stellar structures obtained by wavelet transform applied for $N_{MC}=2000$ synthetic data samples for levels $J=2,3,4$.
   \label{_uv_all}}
\end{figure}

The results are presented in two different forms. First as a structure's position count map, in which the positions of the detected structures of each of the 2\,000 samples are superimposed (see Fig.~\ref{_uv_all}). The detected structures are marked by black boxes. To quantify the `realness' of each group, the fraction of times each group was detected relative to the total number of simulations is calculated. 

Figure~\ref{_uv_all} shows the position count map for the detected structures at scales $J=2$ (4-8$\kms$, top plot), $J=3$ (8-16$\kms$, middle plot) and $J=4$ (16-32$\kms$, bottom plot). The highest number of individual structures, shown by black rectangles with individual numbers, is for $J=2$. However, as can be seen, scale $J=3$ also includes all significant structures detected at scales $J=2$ and $J=4$, and covers smaller and bigger scales.

\begin{figure}
   \centering
   \resizebox{1.0\hsize}{!}{
   \includegraphics[viewport = 40 80 680 1100,clip]{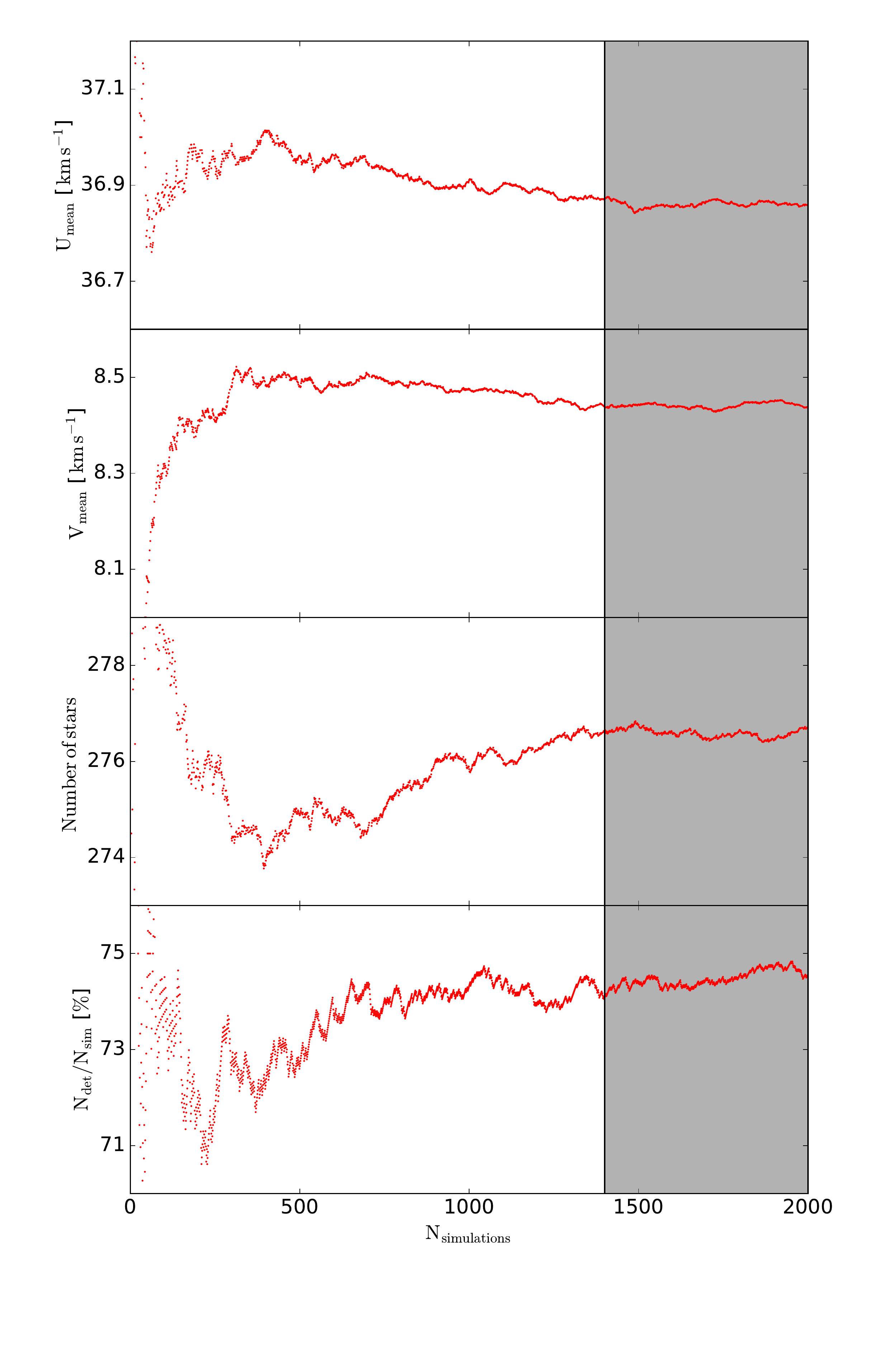}
   }
   \caption{Average position ($U_{\rm{mean}}$, $V_{\rm{mean}}$) (1 and 2 pattern from the top), estimation of the number of stars (3 pattern from the top) and percentage of detection of the moving group 13 from the total sample and $J=3$ (bottom pattern), in function of the number of simulation.
   }
   \label{_convergence}
\end{figure}

How many Monte Carlo simulations are enough for the results to convergence? To explore this, Fig.~\ref{_convergence} shows how the positions for structure number 13 from the $J=3$ map converge as the number of Monte Carlo simulations increases. We introduce four different estimators: The first two are mean positions of the structure $U_{mean}$ and $V_{mean}$ (calculated based on coordinates $U$ and $V$ of all structures inside the rectangle number 13); The third one is the number of stars inside structure number 13 which was calculated as an averaged number of stars from the total number of Monte Carlo simulations ($N_{MC}$ runs); The last estimator is the percentage of structure detection inside the rectangle. Convergence is reached at around 1400 simulations (marked by grey background in Fig.~\ref{_convergence}). We therefore chose to run 2\,000 simulations to have confident results. 

\begin{figure*}
\centering
\resizebox{\hsize}{!}{
\includegraphics[viewport= 5 0 780 340,clip]{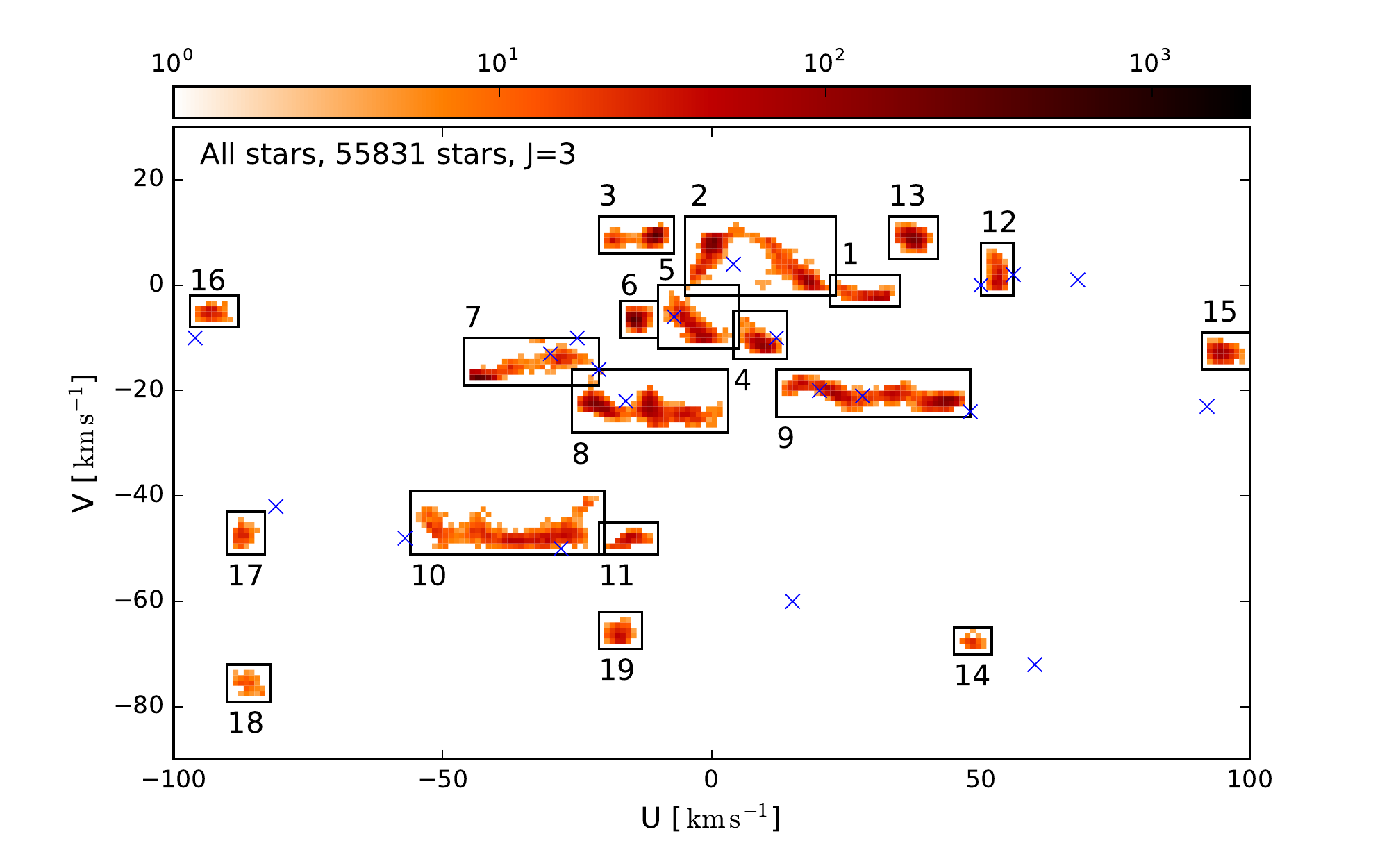}
}
\vspace{-12cm}
\resizebox{\hsize}{!}{
\includegraphics[viewport= 20 0 750 340,clip]{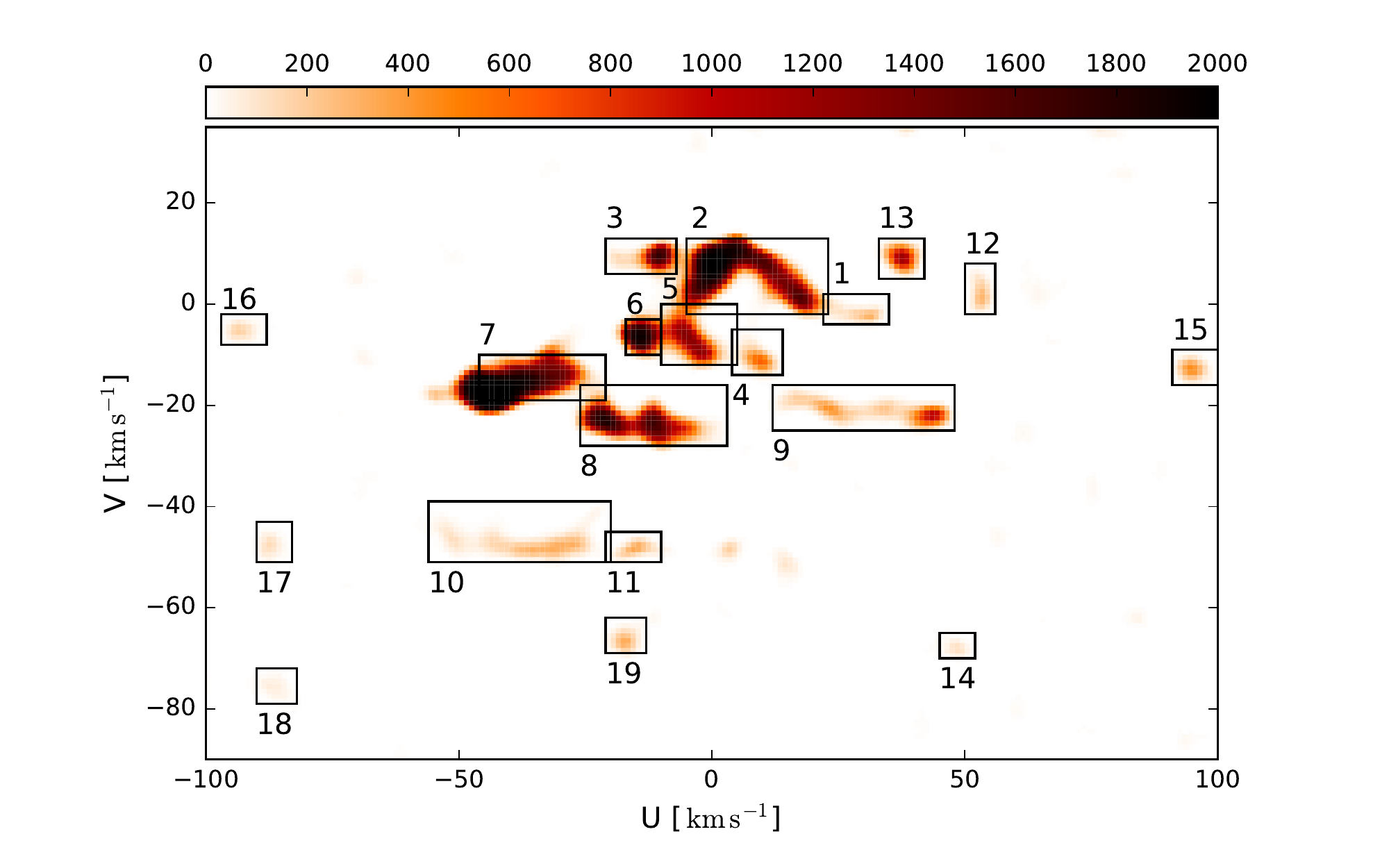}
}
\resizebox{\hsize}{!}{
\includegraphics[viewport= -600 -170 280 400,clip]{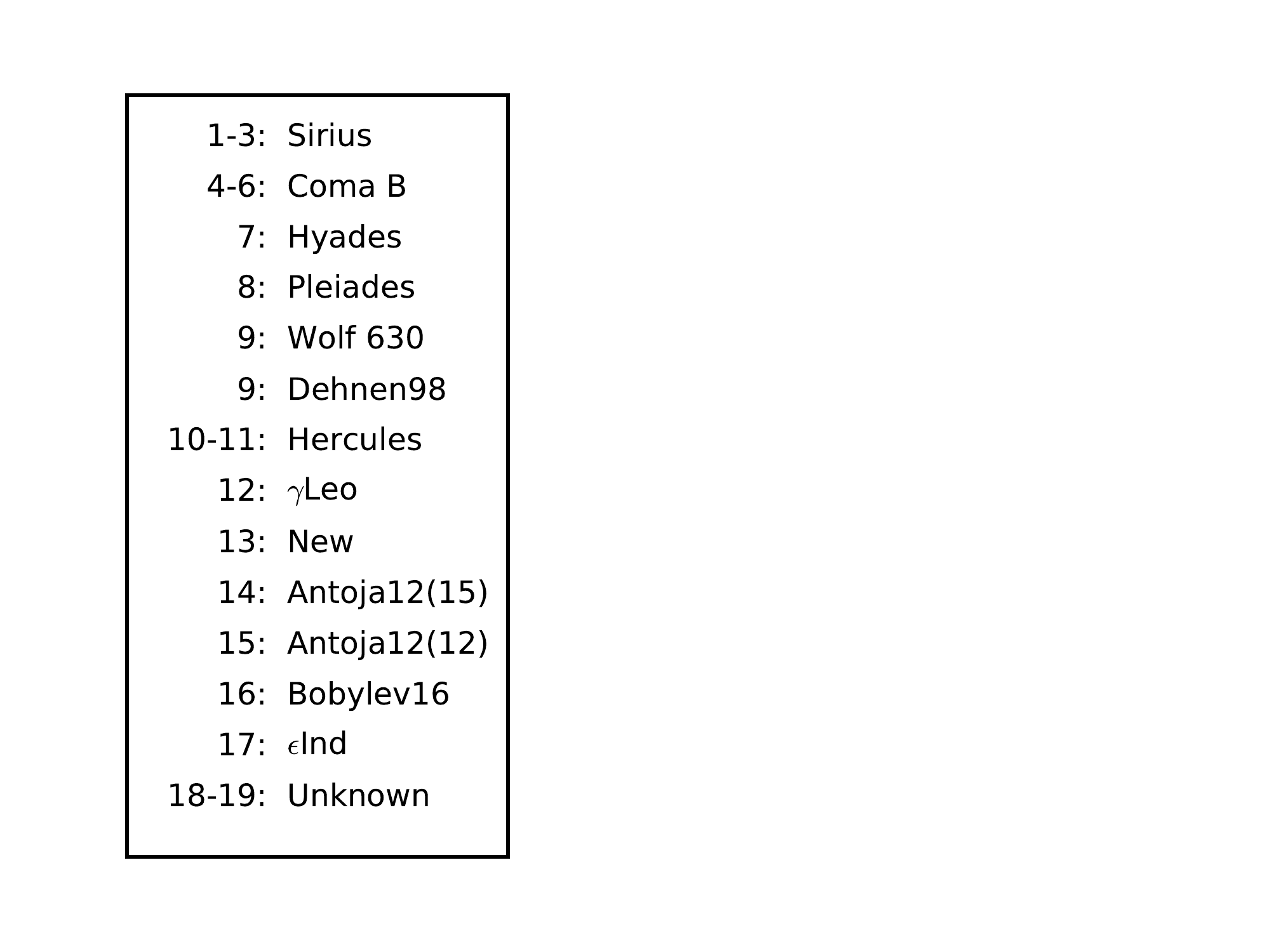}
}
\caption{Top: positions of kinematic stellar structures obtained by wavelet transform applied for $N_{MC}=2\,000$ synthetic data samples for $J=3$ in the $U-V$ plane. Structure counts are shown with the orange colour. Black boxes embrace region of individual structures. Blue crosses show identification of structures in literature if any. 
Bottom: multi-resolution support count map for $N_{MC}=2\,000$ synthetic data samples for $J=3$ in the $U-V$ plane. Black boxes represent the same structures as in the top plot.
\label{_j3}
}
\end{figure*}

The position count map is useful for providing positions of structures. However, one cannot justify if the structures are independent, or are connected to other groups. Hence, another way to represent the results is shown in the bottom plot of Fig.~\ref{_j3}, and is the multi-resolution support for $N_{MC}$ simulations by displaying the quantity M$_{\rm{tot}}$ defined as follows: 
\begin{equation}
     {M_{tot,j}}(i_x,i_y)=\sum_{k=1}^{N_{MC}}M_{j}^{k}(i_x,i_y)
\end{equation}
Thus, if $M_{tot,j}(i_x,i_y)=N_{MC}$, it means that $w_{j}(i_x,i_y)$ is significant for all the simulations. Conversely, if $M_{tot,j}(i_x,i_y)=0$, it means that $w_{j}(i_x,i_y)$ is never significant. We explain in more details results that can be gained from Fig.~\ref{_j3} in Sect.~\ref{sec:results}.

\begin{table*}
\centering
\caption{
$3\sigma$-significant kinematic structures detected for level $J=3$, 8-16 $\kms$, $N_{MC}=2\,000$.\tablefootmark{$\dagger$}
\label{_groups}
}
\tiny
\setlength{\tabcolsep}{3mm}
\begin{tabular}{rlrrrrrr|cccc}
\hline 
\hline
N        &
Name     &
$U$      &
$V$      &
$\Delta U$ &
$\Delta V$ &
$\frac{N_d} {N_{MC}}, _{\%}$ &
$N_{*}$ &
$SN$ &
$BSN$&
$D$  &
$TD$ \\
\noalign{\smallskip}
(1) &
(2) &
(3) &
(4) &
(5) &
(6) &
(7) &
(8) &
(9) &
(10) &
(11) &
(12) \\
\noalign{\smallskip}
\hline 
\multicolumn{8}{c|}{55\,831}       &
31\,533                            & 
24\,298                            &
36\,439                            &
11\,410                           \\
\hline 
\noalign{\smallskip}
1   &Sirius            &30   &    $-$3   &3  &1  &31   &434  &$+$  &     &$+$  &$+$ \\
2   &Sirius            &0    &    8      &7  &3  &154  &4821 &$+$  &$+$  &$+$  &$+$ \\
3   &Sirius            &$-$11&    9      &3  &1  &83   &879  &$+$  &     &$+$  &    \\
4   &Coma B            &9    &    $-$12  &2  &2  &52   &1102 &     &$+$  &     &$+$ \\
5   &Coma B            &$-$2 &    $-$11  &3  &3  &70   &2753 &$+$  &$+$  &$+$  &    \\
6   &Coma B            &$-$15&    $-$7   &1  &1  &79   &673  &$+$  &$+$  &$+$  &    \\
7   &Hyades            &$-$44&    $-$18  &6  &2  &90   &2344 &$+$  &     &$+$  &    \\
8   &Pleiades          &$-$22&    $-$23  &7  &3  &170  &4257 &$+$  &$+$  &$+$  &    \\
9   &Wolf630+Dehnen98  &43   &    $-$22  &9  &2  &168  &1777 &$+$  &     &$+$  &    \\
10  &Hercules          &$-$38&    $-$49  &9  &3  &116  &1451 &$+$  &     &$+$  &    \\
11  &Hercules          &$-$16&    $-$48  &2  &1  &22   &197  &$+$  &     &$+$  &    \\
12  &$\gamma$Leo       &52   &    0      &1  &2  &27   &96   &$+$  &     &$+$  &    \\
13  &New               &37   &    8      &2  &2  &74   &201  &$+$  &     &$+$  &$+$ \\
14  &Antoja12(15)      &48   &    $-$68  &1  &1  &6    &8    &$+$  &     &$+$  &    \\
15  &Antoja12(12)      &94   &    $-$13  &1  &1  &38   &10   &$+$  &     &$+$  &    \\
16  &Bobylev16         &$-$94&    $-$5   &1  &1  &17   &14   &$+$  &$+$  &$+$  &$+$ \\
17  &$\epsilon$Ind     &$-$88&    $-$48  &2  &2  &12   &24   &$+$  &     &     &    \\
18  &Unknown           &$-$86&    $-$76  &2  &1  &8    &12   &$+$  &     &     &    \\
19  &Unknown           &$-$18&    $-$67  &1  &1  &22   &70   &$+$  &     &$+$  &    \\
\noalign{\smallskip}
\hline
\end{tabular}
\tablefoot{
\tablefoottext{$\dagger$}{Columns 1-8 are for the total sample. Column 1 gives the order of positions of wavelet coefficients for $J=3$ obtained for 2\,000 synthetic data samples. Column 2 is the name of the structure if available in literature. Columns 3 and 4 are central positions of kinematic structures in $\kms$, their uncertainties (standard deviations) are given in columns 5 and 6 respectively also in $\kms$. Column 7 is a percentage showing how many times the structure obtained by MC simulations appears in the wavelet space. The estimated number of stars in each group is given in column 8. Columns 9-12 show the presence of the structure in the SN, BSN, D and TD samples for $J=3$ with $+$ sign. Number of stars of each data sample is indicated in the row 3. Question marks correspond to tentatively new structures with a small detection percentage in MC simulation.}
}
\end{table*}

\begin{table*}
\centering
\caption{
Kinematic structures detected in the Solar neighbourhood for levels $J=2$ and $J=4$, $N_{MC}=2\,000$. For details see Table~\ref{_groups}.
\label{_groups2}
}
\tiny
\setlength{\tabcolsep}{5mm}
\begin{tabular}{rlrrrrrr}
\hline 
\hline
N        &
Name     &
$U$      &
$V$      &
$\Delta U$ &
$\Delta V$ &
$\frac{N_d} {N_{MC}}, _{\%}$ &
$N_{*}$ \\
(1) &
(2) &
(3) &
(4) &
(5) &
(6) &
(7) &
(8)  \\
\noalign{\smallskip}
\hline 
\noalign{\smallskip}
\multicolumn{8}{c}{$J=2$, 4-8 $\kms$} \\
\hline
\noalign{\smallskip}
1  &   Sirius        &    31   &    $-$3 &    2&    1&  12 & 285  \\
2  &   Sirius        &    17   &       0 &    1&    1&  10 & 377  \\
3  &   Sirius        &    8    &       9 &    3&    2&  23 & 1105 \\
4  &   Sirius        &    $-$1 &       7 &    1&    1&  18 & 511  \\
5  &   Sirius        &    $-$11&       9 &    1&    1&  23 & 496  \\
6  &   Sirius        &    $-$3 &       2 &    1&    1&  7  & 459  \\
7  &   Coma B        &    $-$2 &    $-$11&    1&    1&  7  & 377  \\
8  &   Coma B        &    $-$15&    $-$8 &    3&    2&  26 & 1692 \\
9  &   Hyades        &    $-$43&    $-$19&    8&    3&  108& 3392 \\
10 &   Hyades        &    $-$41&    $-$11&    1&    1&  5  & 257  \\
11 &   Hyades        &    $-$33&    $-$10&    1&    1&  7  & 243  \\
12 &   Pleiades      &    $-$19&    $-$25&    3&    1&  30 & 1262 \\
13 &   Pleiades      &    $-$12&    $-$22&    1&    3&  20 & 1181 \\
14 &   Pleiades      &    $-$6 &    $-$24&    1&    1&  8  & 342  \\
15 &   Wolf 630      &      16 &    $-$18&    2&    1&  6  & 430  \\
16 &   Wolf 630      &      22 &    $-$21&    1&    1&  6  & 153  \\
17 &   Dehnen98      &      34 &    $-$20&    2&    1&  6  & 252  \\
18 &   Dehnen98      &      45 &    $-$22&    2&    1&  10 & 193  \\
19 &   Hercules      &    $-$36&    $-$49&    2&    1&  9  & 178  \\
20 &   Hercules      &    $-$19&    $-$51&    2&    2&  10 & 266  \\
21 &   $\gamma$ Leo  &      66 &    $-$8 &    1&    1&  5  & 14   \\
22 &   $\gamma$ Leo  &      53 &       0 &    1&    1&  7  & 40   \\
23 &   $\gamma$ Leo  &      52 &       5 &    1&    1&  5  & 40   \\
24 &   New           &      38 &       6 &    2&    2&  12 & 201  \\
25 &   Antoja12(15)  &      49 &    $-$69&    1&    1&  9  & 12   \\
26 &   Antoja12(15)  &      63 &    $-$64&    1&    1&  7  & 5    \\
27 &   Antoja12(15)  &      83 &    $-$62&    1&    1&  9  & 5    \\
28 &   Antoja12(15)  &      56 &    $-$48&    1&    1&  5  & 20   \\
29 &   Antoja12(12)  &      88 &    $-$32&    1&    1&  6  & 5    \\
30 &   Antoja12(12)  &      93 &    $-$13&    1&    1&  14 & 10   \\
31 &   Bobylev16     &    $-$95&    $-$6 &    1&    1&  8  & 11   \\
32 &   $\epsilon$Ind &    $-$89&    $-$52&    1&    2&  7  & 20   \\
\hline
\noalign{\smallskip}
\multicolumn{8}{c}{$J=4$, 16-32 $\kms$} \\
\hline
\noalign{\smallskip}
1 &Sirius      &$-$3   &3      &2    &3     &97 &1767\\
2 &New         &38     &7      &1    &1     &60 &106 \\
3 &$\gamma$Leo &55     &2      &2    &1     &8  & 29 \\
4 &Hercules    &$-$32  &$-$48  &7    &1     &100&437 \\
\noalign{\smallskip}
\hline
\end{tabular}
\end{table*}

\section{Results}
\label{sec:results}

In this section we present the detected structures in the $U-V$ plane for the following samples: the full TGAS-RAVE sample, the sample split into a nearby and a distant sample, and two chemically defined samples that to a first degree represent the stars belonging to the Galactic thin and thick disks.

\subsection{Full sample}
\label{sec:sec:full}

Figure~\ref{_uv_all} shows detected structures in the $U-V$ planes for three different scales, $J=2$ (4-8 $\kms$): 32 structures, $J=3$ (8-16 $\kms$): 19 structures, and $J=4$ (16-32 $\kms$): 4 structures. The $J=3$ structures are listed in Table~\ref{_groups}, and the structures from the $J=2$ and $J=4$ scales in Table~\ref{_groups2}. As can be seen, $J=3$ appears to cover all the detected features, including smaller structures at $J=2$, as well as larger groups at $J=4$. Therefore, we will from now on consider $J=3$ as the main scale since it covers the a range around the typical sizes of kinematic structures found in the Solar neighbourhood (both small- and big-scale structures), and secondly focus on $J=2$ and $J=4$ that covers even smaller and larger structures, respectively.


The top plot of Fig.~\ref{_j3} shows again the detected kinematic structures in the $U-V$ plane for $J=3$ (as in the Fig.~\ref{_uv_all} middle plot), but now with previously detected structures found in the literature (\citet{_eggen96, _antoja08, _antoja12, _bobylev16}) marked with blue crosses. Classical structures such as Sirius (structures number $1-3$ in Fig.~\ref{_j3}), Coma Berenices (structures $4-6$), Hyades (structure 7), Pleiades (structure 8) and Hercules (structures $10-11$), and some smaller structures like Wolf 630 (structure 9), Dehnen98 (structure 9), $\gamma$Leo (structure 12) can be easily recognised. They all have a comparably high percentage of detection (column 7 in Tab.~\ref{_groups}) and big number of stars (column 7 in Tab.~\ref{_groups}). The two structures from \cite{_antoja12} (structures 14 and 15) and one structure from \citet{_bobylev16} (structure 16) are confirmed. We also present evidence for a new structure (number 13) that is detected with 74\% significance. Structures $18-19$ have low percentages of detection, less than$15\,\%$, and might be insignificant. In Sect.~\ref{sec:sec:individual} we will discuss how our results agree with those from the literature.

The way the detected structures are split into groups is motivated with the bottom of Figure~\ref{_j3} which shows the multi-resolution support obtained for $J=3$ for all stars and 2\,000 MC simulations. In other words, this is the same plot as the top Figure~\ref{_j3}, but instead of structure counts we show multi-resolution support counts. This representation allows to see whether structures are bound or separated. Structures $1-3$ seem to be connected and thus are united into Sirius stream. Group 5 is bound to structure 2 in the wavelet case. It should not be associated with the Sirius stream as its most significant part is located slightly aside Sirius, but lays on one line with structures 4 and 6, therefore grouping structures $4-6$ into the Coma Berenices stream. Groups that have percentage detection higher than 100\% (8, 9, 10) show a few distinct peaks in this plot, supporting the statement that these groups consist of a few smaller structures that overlap in the structure count map. Based on that we split group 9 into Wolf 630 (to the left) and Dehnen98 (to the right). Group 11 is a part of the Hercules stream. Structures 12-19 are not connected to other groups.

\subsection{Solar neighbourhood and beyond samples}
\label{sec:sec:sn}

The detected structures are found in velocity space. The question is if they depend on the distance from the Sun? We divide the sample into a nearby Solar neighbourhood sample with 31\,533 stars that have distances $d<300$\,pc (SN), and a beyond the Solar neighbourhood sample (BSN), with 24,\,298 stars that have $d>300$\,pc (most distant star at 2\,kpc). Distance $d=300$\,pc that is arbitrarily chosen to split the sample, is also a reasonable value, because it allows to have almost equal number of stars in both samples. Both samples are then independently analysed in the same way as for the full sample: applying the wavelet transform, filtering, and structure detection procedure for 2\,000 synthetic data samples.

\begin{figure*}
   \centering
   \resizebox{\hsize}{!}{
   \includegraphics[viewport = 0  30 550 360,clip]{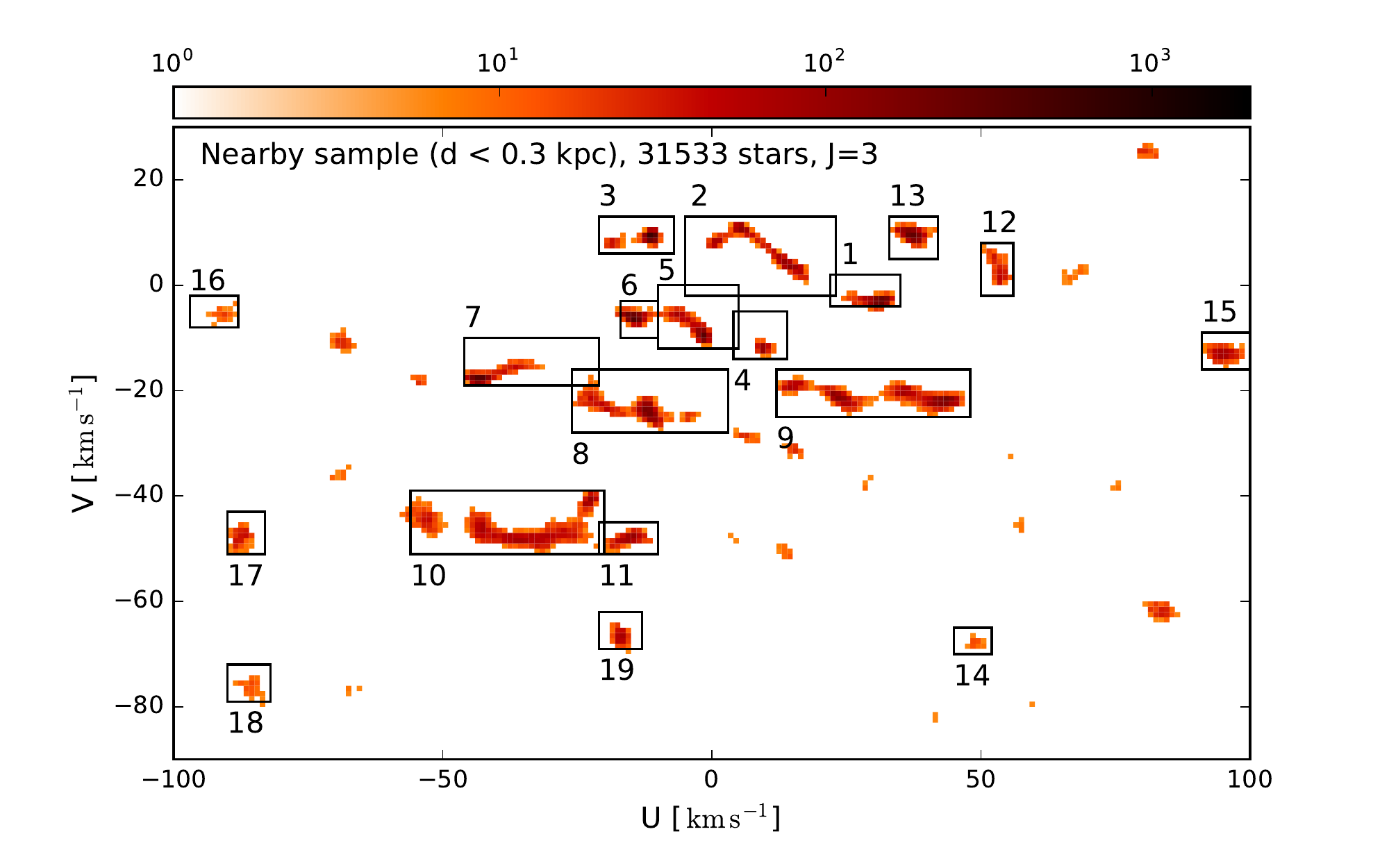}
   \includegraphics[viewport = 45 30 550 360,clip]{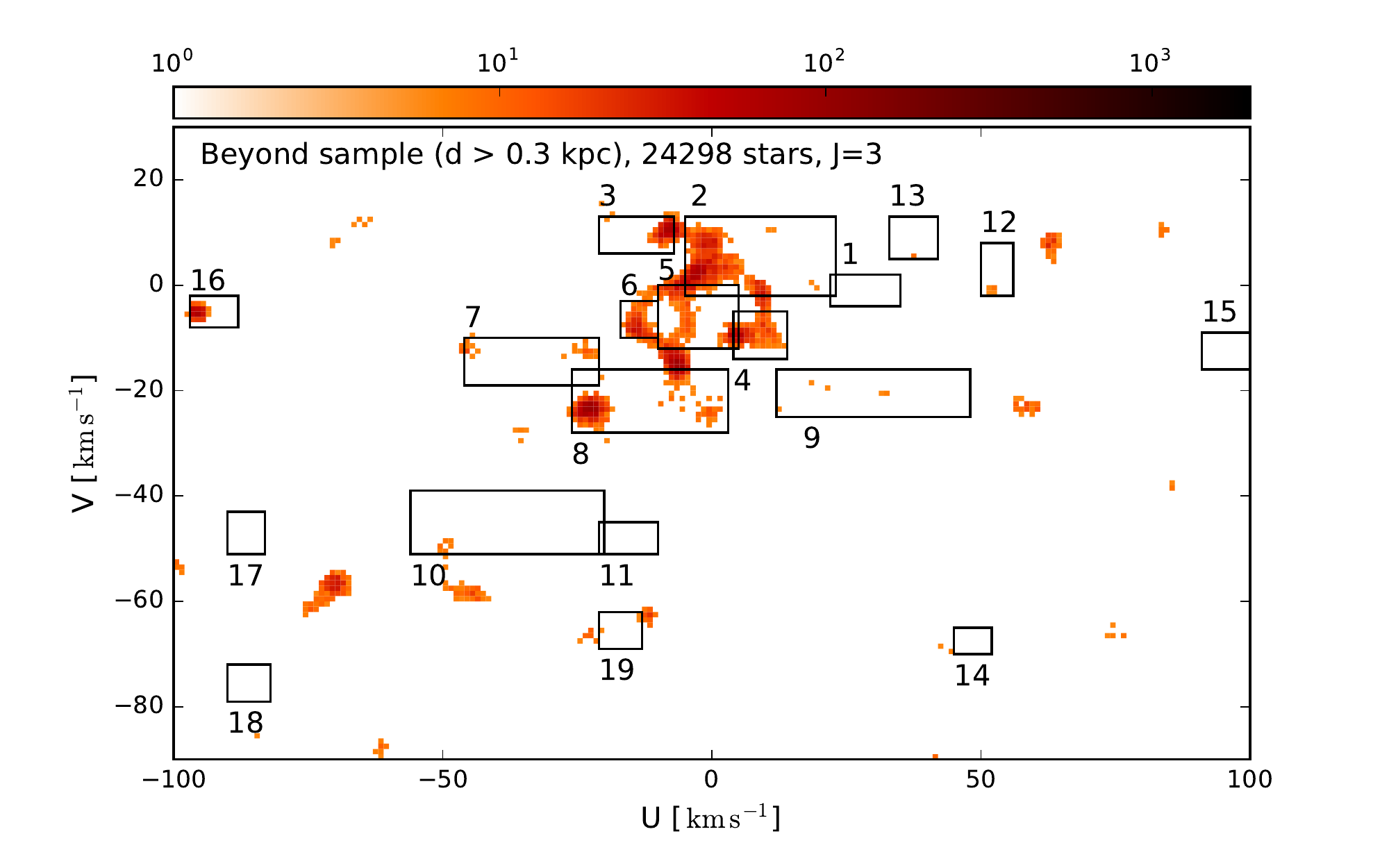}
   }
   \resizebox{\hsize}{!}{
   \includegraphics[viewport = 0  0 550 360,clip]{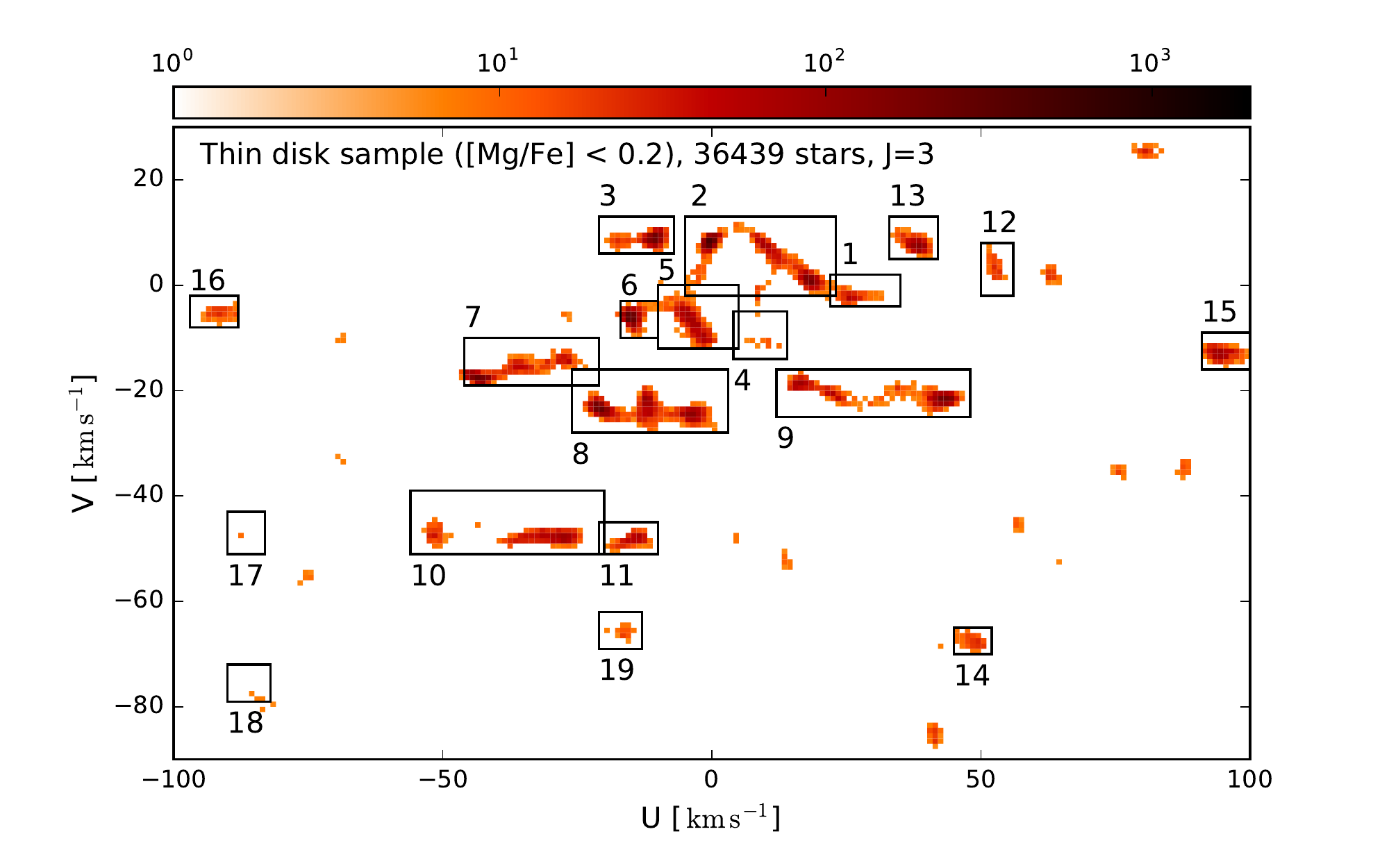}
   \includegraphics[viewport = 45 0 550 360,clip]{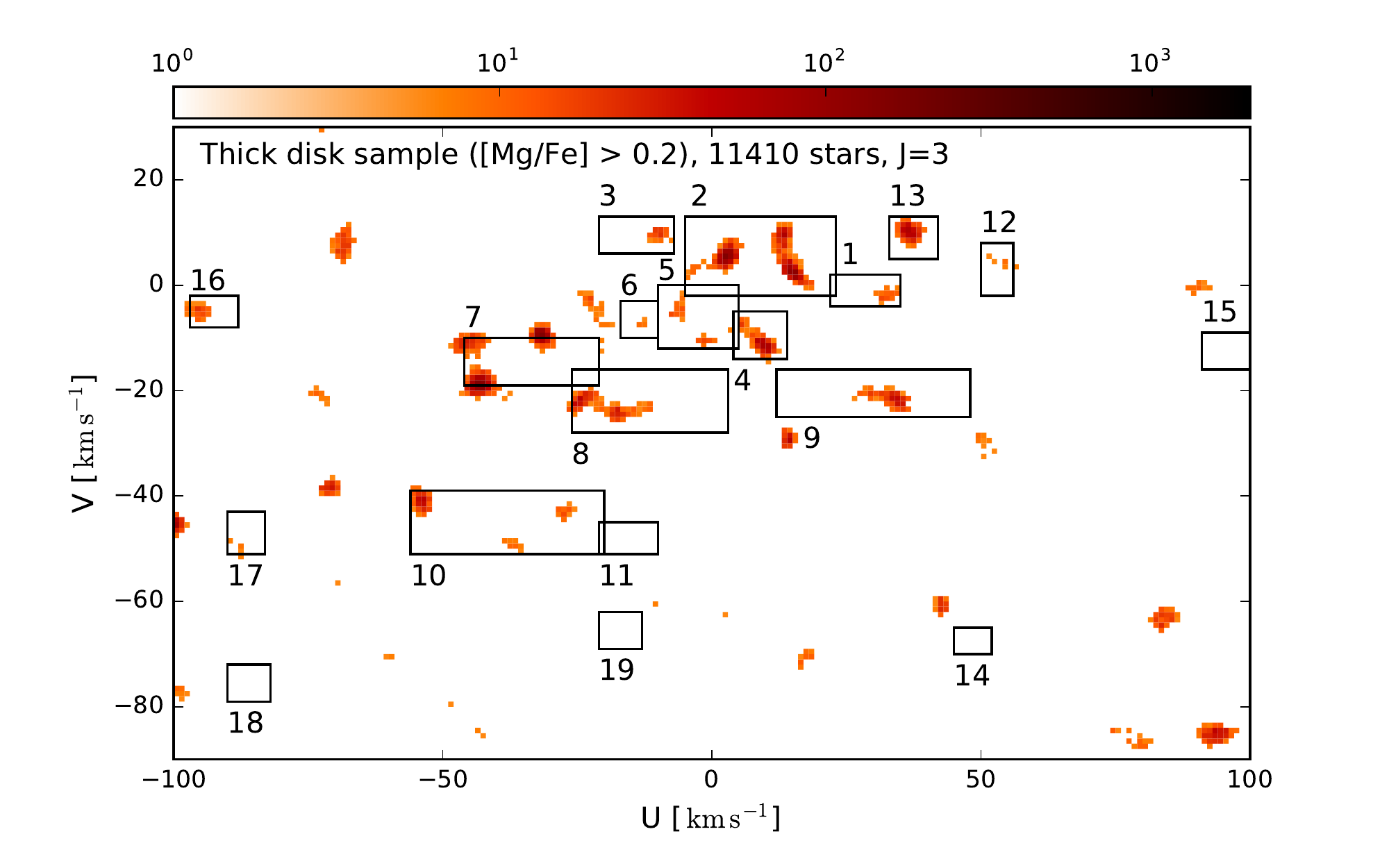}
   }
   \caption{
   Positions of kinematic stellar structures obtained by wavelet transform applied for $N_{MC}=2\,000$ synthetic data samples for level $J=3$. Top left plot is for SN sample (31\,533 stars), top right is for BSN sample (24\,298 stars), bottom left is for D sample (36\,439 stars) and bottom right is for TD sample (11\,410 stars). Black boxes embrace regions of detected structures for the total sample and $J=3$.
   \label{_samples}
   }
\end{figure*}

Figure~\ref{_samples} shows the detected structures associate with the SN sample (top left plot), and the BSN sample (top right plot) for the scale $J=3$. The rectangles mark the borders for the structures that were detected for the full sample (see Fig.~\ref{_j3}). This allows an easier comparison how the shapes on kinematic structures change with the respect to the full sample.  

In Table~\ref{_groups} we have indicated in columns 9 and 10 with ``$+$'' signs if the structure is present in SN and BSN samples. Almost of all of the full sample structures are observed in the SN except two weak Hyades peaks (groups 10, 11). So, the SN results almost completely reproduce the results from the full sample. For the BSN sample that has 7\,000 less stars than the SN sample, most of structures appear to have slightly changed their positions relative to the SN case. Similar result was obtained by \citet{_antoja12} where the structures detected in distant regions were shifted on the velocity plane. Hence, only 6 of 19 kinematic groups can be recognised: strong Sirius peaks 2, all Coma B peaks (4-6), Bobylev16 peak 16, Pleiades peak 8 is slightly shifted. 

In summary, it appears that some kinematic structures are located only in the SN sample as a few significant groups are not detected in the BSN sample at all (groups 1, 3, 7, 9-15, 17-19). These changes in the number of structures, their positions and shapes in the respect to distance can be due to a different numbers of stars that fall into SN and BSN samples with the SN sample containing 10\,000 stars more. The technique of wavelet analysis is sensitive to the number of stars in the initial sample, the more stars we have, the more realistic picture of structures we can get. Mean values of velocity uncertainties for two samples are also slightly different and are bigger in the case of the BSN sample: $\langle\sigma_{U}\rangle_{SN}=1.7$, $\langle\sigma_{V}\rangle_{SN}=1.6$ for the SN sample; $\langle\sigma_{U}\rangle_{BSN}=2.5$, $\langle\sigma_{V}\rangle_{BSN}=2.2$ for the BSN sample. So that for the BSN sample, which is at the same time smaller, velocity uncertainties are slightly higher and this can lead to some displacements of the structures. This issue can be investigated further with the availability of the {\it Gaia} DR2 in April 2018 which will provide precise astrometric parameters for $10^9$ stars and first radial velocities for bright stars. 

\subsection{Thin and thick disk structures}
\label{sec:disks}

Several high-resolution spectroscopic studies of nearby stars have identified and characterised the thin and thick disks as distinct stellar populations, not only in terms of kinematics, but also in terms of elemental abundances and stellar ages \citep[e.g.][]{reddy2006,fuhrmann2008,adibekyan2012,_bensby14}. The two co-existing and largely overlapping disk populations points to a complex formation history for the Milky Way. A process which is currently is not well understood. The question is if we can gain further insights into the nature and origin of this two-disk structure from the kinematic structures seen in the Solar neighbourhood?

As shown in \cite{_bensby14} and \cite{haywood2013} stellar ages appear to be the best discriminator between the thin and thick disks. However, stellar ages are not available for the stars in our sample. Another way would be to use kinematics, but as this is exactly the property that we want to investigate. Another approach which can reveal more features of kinematic group associated with the thin and thick disks is to use their chemical compositions. Several papers have shown that the two disks follow distinct and well separated abundance trends both in the Solar neighbourhood \cite[e.g.][]{_bensby14,adibekyan2012,fuhrmann2008} and also further away \citep{bensby2011letter,hayden2015}. All these studies show that thick disk stars, at a given metallicity, are more $\alpha$-enhanced than thin disk stars. 

In this paper we do not perform any spectroscopic analysis of our structures. Instead we separate our stellar sample by magnesium [Mg/Fe] abundances provided by RAVE in order to study our sample in terms of thick (metal-poor) and thin (metal-rich) disks. 


\begin{figure*}
   \centering
   \resizebox{0.95\hsize}{!}{
   \includegraphics[viewport = -21 64 430 160,clip]{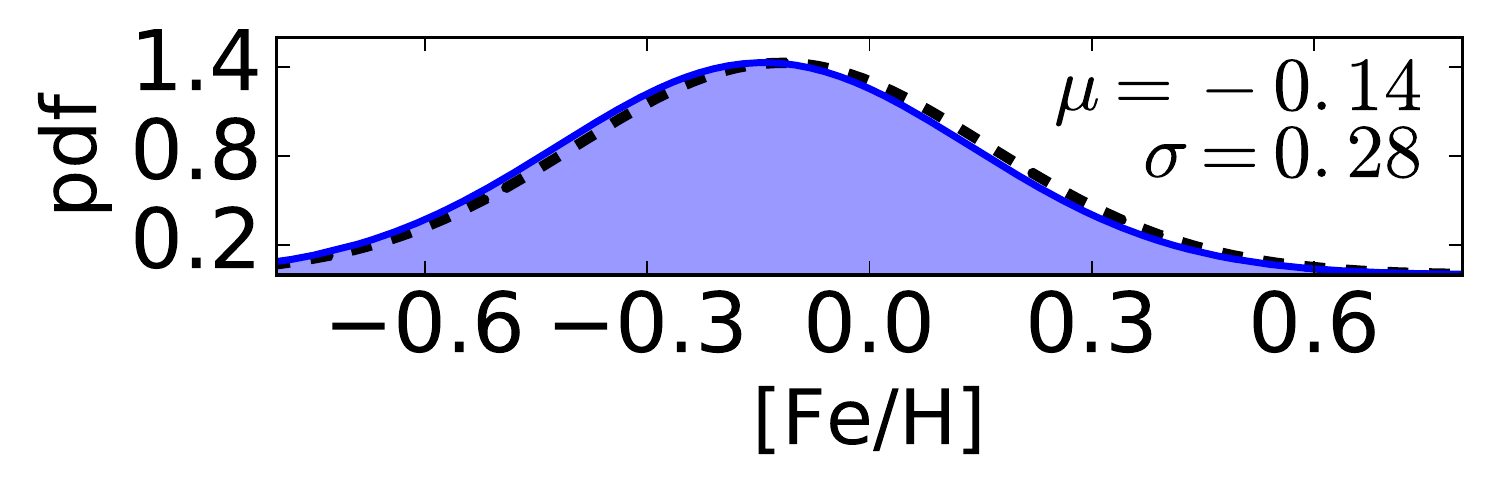}
   \includegraphics[viewport = 77  64 430 160,clip]{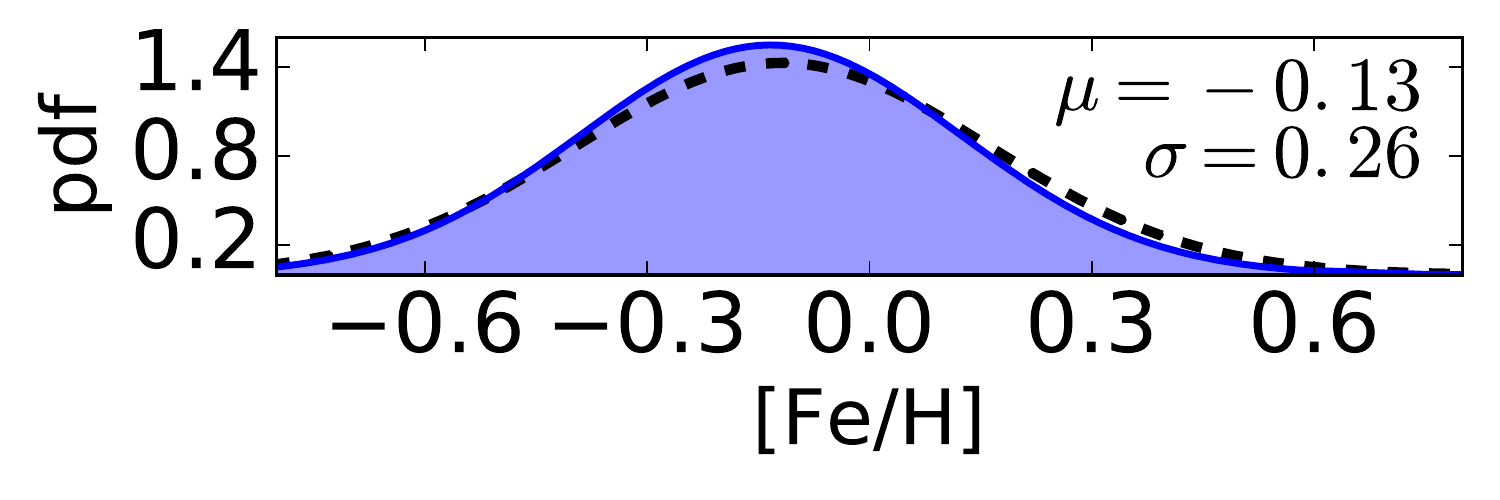}
   \includegraphics[viewport = 77  64 430 160,clip]{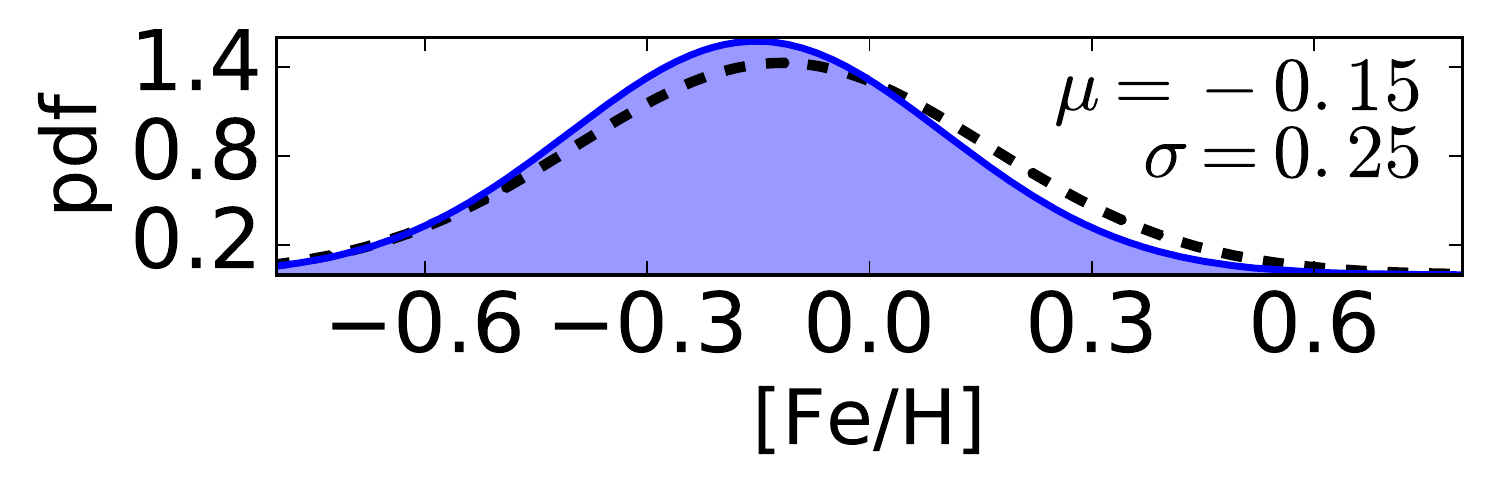}
   \includegraphics[viewport = 77  64 430 160,clip]{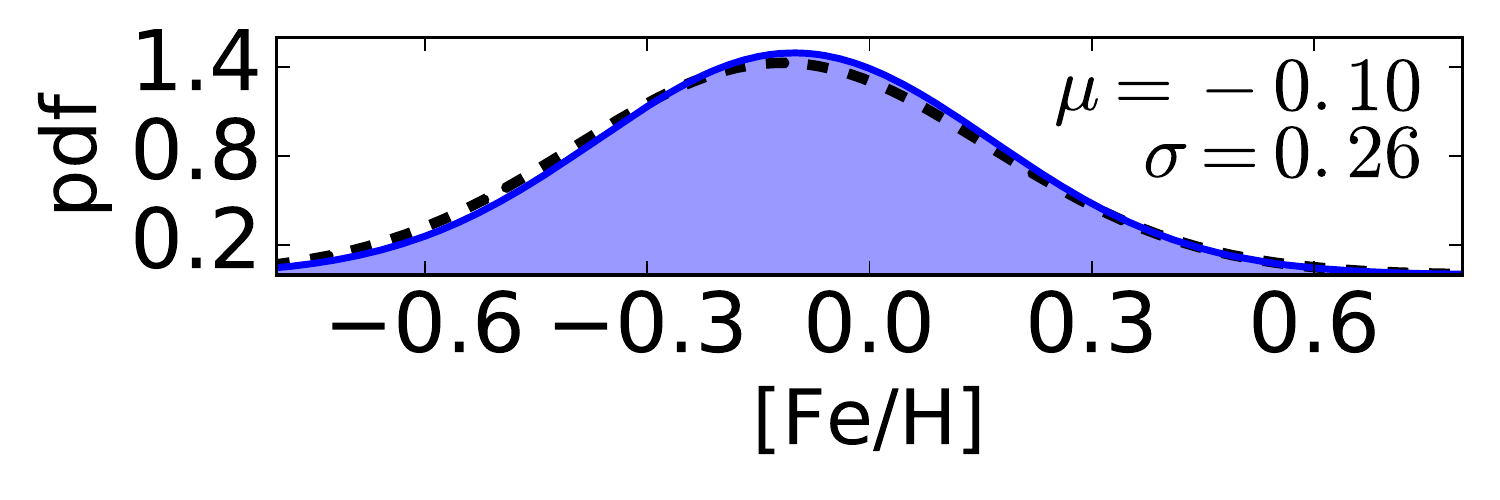}
   }
   \resizebox{0.95\hsize}{!}{
   \includegraphics[viewport = 0  59.5 430 279,clip]{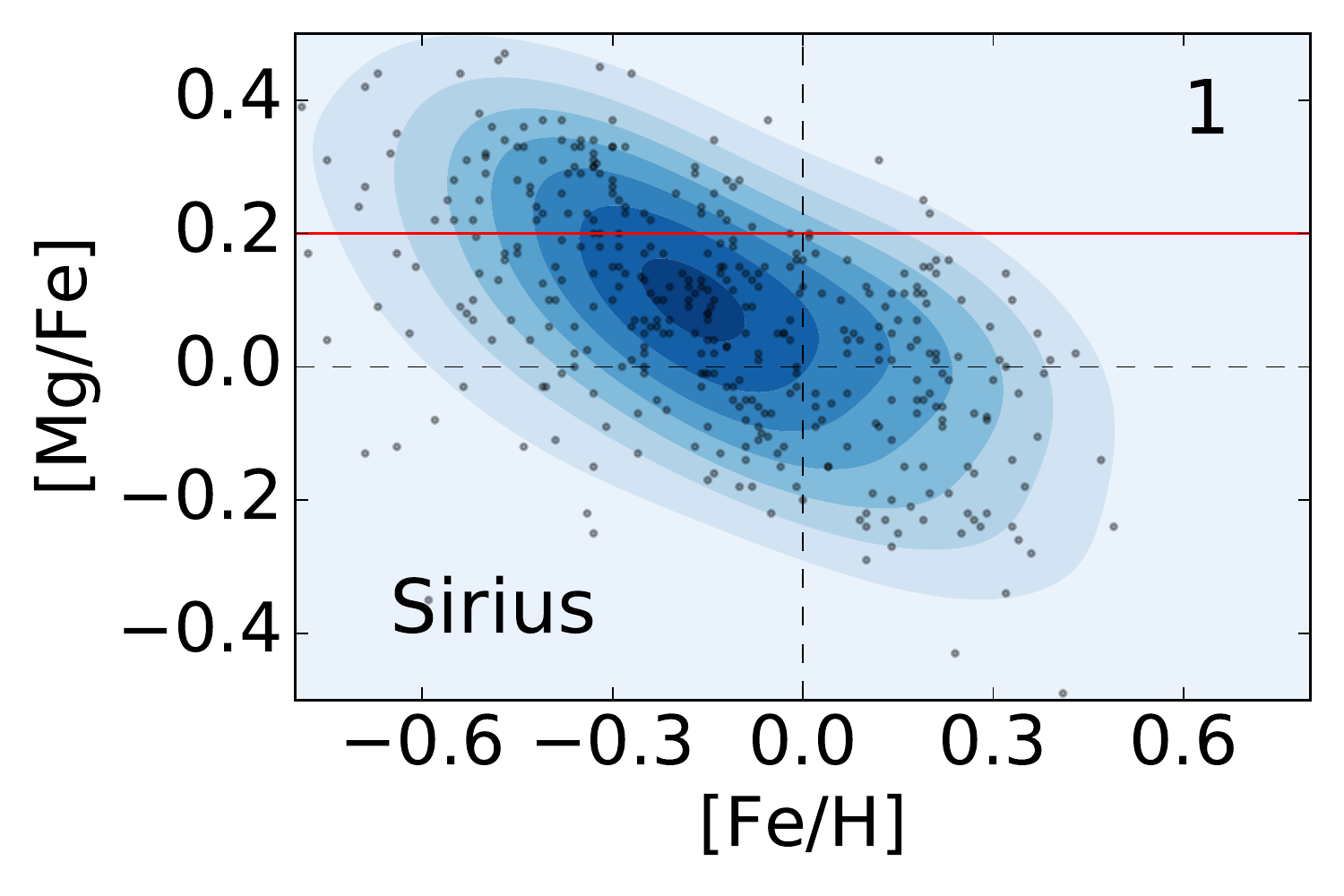}
   \includegraphics[viewport = 93 59.5 430 279,clip]{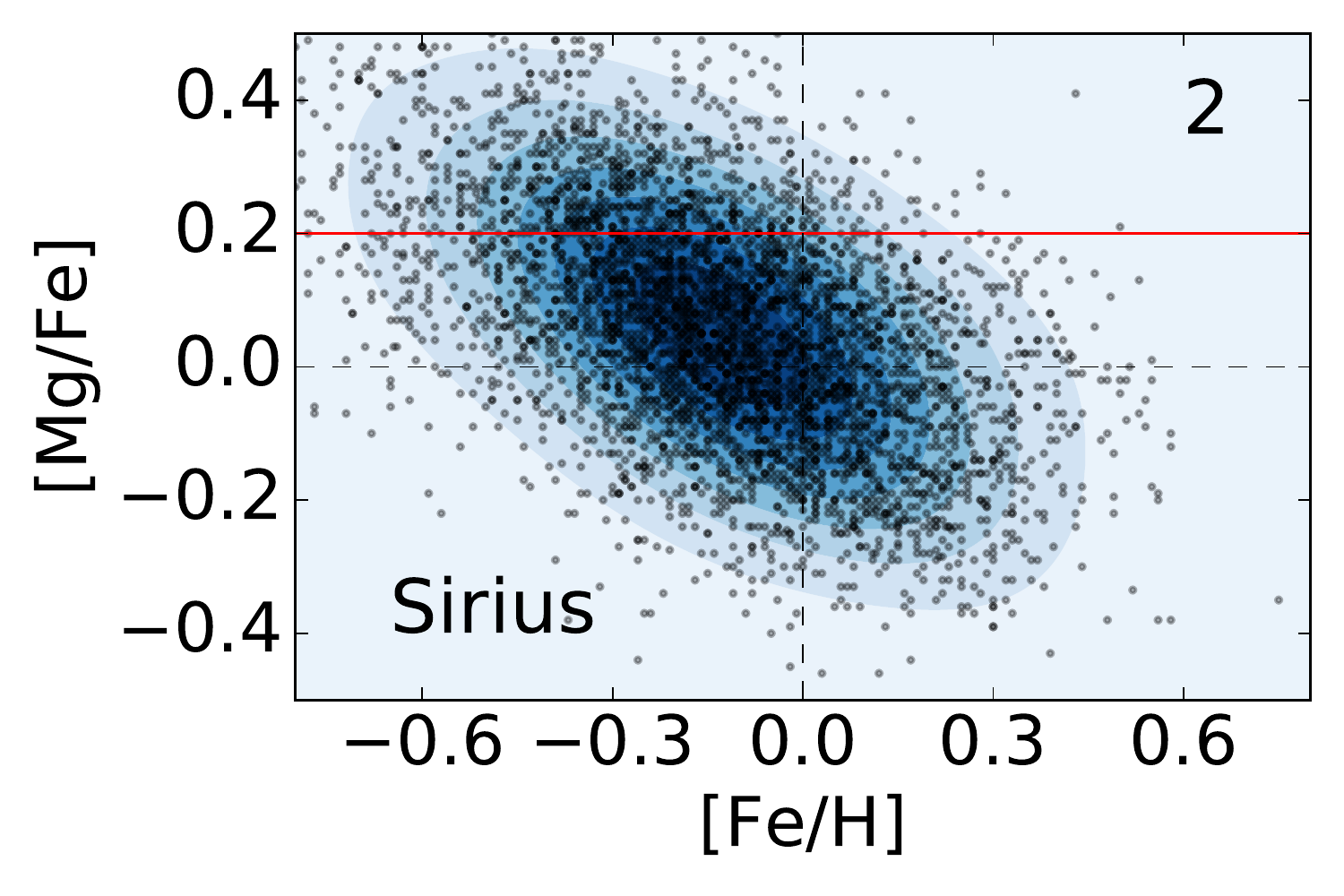}
   \includegraphics[viewport = 93 59.5 430 279,clip]{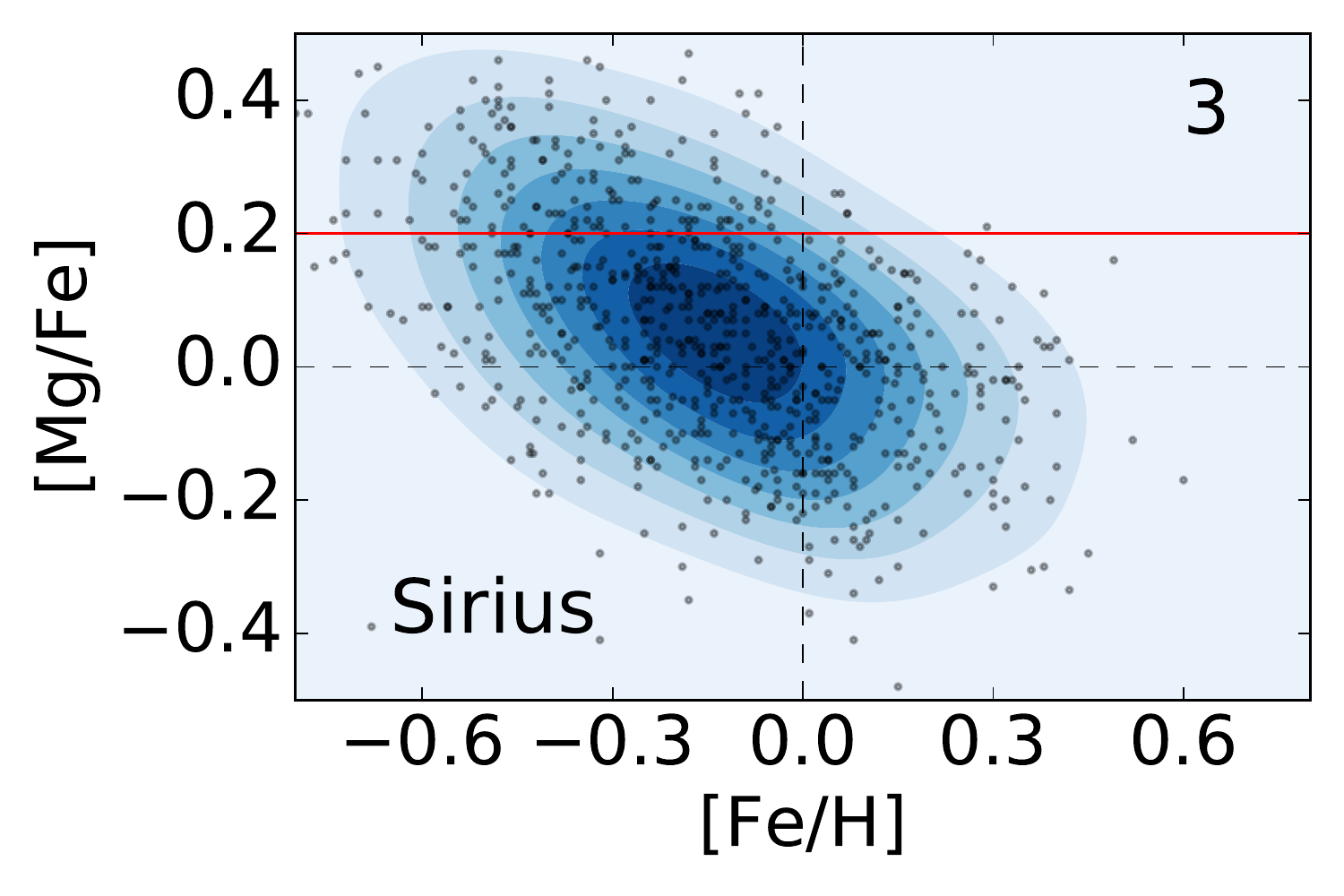}
   \includegraphics[viewport = 93 59.5 430 279,clip]{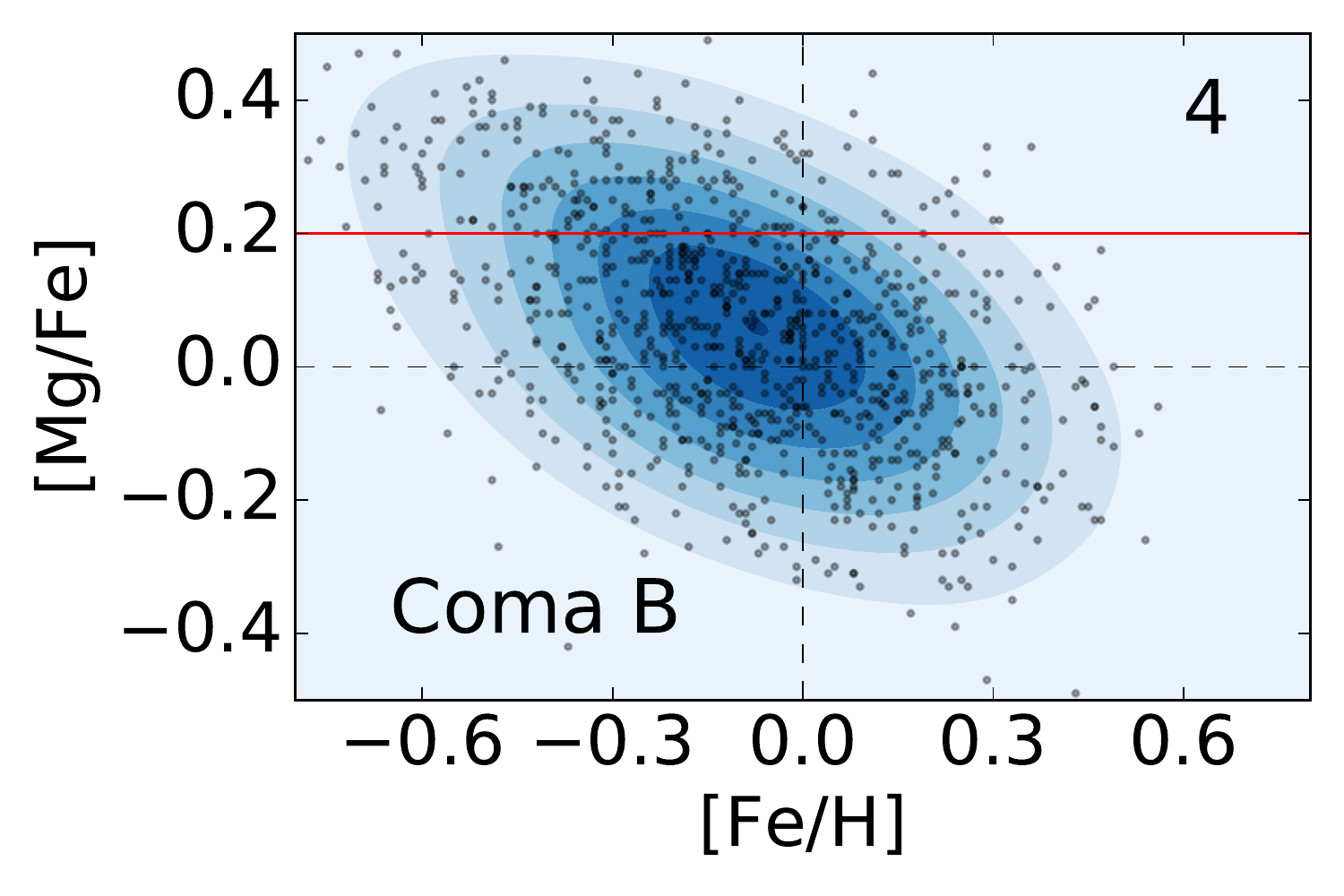}
   }
   \resizebox{0.95\hsize}{!}{
   \includegraphics[viewport = -21 64 430 160,clip]{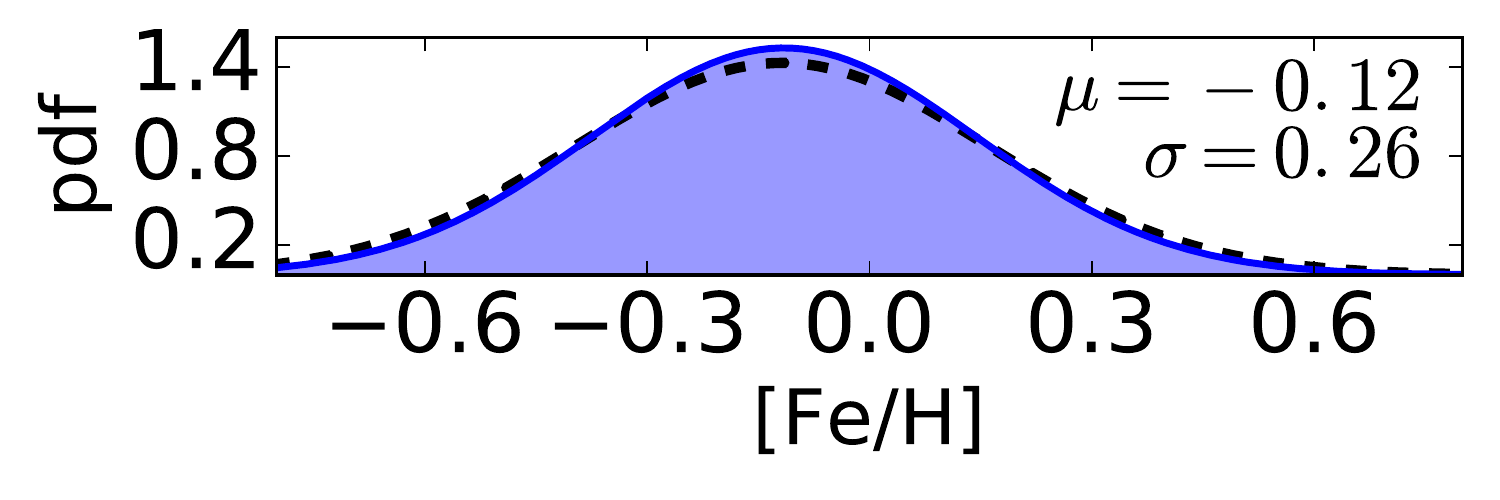}
   \includegraphics[viewport = 77  64 430 160,clip]{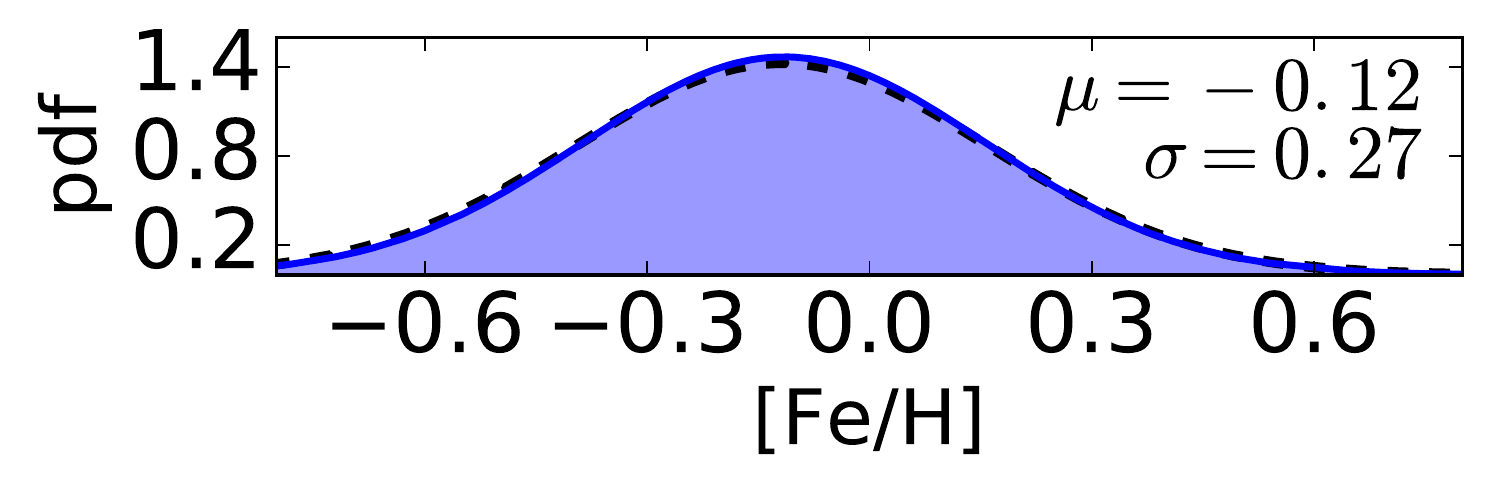}
   \includegraphics[viewport = 77  64 430 160,clip]{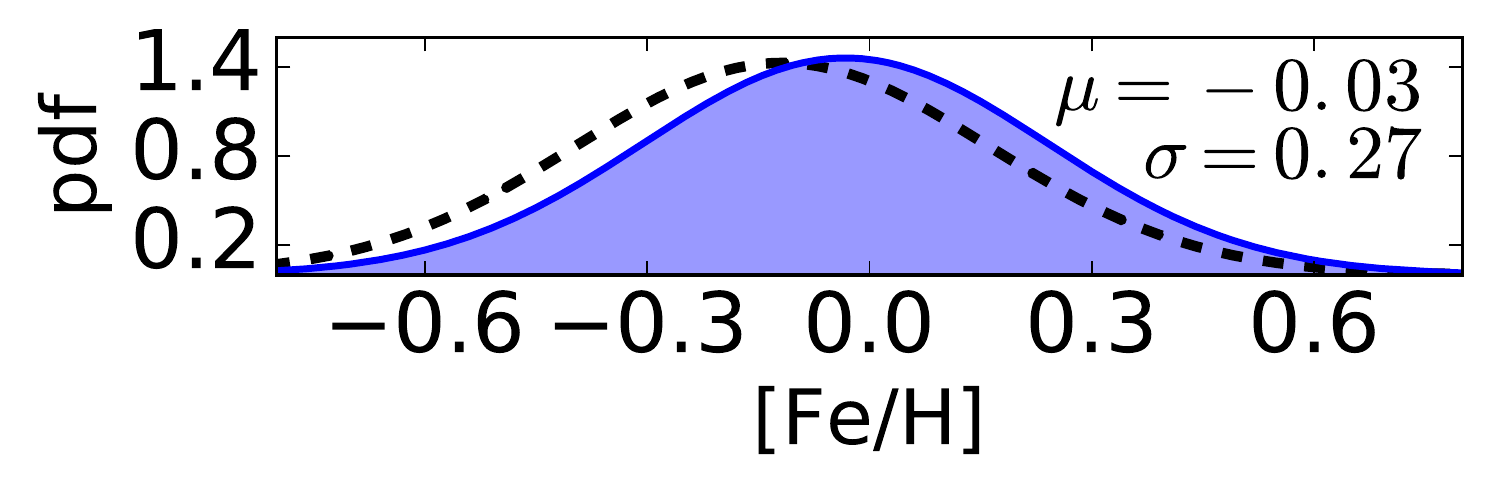}
   \includegraphics[viewport = 77  64 430 160,clip]{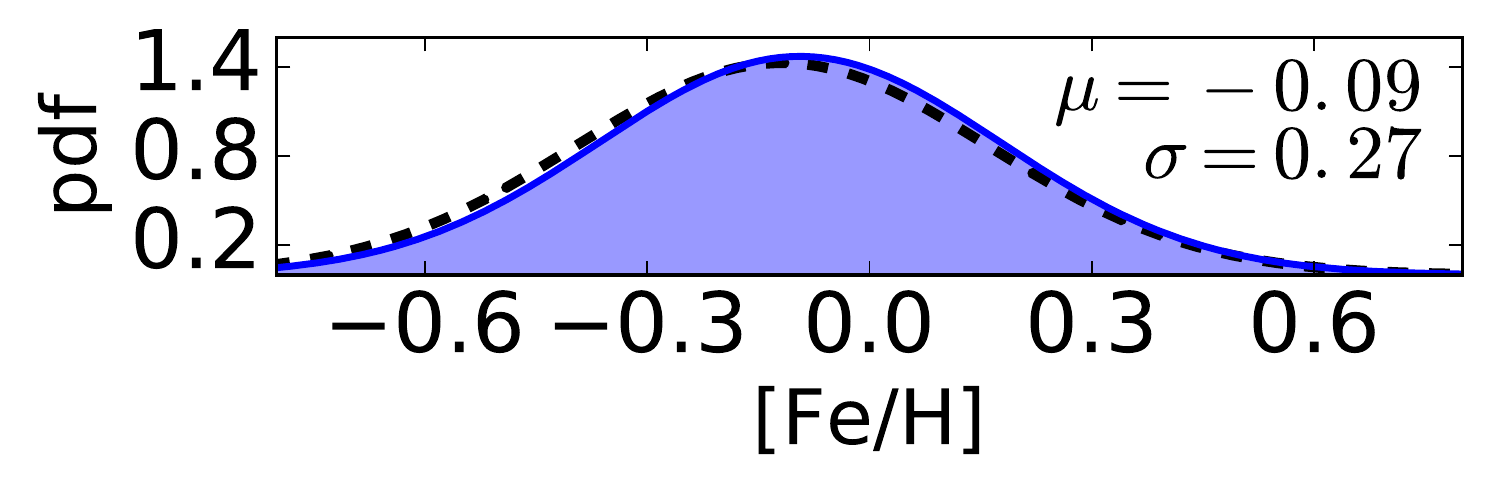}
   }
   \resizebox{0.95\hsize}{!}{
   \includegraphics[viewport = 0  59.5 430 279,clip]{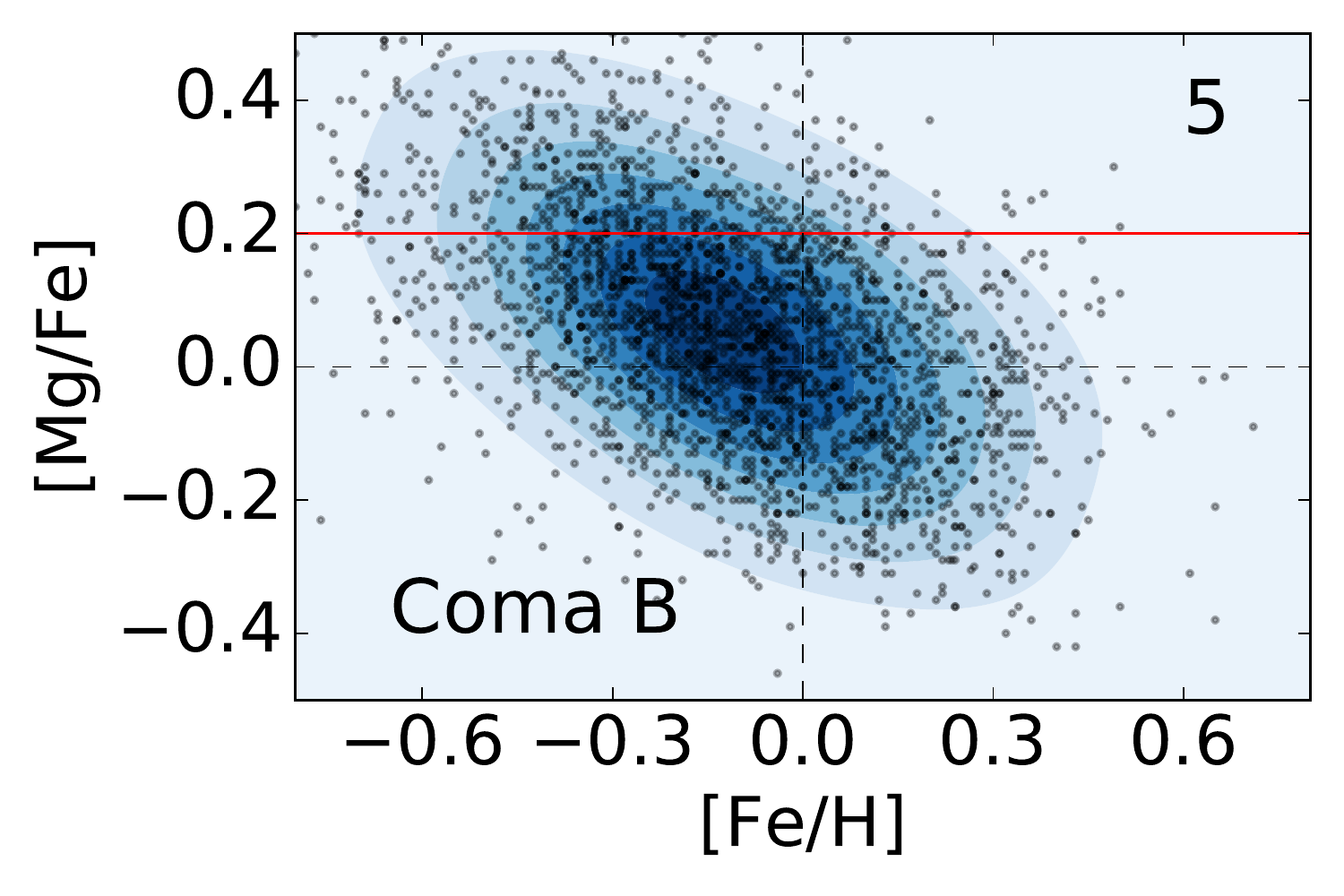}
   \includegraphics[viewport = 93 59.5 430 279,clip]{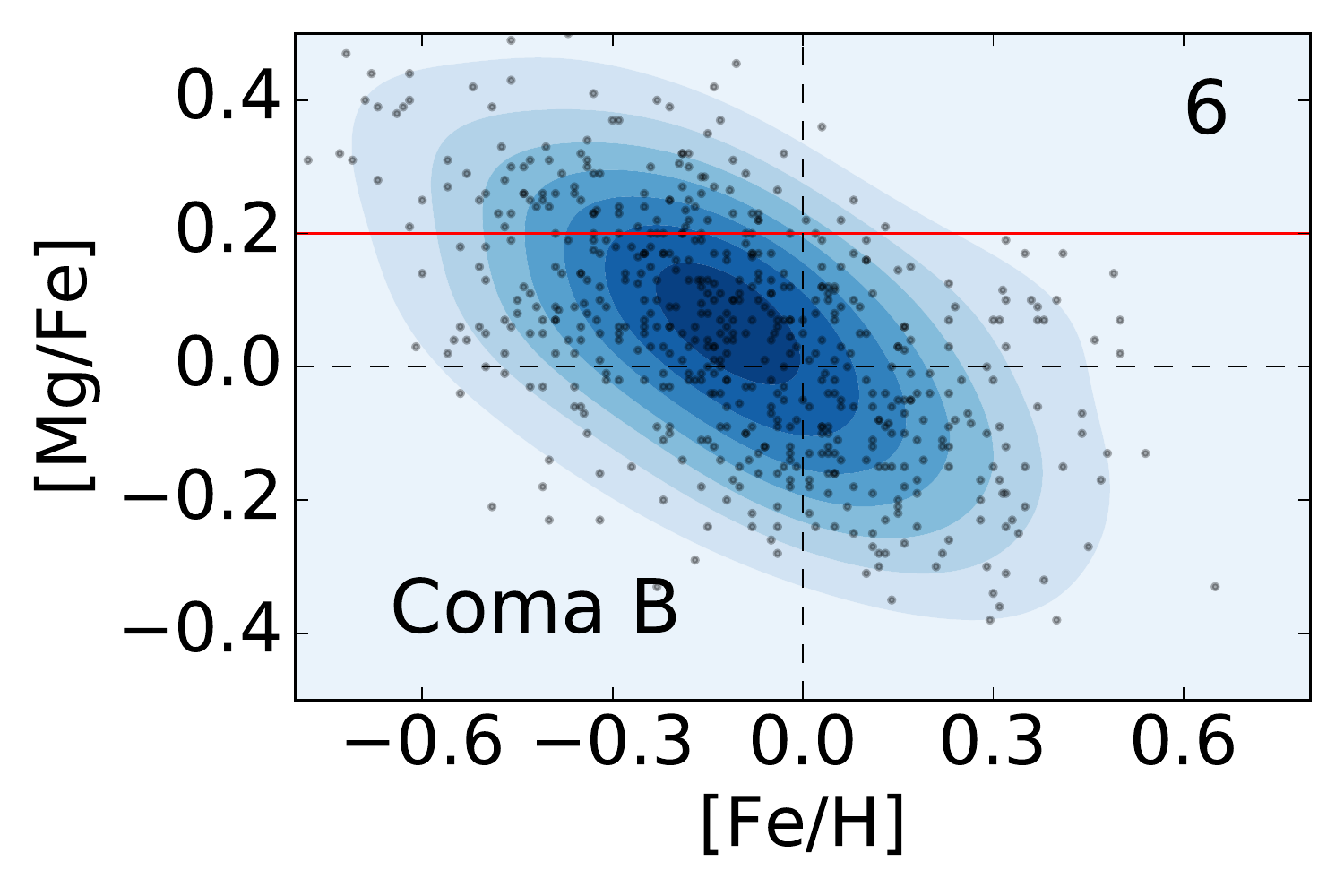}
   \includegraphics[viewport = 93 59.5 430 279,clip]{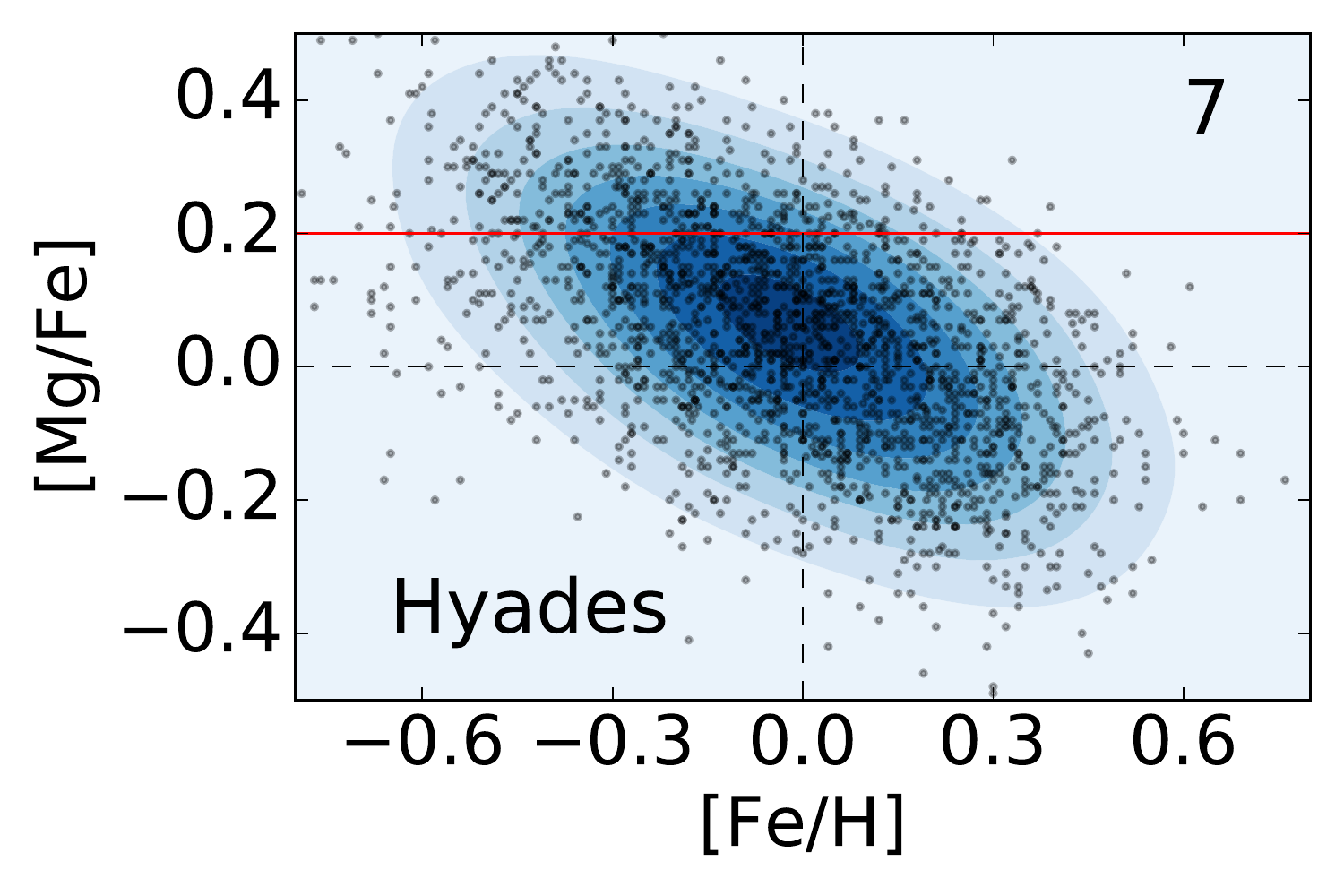}
   \includegraphics[viewport = 93 59.5 430 279,clip]{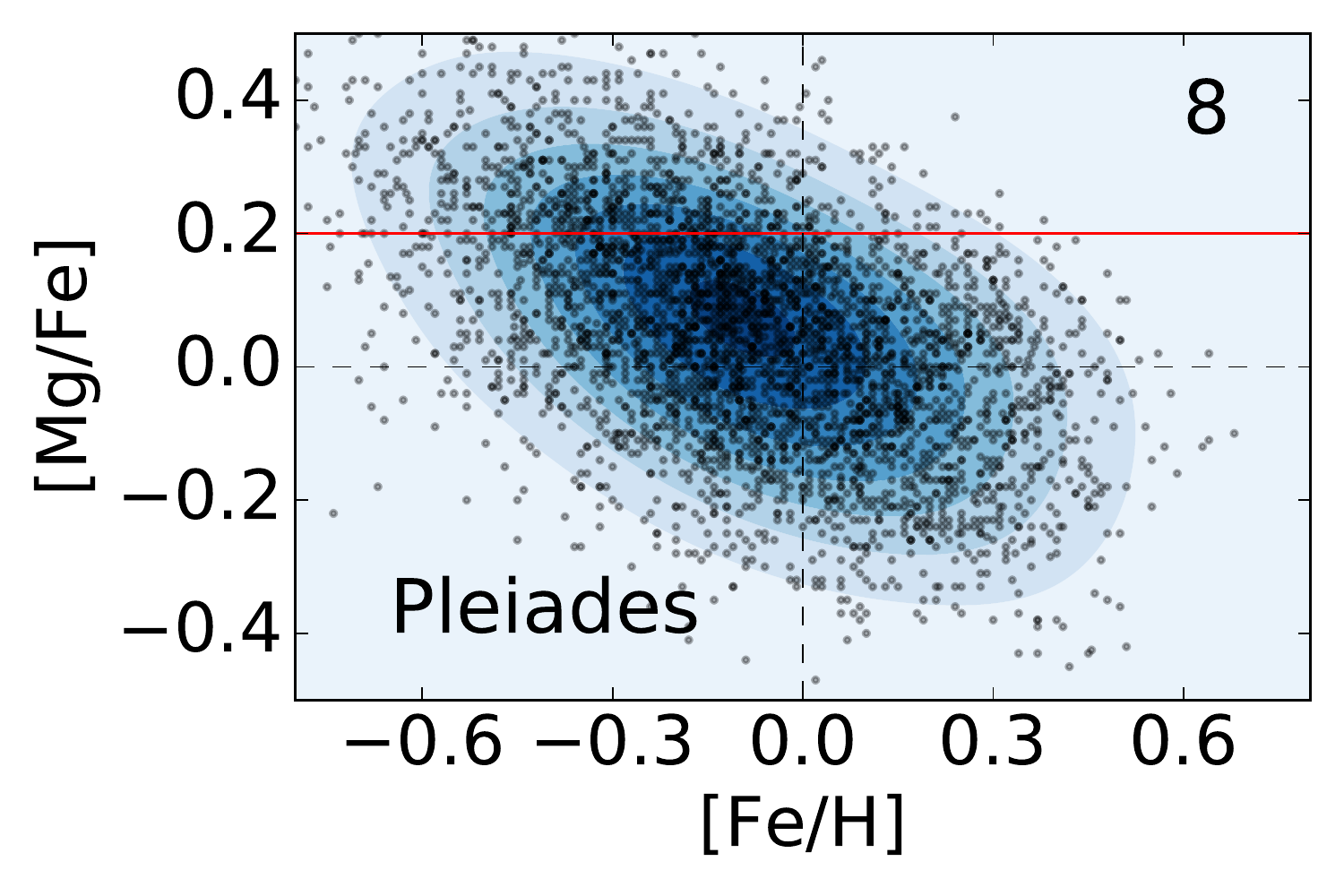}
   }
   \resizebox{0.95\hsize}{!}{
   \includegraphics[viewport = -21 64 430 160,clip]{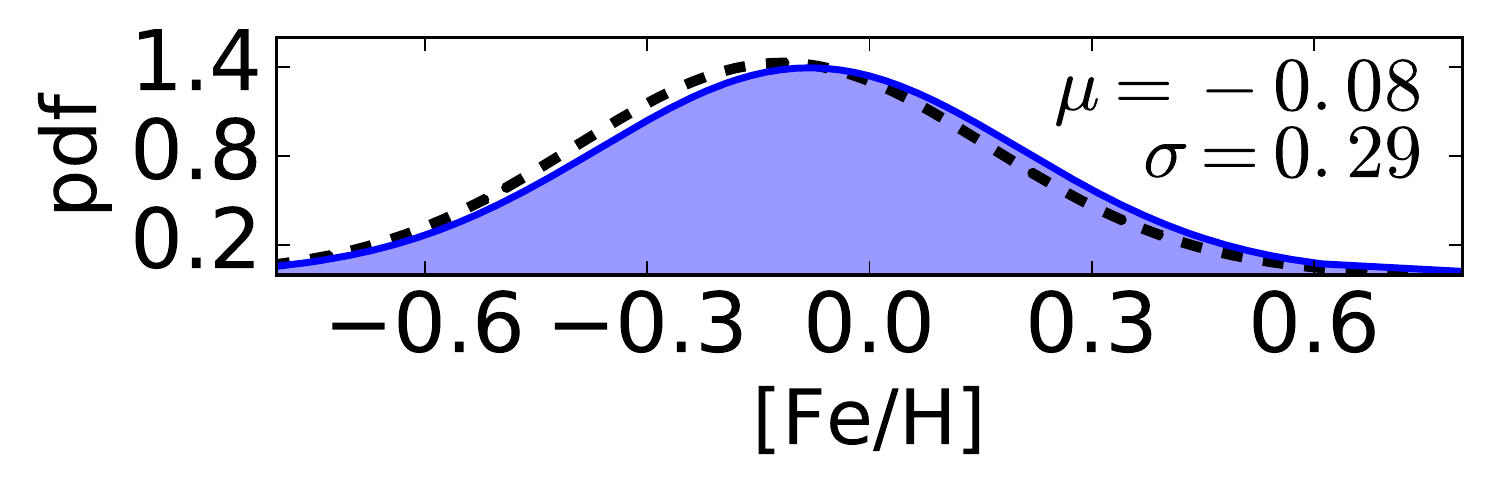}
   \includegraphics[viewport = 77  64 430 160,clip]{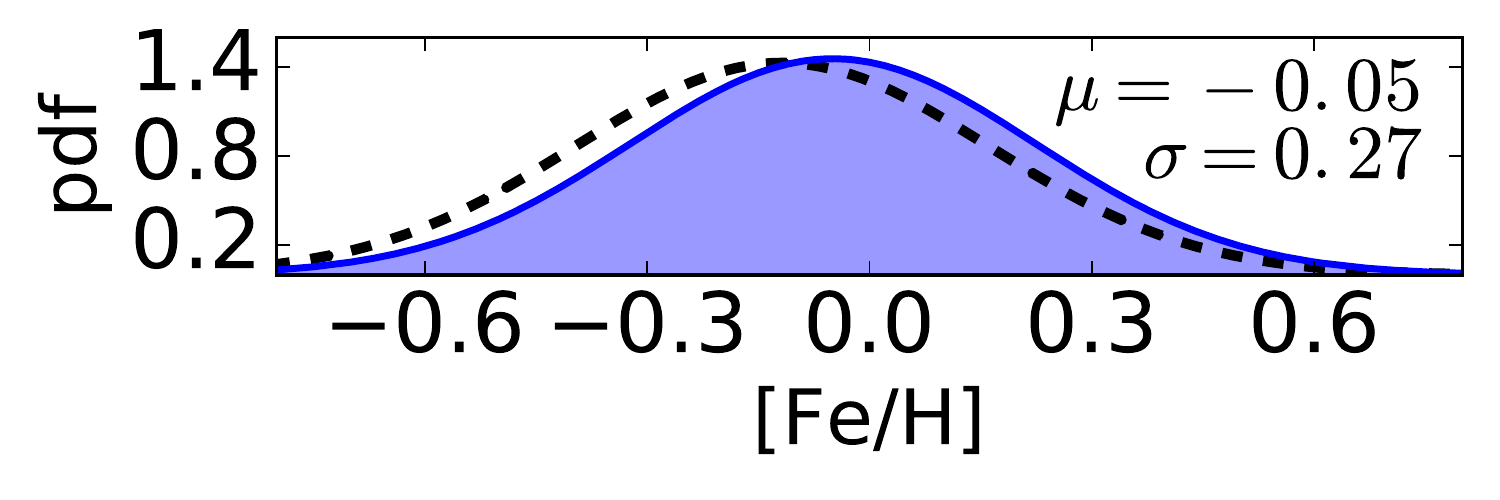}
   \includegraphics[viewport = 77  64 430 160,clip]{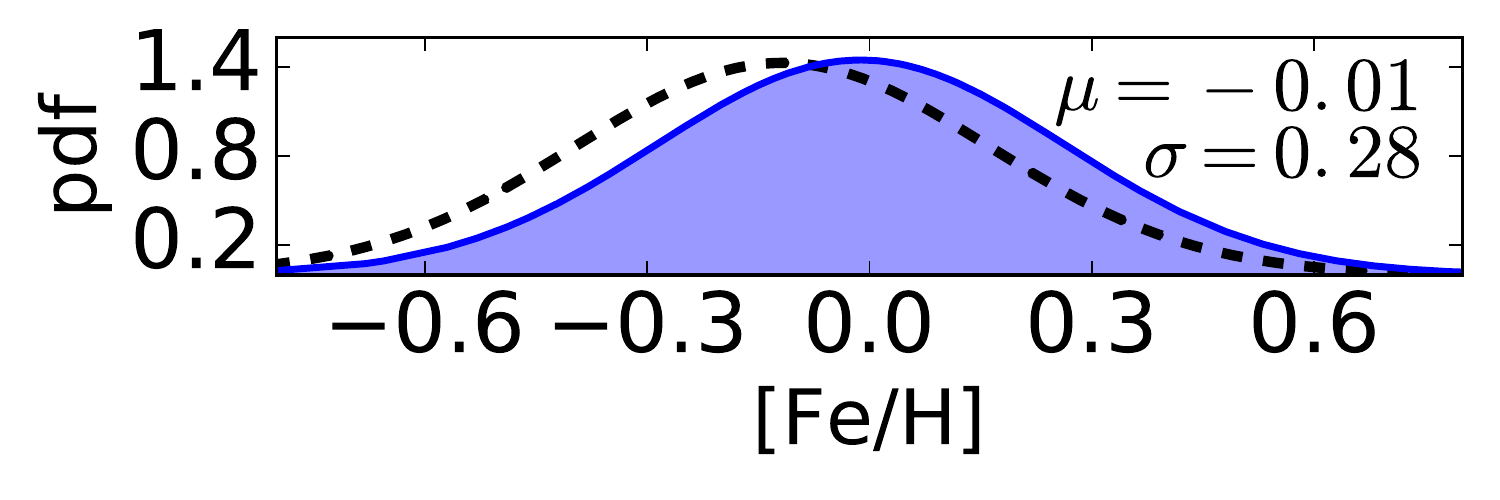}
   \includegraphics[viewport = 77  64 430 160,clip]{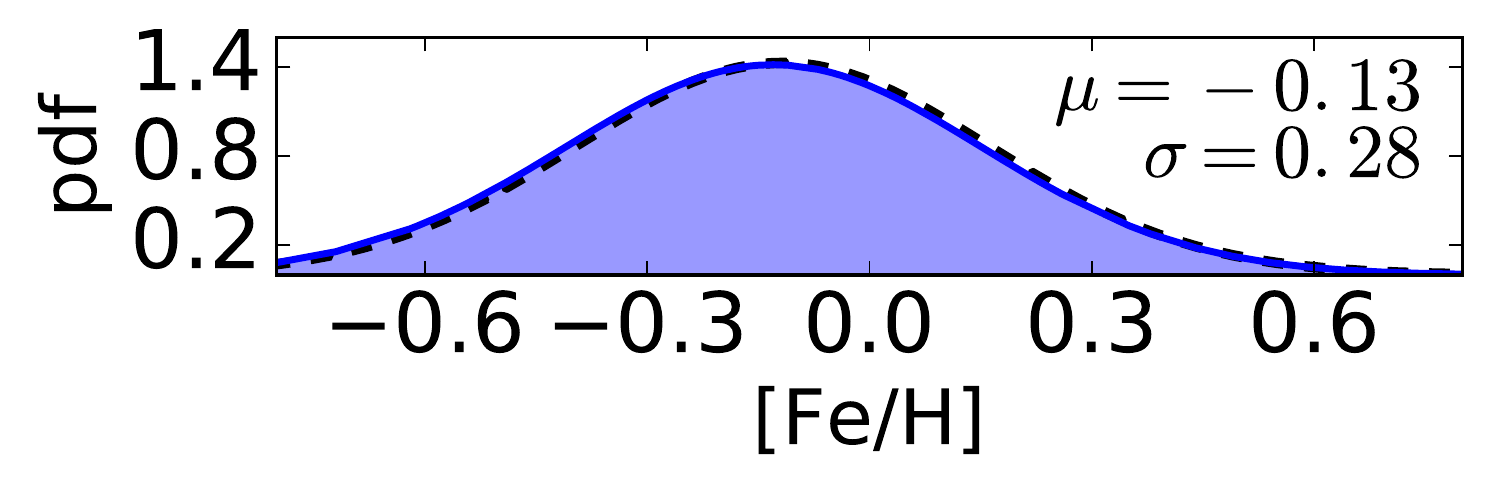}
   }
   \resizebox{0.95\hsize}{!}{
   \includegraphics[viewport = 0  59.5 430 279,clip]{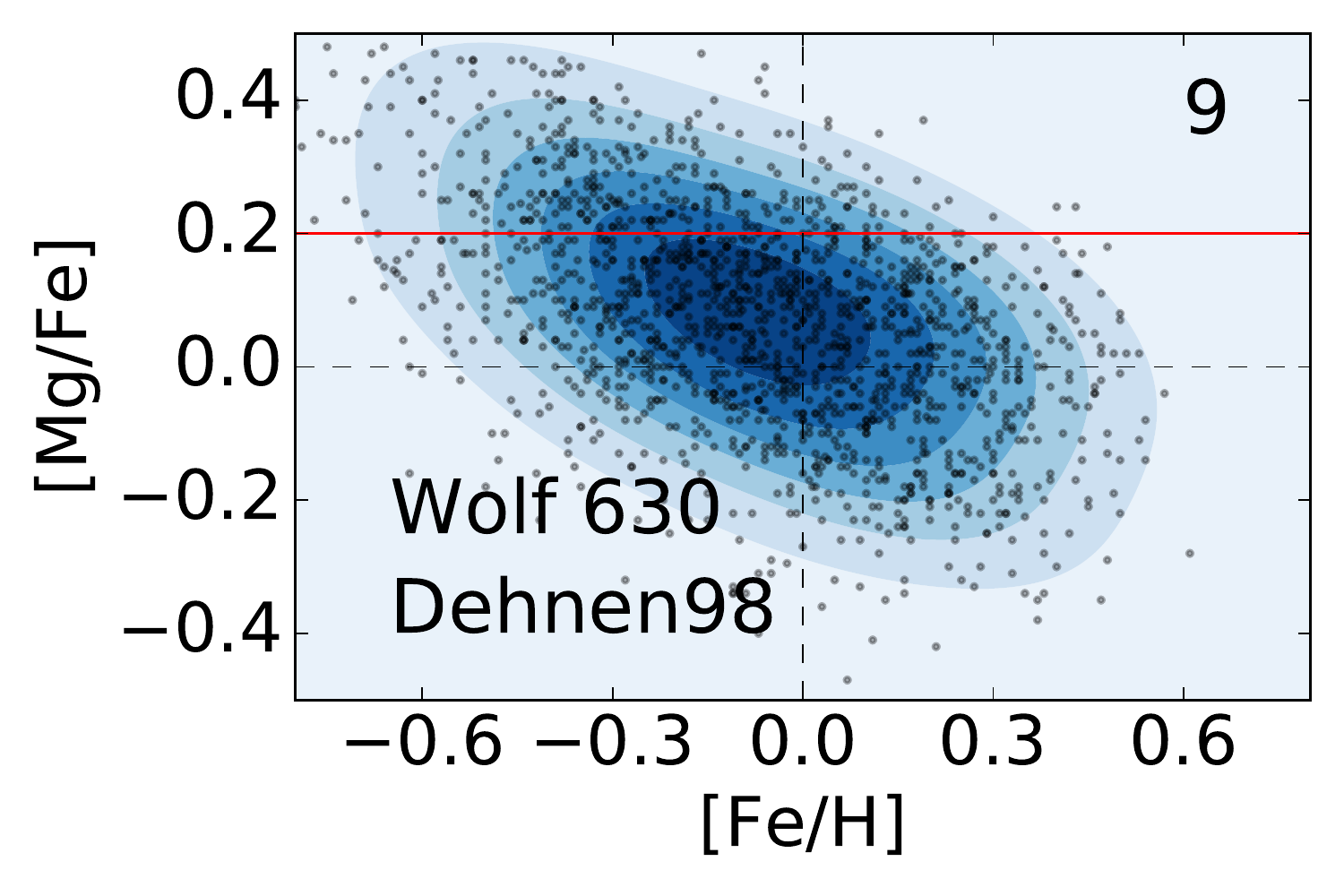}
   \includegraphics[viewport = 93 59.5 430 279,clip]{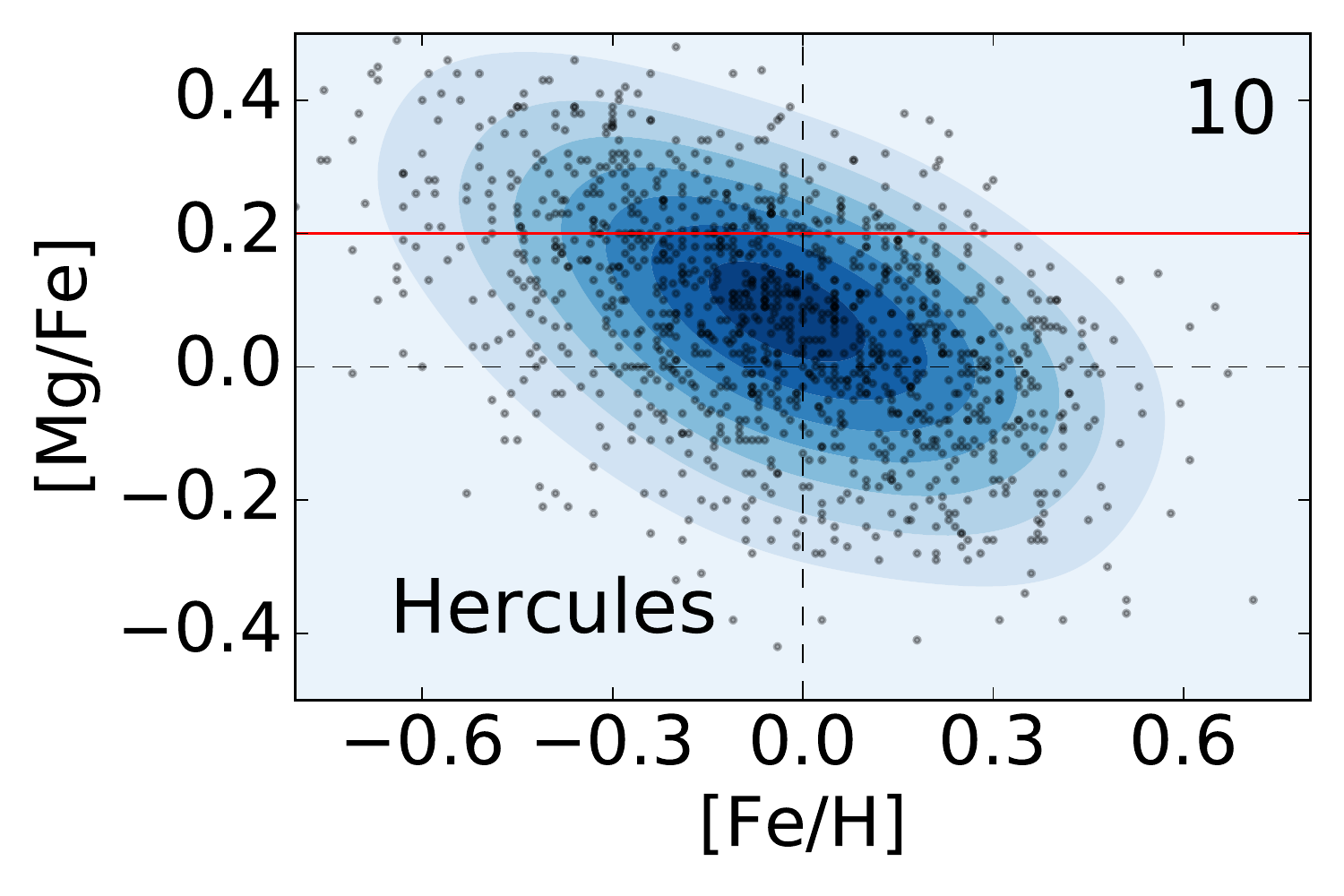}
   \includegraphics[viewport = 93 59.5 430 279,clip]{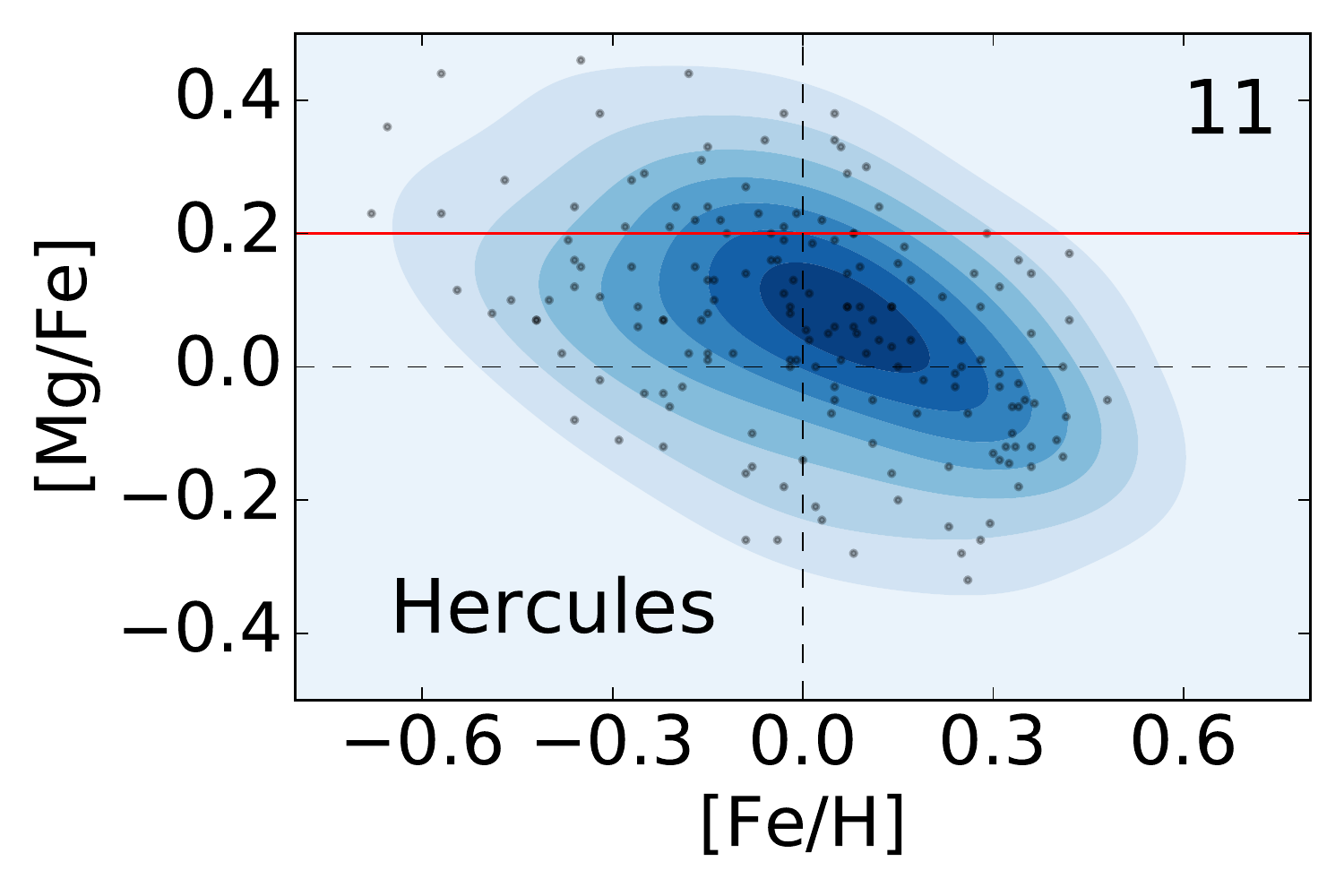}
   \includegraphics[viewport = 93 59.5 430 279,clip]{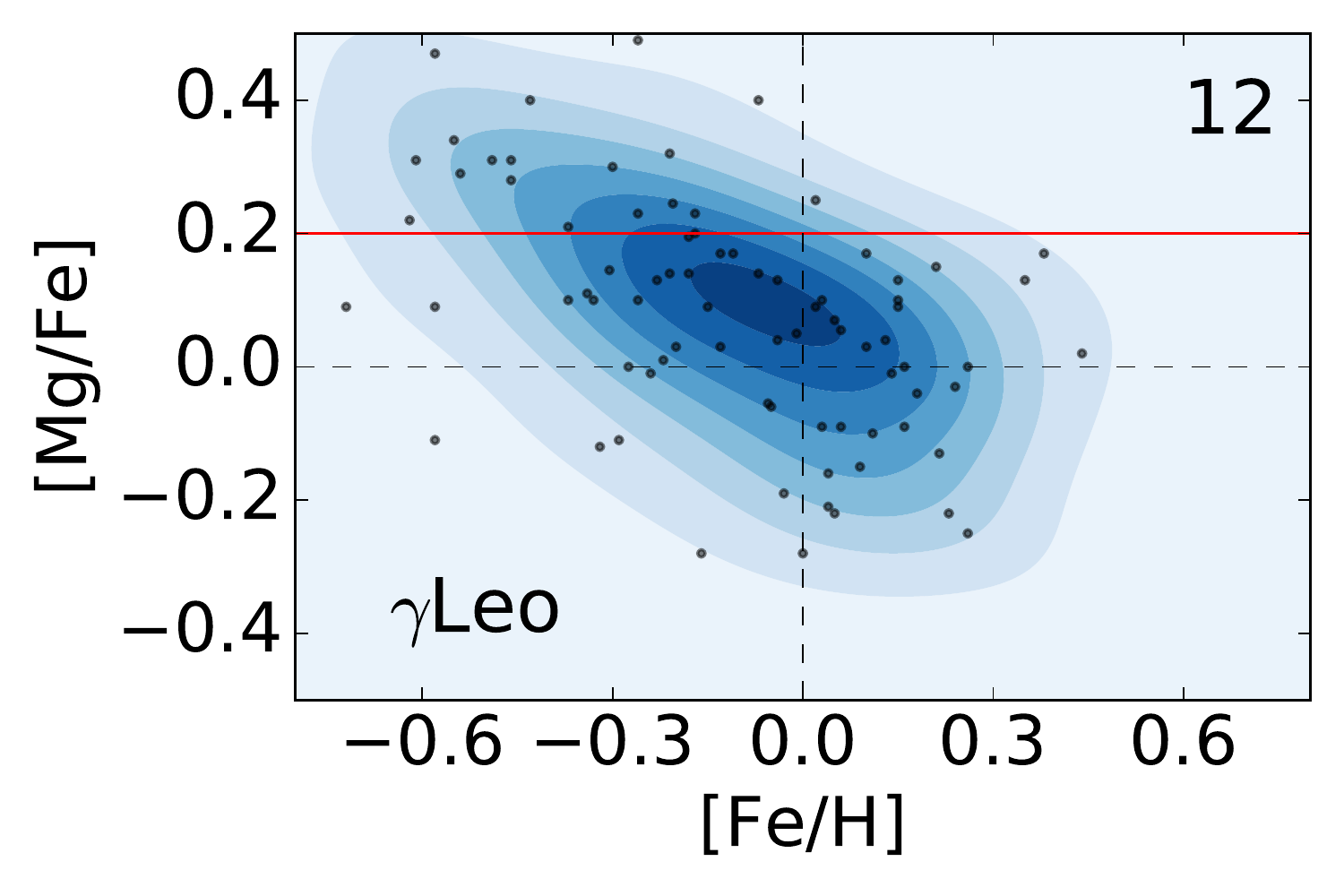}
   }
   \resizebox{0.95\hsize}{!}{
   \includegraphics[viewport = -21 64 430 160,clip]{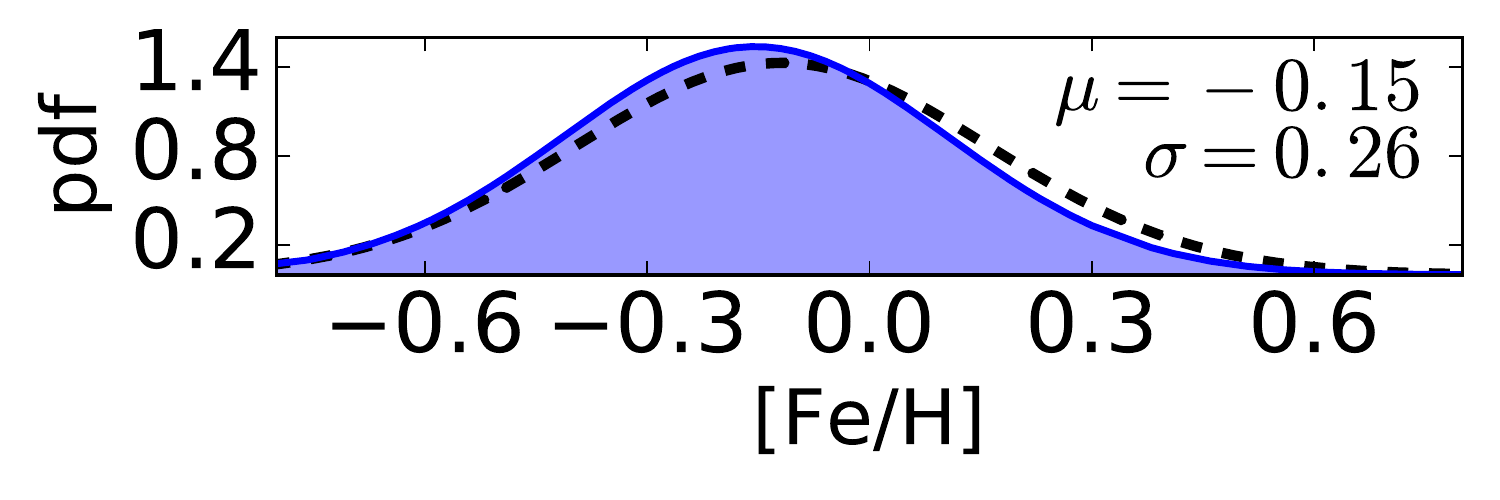}
   \includegraphics[viewport = 77  64 430 160,clip]{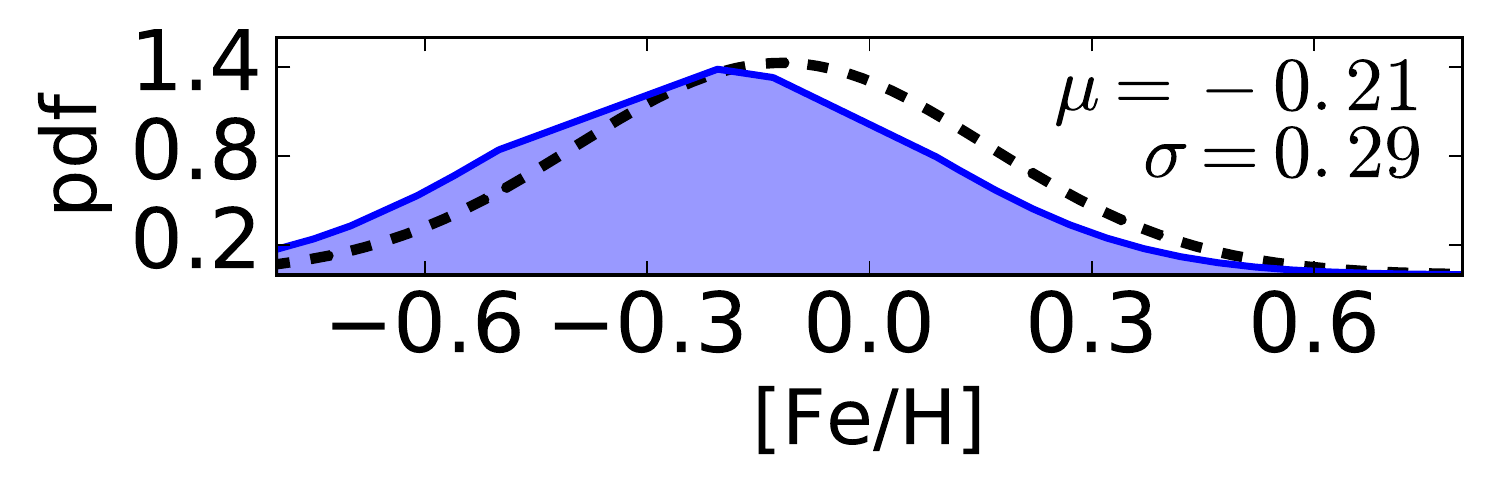}
   \includegraphics[viewport = 77  64 430 160,clip]{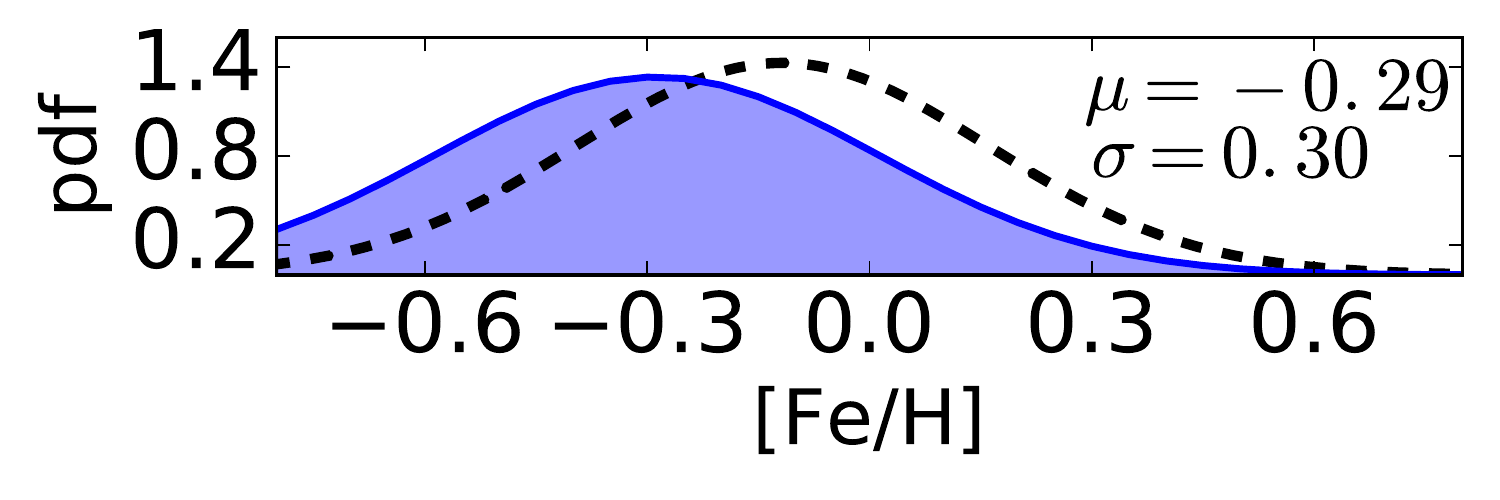}
   \includegraphics[viewport = 77  64 430 160,clip]{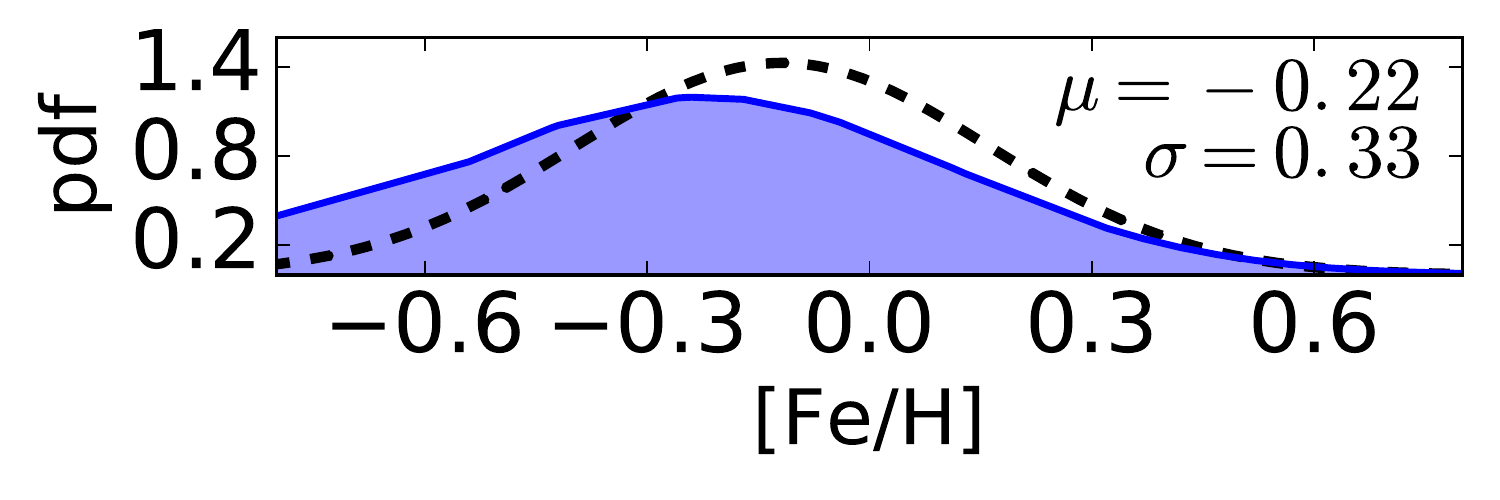}
   }
   \resizebox{0.95\hsize}{!}{
   \includegraphics[viewport = 0  59.5 430 279,clip]{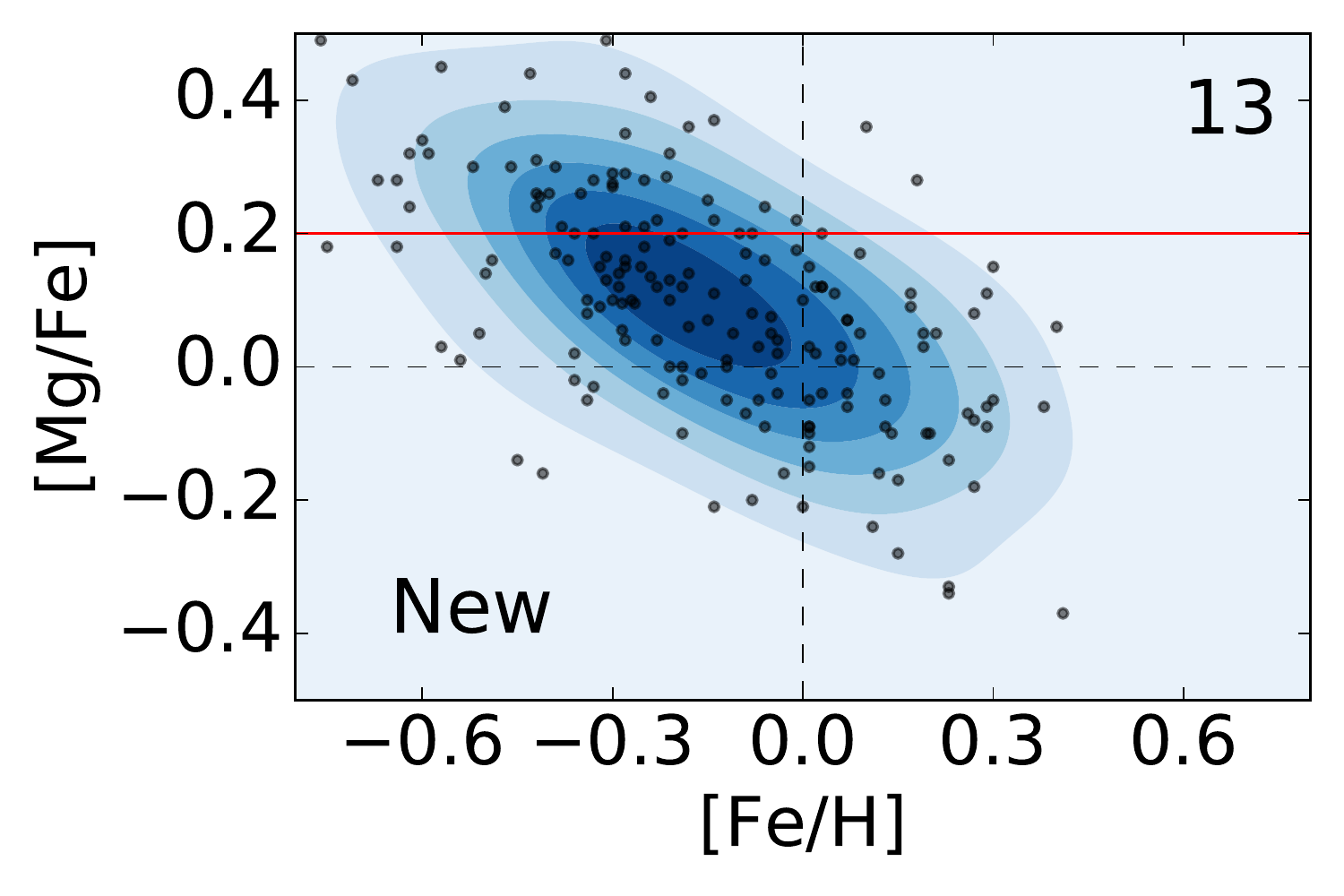}
   \includegraphics[viewport = 93 59.5 430 279,clip]{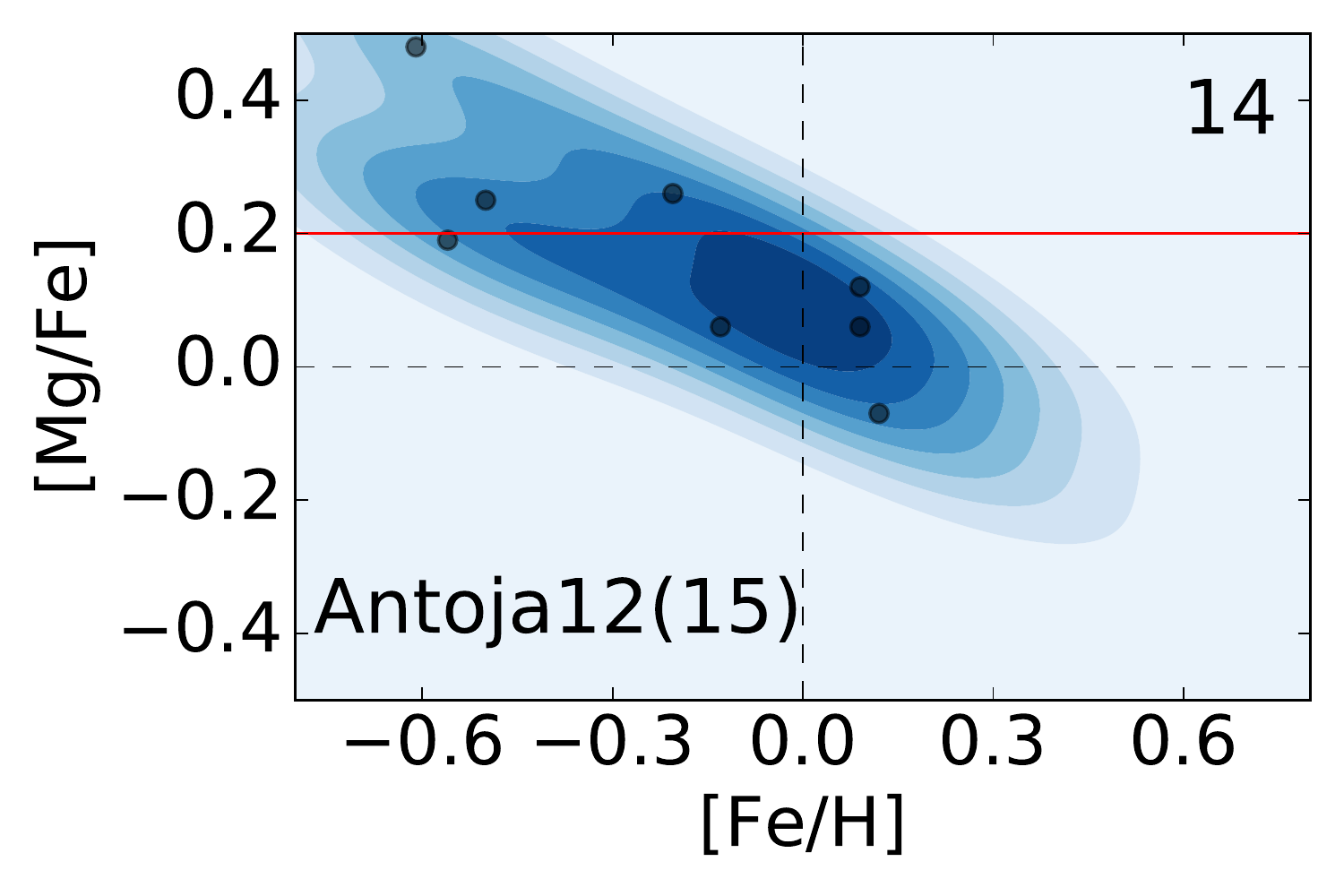}
   \includegraphics[viewport = 93 59.5 430 279,clip]{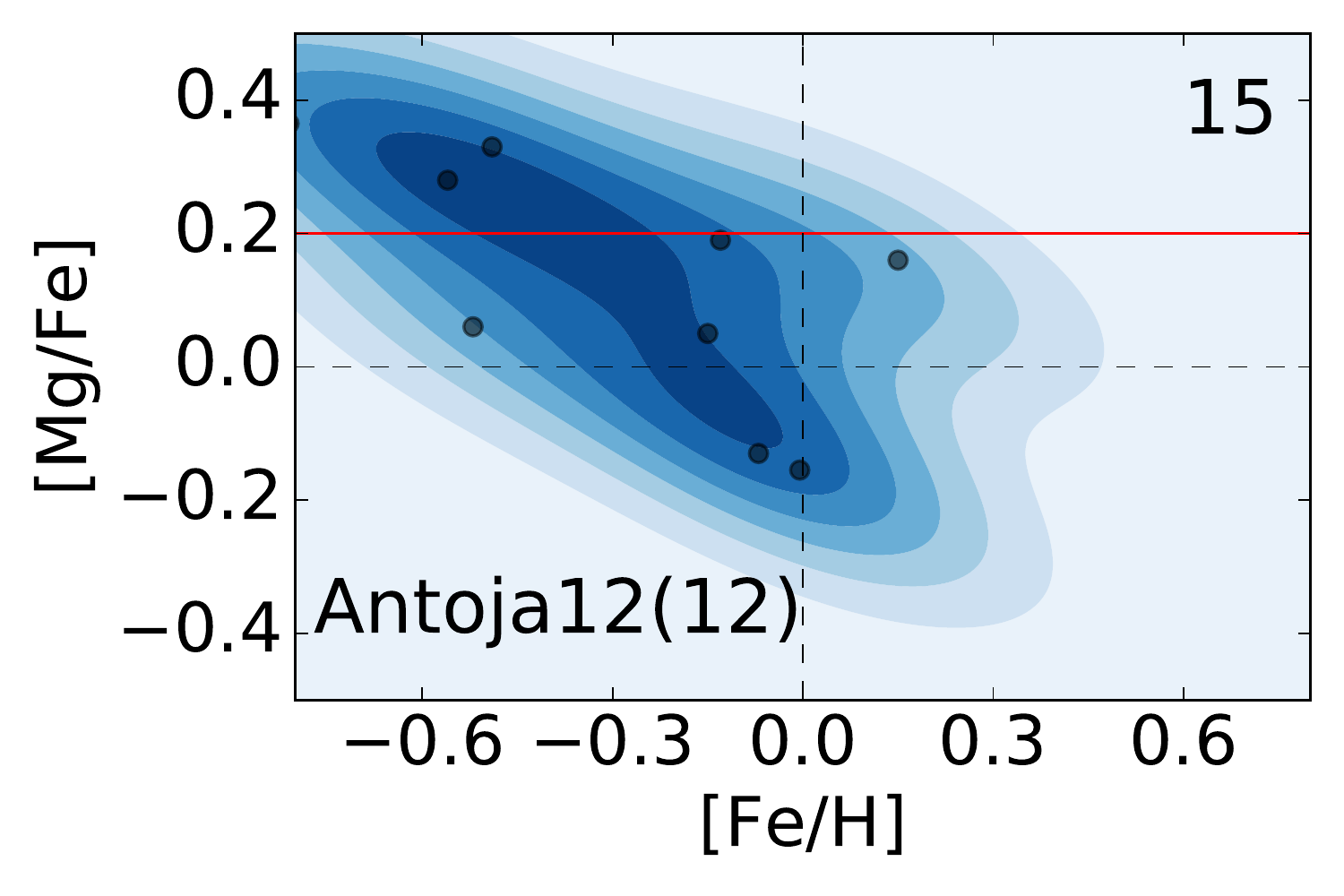}
   \includegraphics[viewport = 93 59.5 430 279,clip]{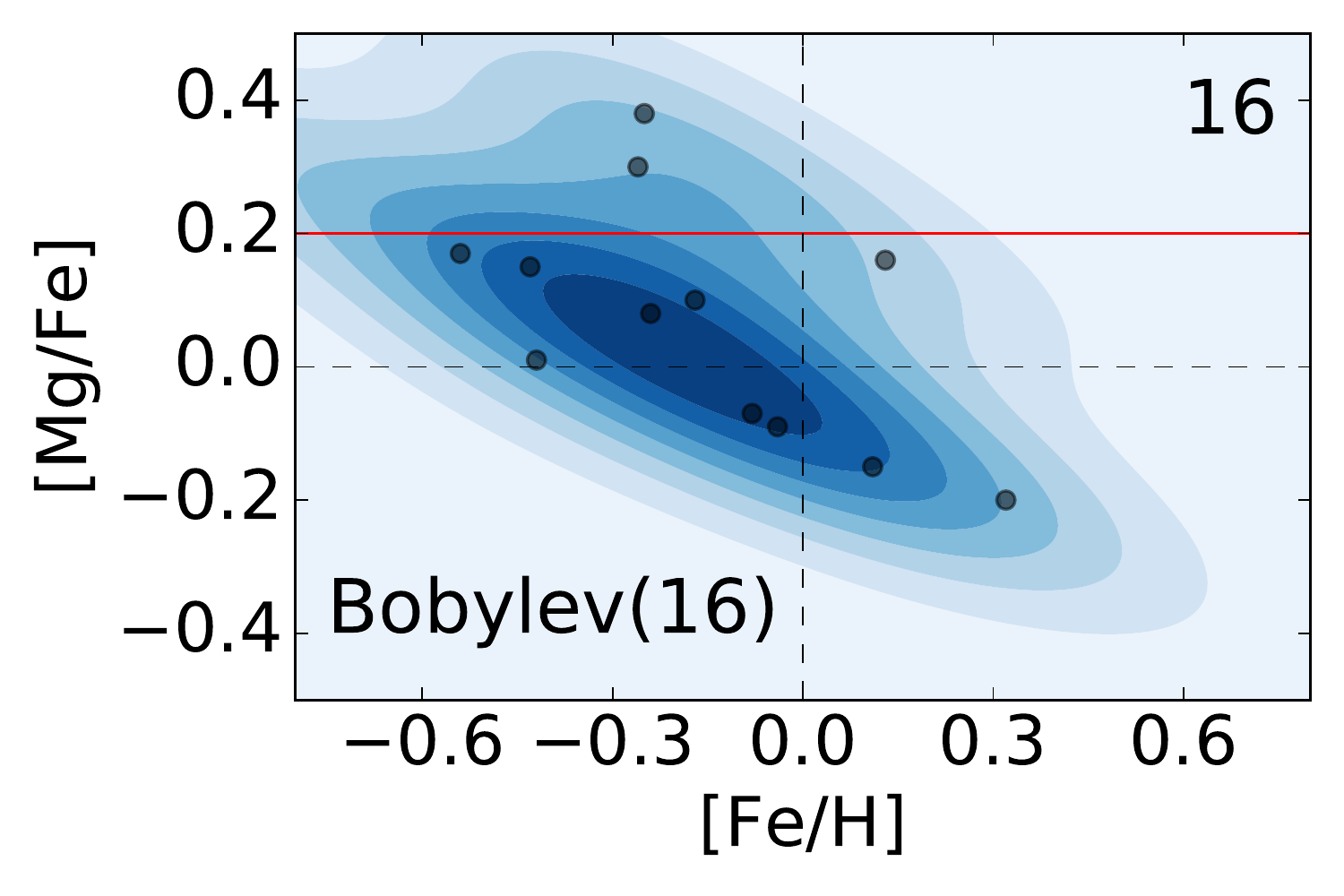}
   }
   \resizebox{0.95\hsize}{!}{
   \includegraphics[viewport = -21 64 430 160,clip]{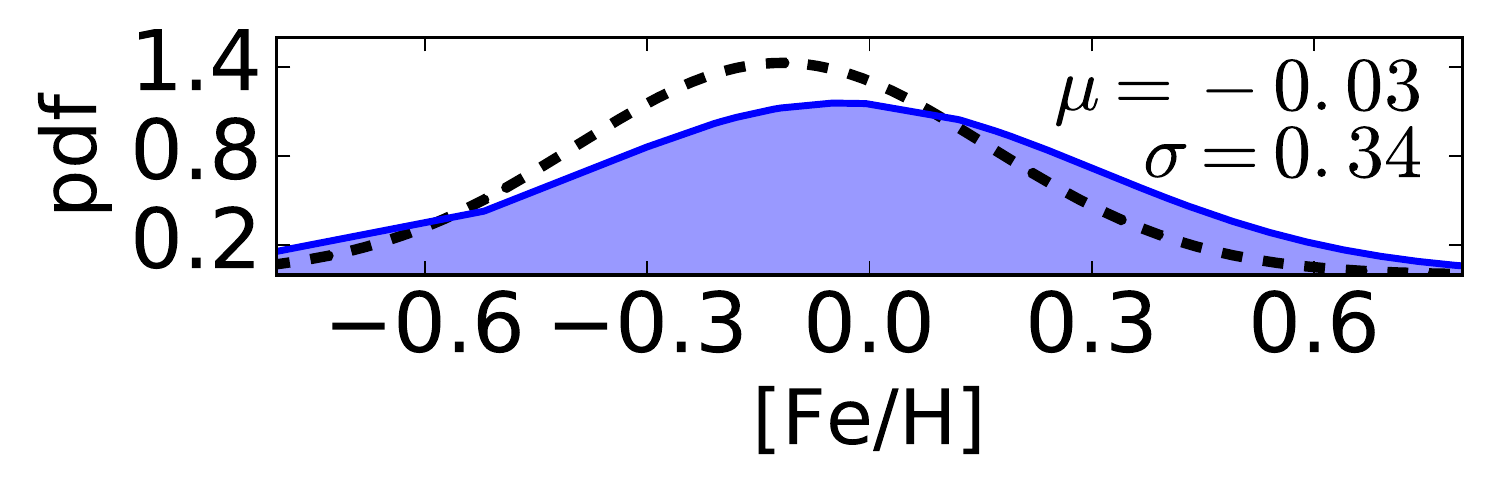}
   \includegraphics[viewport = 77  64 430 160,clip]{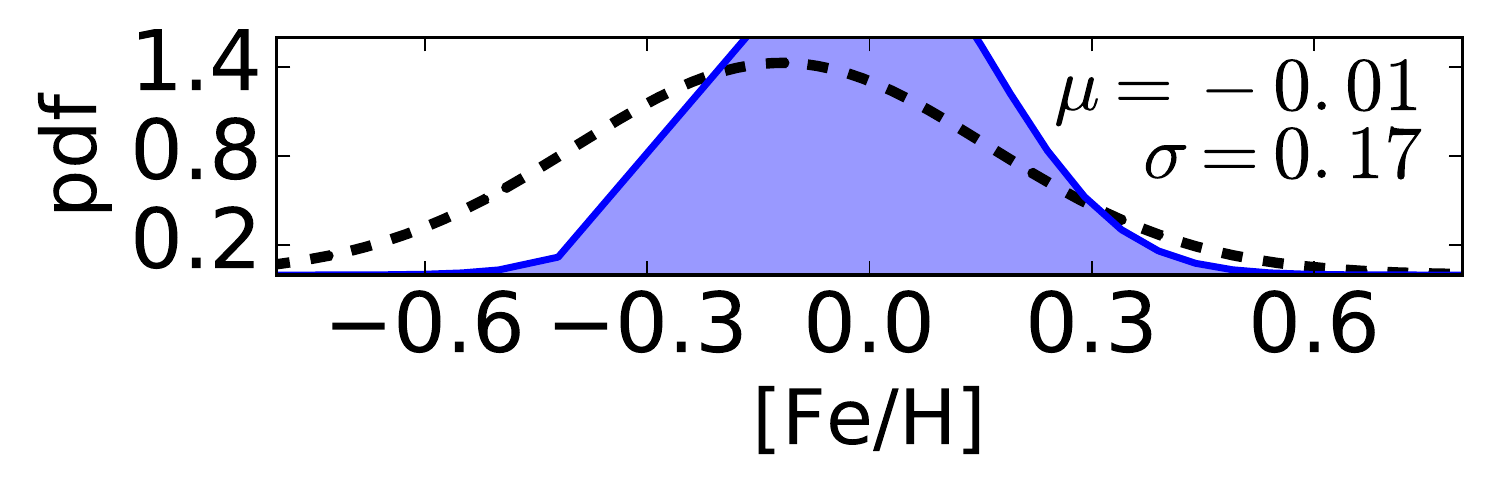}
   \includegraphics[viewport = 77  64 430 160,clip]{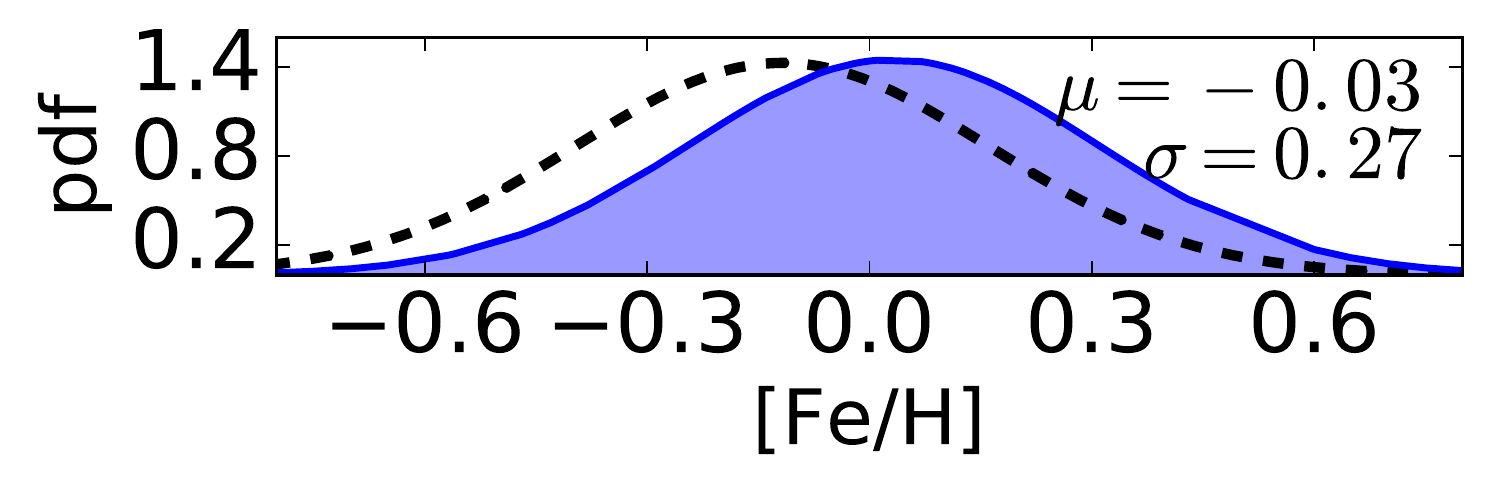}
   \includegraphics[viewport = 77  64 430 160,clip]{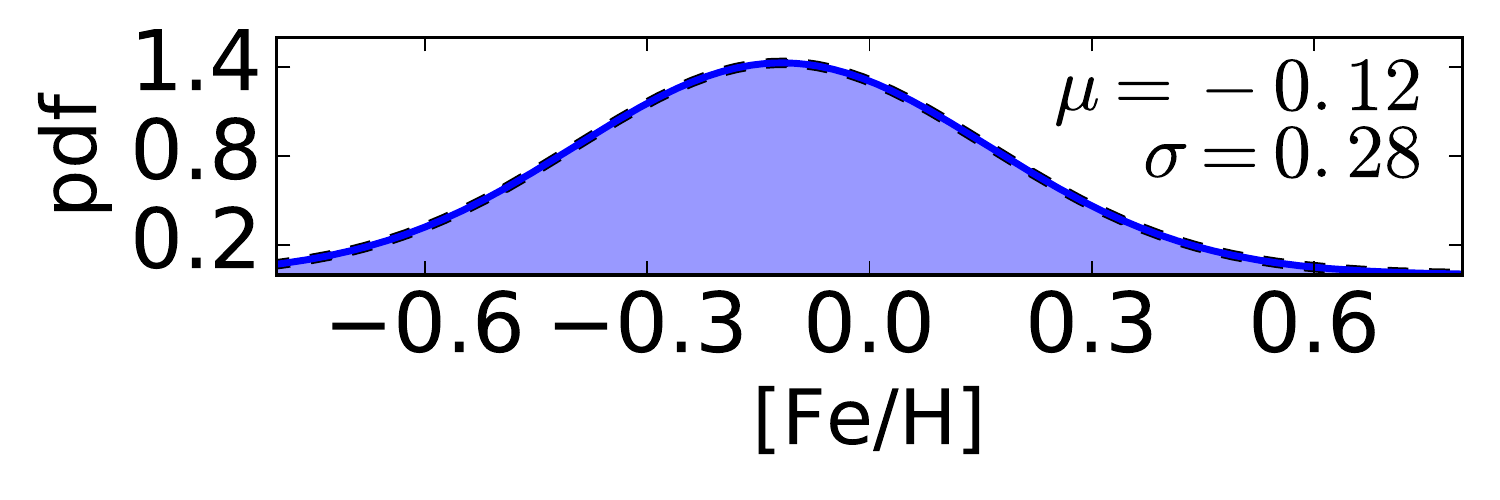}
   }
   \resizebox{0.95\hsize}{!}{
   \includegraphics[viewport = 0  10 430 279,clip]{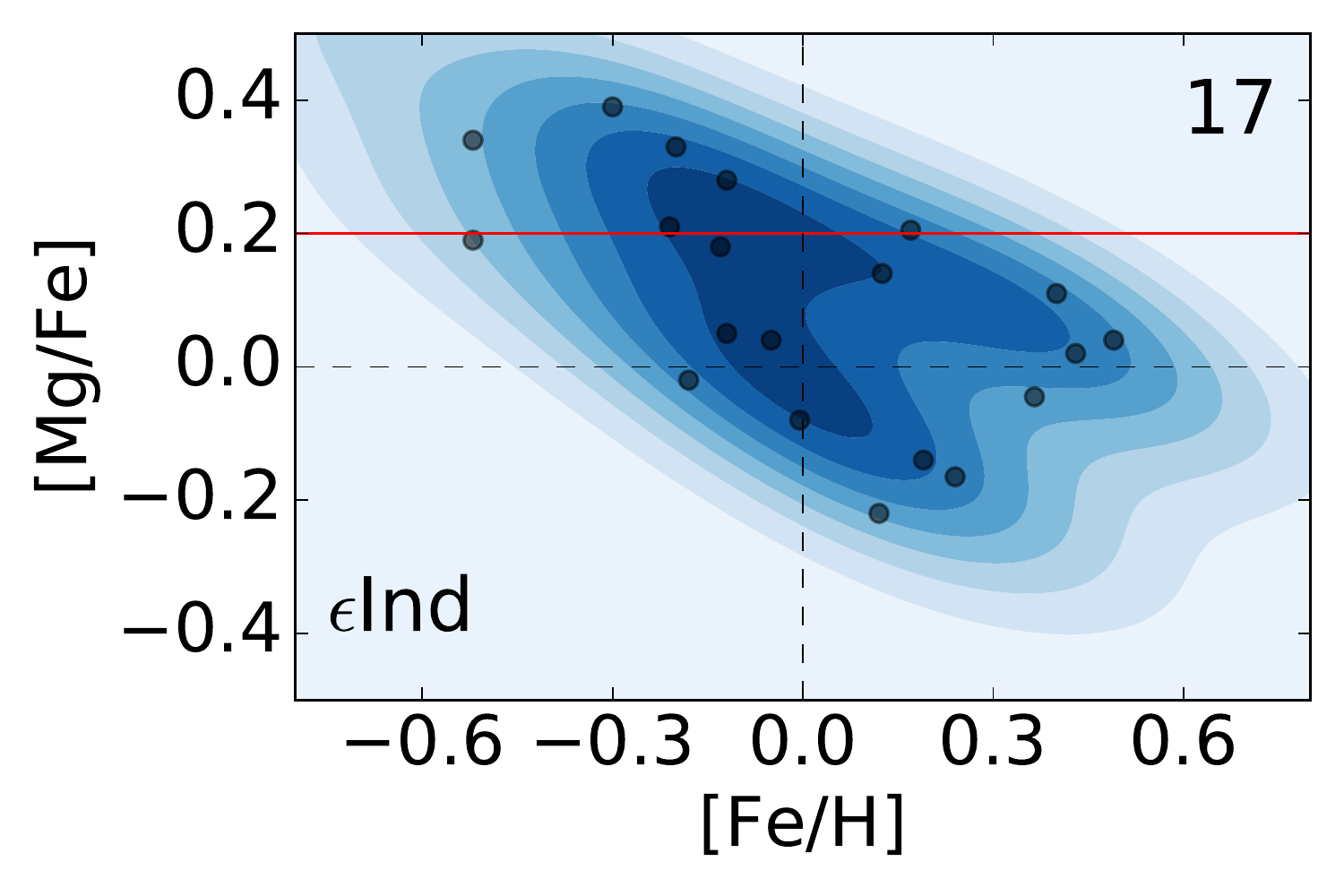}
   \includegraphics[viewport = 93 10 430 279,clip]{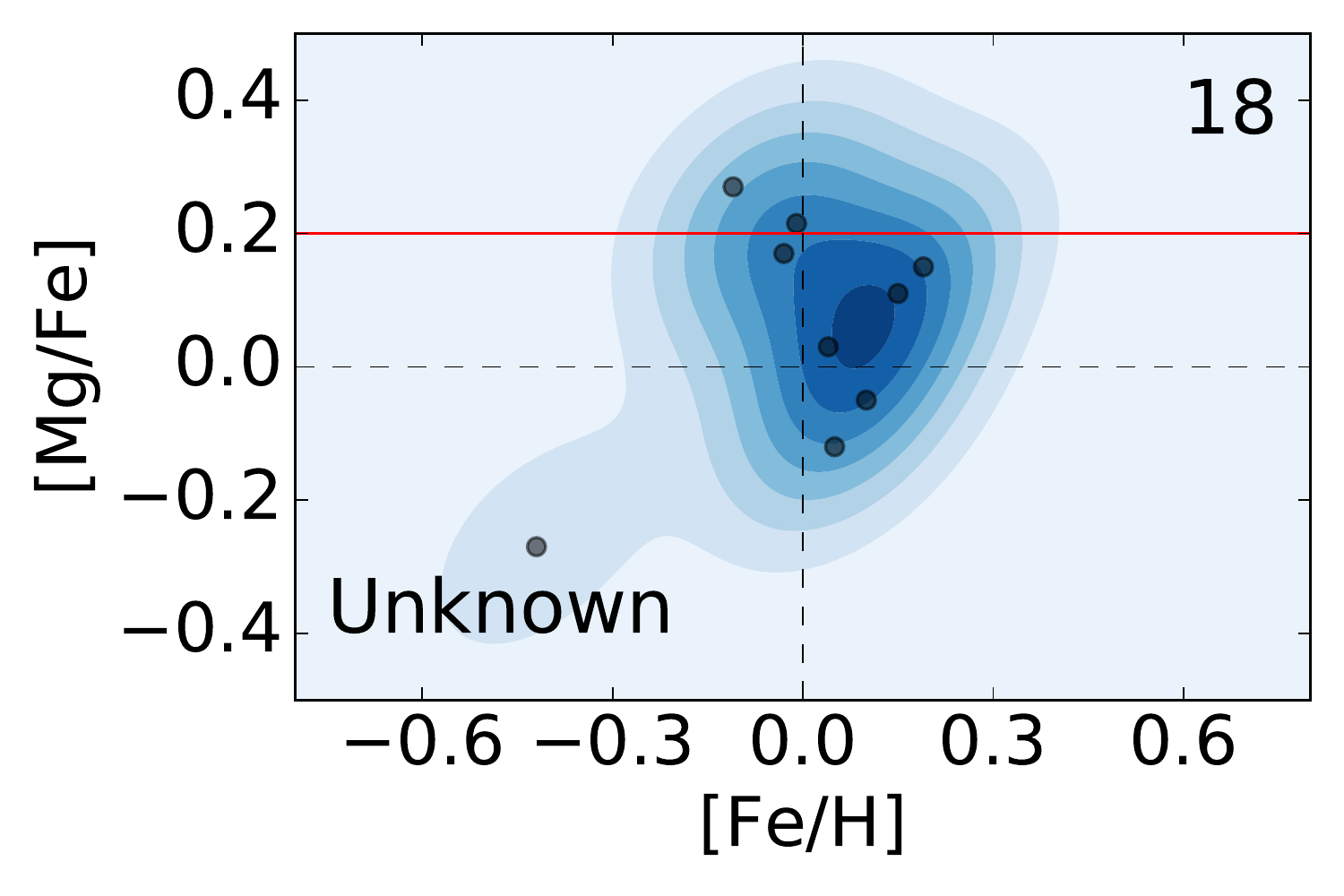}
   \includegraphics[viewport = 93 10 430 279,clip]{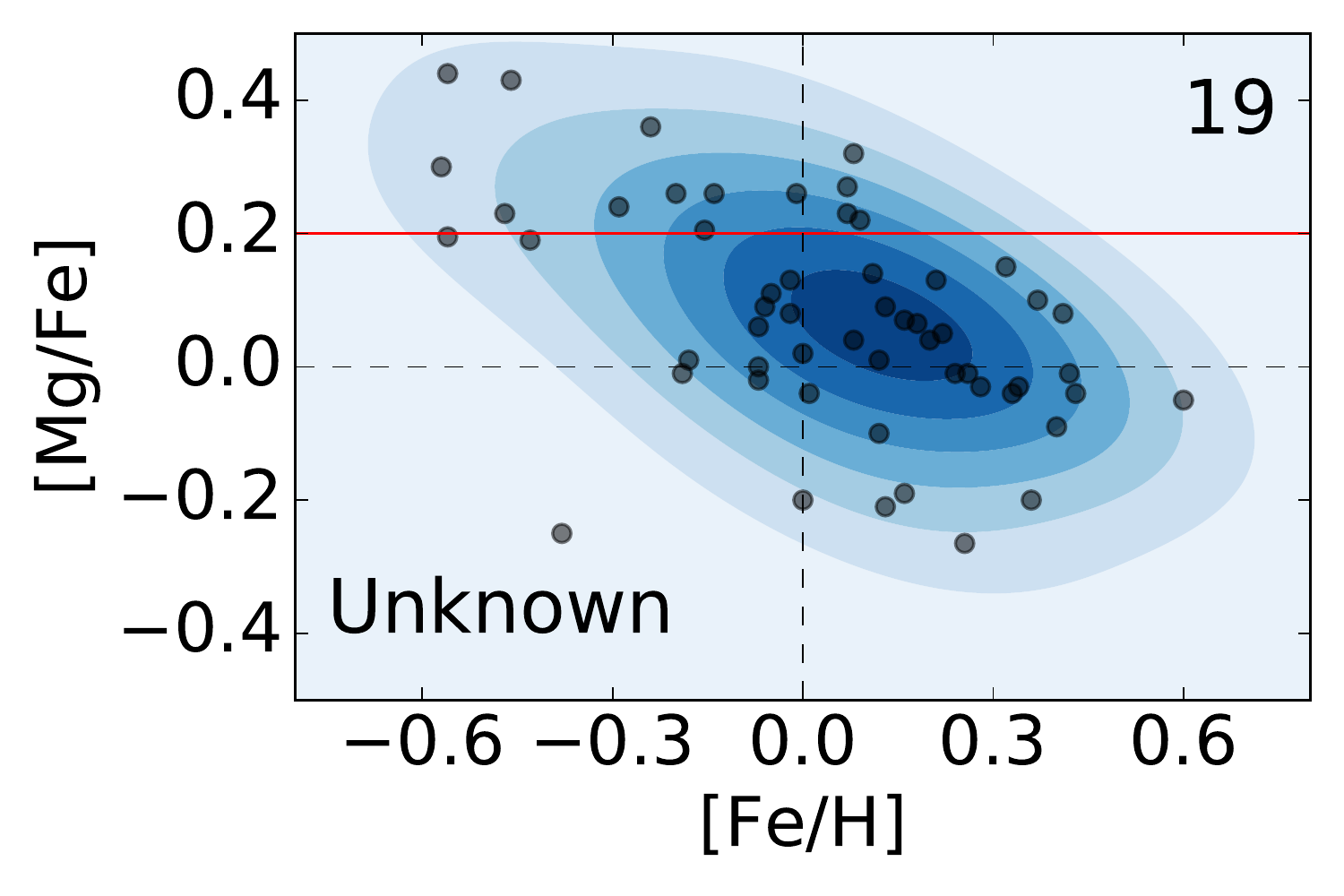}
   \includegraphics[viewport = 93 10 430 279,clip]{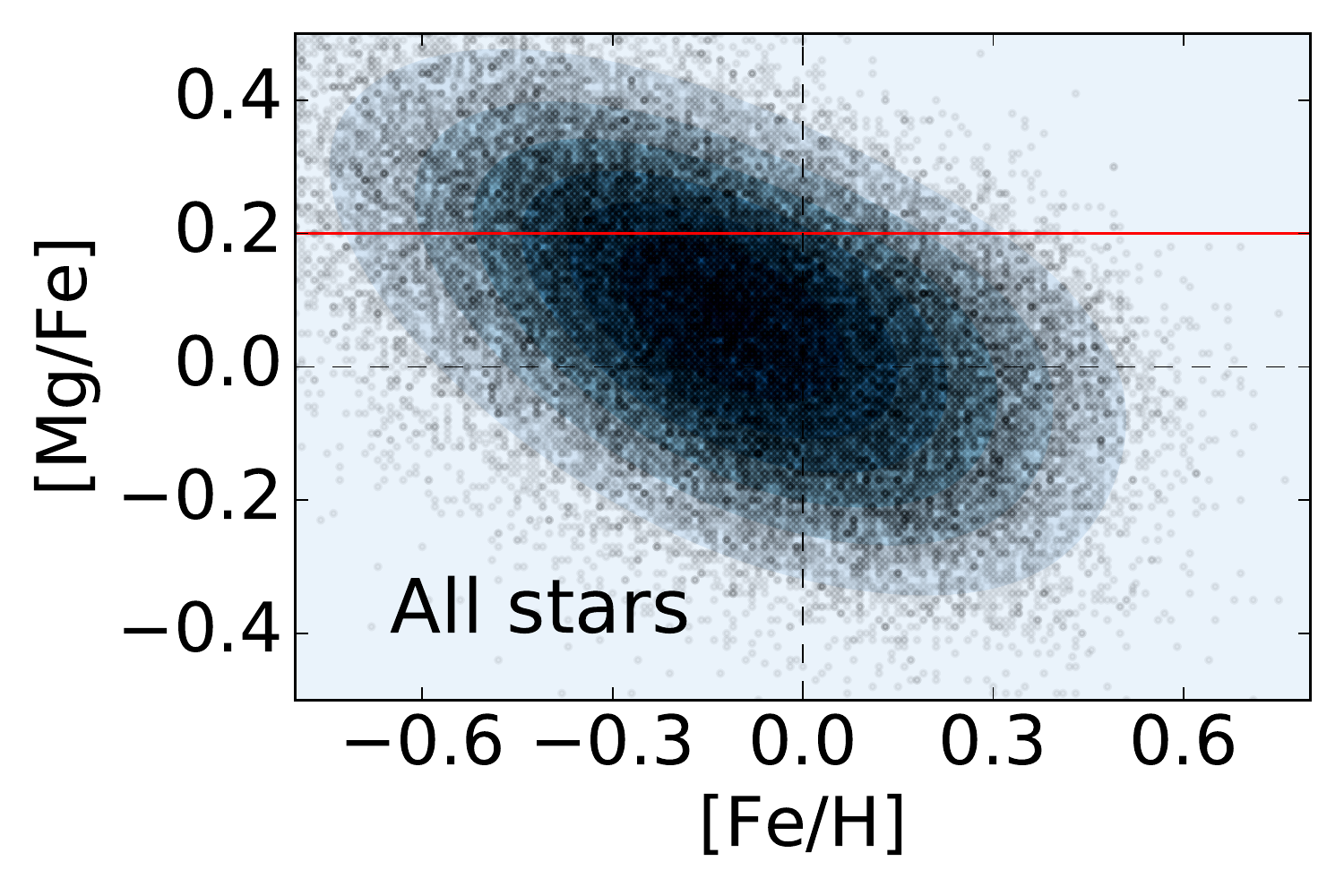}
   }
   \caption{
  [Mg/Fe]-[Fe/H] diagram for kinematic structures detected for the scale $J=3$. Numbers on each panel correspond to structures 1-19 as stated in the legend to Fig.~\ref{_j3}. The last plot corresponds to the total sample with limits on $S/N > 40$ (47\,849 stars). The red line at each density plot corresponds to $\rm [Mg/Fe]=0.2$ and divides the total sample into thick (above the line) and thin (below the line) disks. Dashed black lines show $\rm [Fe/H]=0$ and $\rm [Mg/Fe]=0$. Scatter plots at each density map show positions of individual stars at the diagram. Histograms at the top of each panel show the metallicity distribution for the total sample (black dashed line) and for each group (solid violet distribution). Values of $\mu$ and $\sigma$ represent the violet distribution.
   \label{_kde}
   }
\end{figure*}



In Fig.~\ref{_kde} (see the last plot on the right-hand side on the bottom line) we show [Mg/Fe]-[Fe/H] diagram for the total RAVE sample of 47\,849 stars that have RAVE signal to noise ratio $S/N > 40$. The last limit is needed to get abundances with a higher precision (abundances uncertainties less that 0.2\,dex, \citealt[see][]{_kunder17}). A chemical separation of thick and thin disks with RAVE based on probability density approach has been done in \citet{_wojno16} and we define a thin disk sample (D) and a thick disk sample (TD) samples according to \citet{_wojno16}: thin disk $\rm [Mg/Fe] < 0.2$, thick disk $\rm [Mg/Fe] > 0.2$. This separation is shown by the red horizontal line in all plots of Fig.~\ref{_kde}. 
Effective ranges of disk metallicities obtained for a RAVE sample by \citet{_wojno16} are the following: $\rm -0.27<[Fe/H]<0.38$ for an $\alpha$-poor component (thin disk) and $\rm -1.15<[Fe/H]<-0.05$ for an $\alpha$-enhanced component (thick disk). The metallicity distribution function for the total sample reaches the maximum value at $\rm [Fe/H]\approx-0.1$, which is close to the disk separation values, hence, the total sample represents a mixture disk populations. 
In the thin disk sample we have 36\,439 stars, and in the thick disk sample 11\,410 stars. As in the case with SN and BSN samples, we run the same procedure as for the full sample and the SN and BSN samples (i.e. applying the wavelet transform, filtering, and structure detection procedure for 2\,000 synthetic data samples.) 

Uncertainties for both [Fe/H] and [Mg/Fe] from RAVE are stated to be around 0.2\,dex \citep[see][]{_kunder17}, which is comparably large to make a clear separation of the two disks. The separation line shown in Fig.~\ref{_kde} is therefore only a first approximation to represent thick and thin disks. A better precision could be achieved with a detailed spectroscopic analysis of stars associated with kinematic structures to investigate which disk population do they belong to.

The bottom panels of Fig.~\ref{_samples} show the structures that were detected by applying wavelet transform to the 2\,000 synthetic samples associated with thick and thin disks, respectively. The rectangles correspond to the structures detected for the full sample at scale $J=3$. In Table~\ref{_groups} columns 11 and 12 show a clear presence of the structure in T and TD with ``$+$'' sign.

Similarly to the SN sample, the thin disk sample (D) contains more stars, so most of the structures detected for the full sample can be recognised. Only groups 17 and 18 appear to be missing. Hyades and Pleiades groups 7 and 8 are more distinct in the D sample, but a few stars are also detected in the TD sample, so they could be a mixture of the two stellar populations. The Hercules stream is almost missing in the TD map, so that probably is constructed mostly of thin disk stars. The same result was obtained by \citet{_bensby14} and \citet{_ramya16} from a chemical abundances analysis of stars that belong to the Hercules (for more discussions see Sect.~\ref{sec:sec:individual}). Coma Berenices slightly changed its location in the TD case, being more significant in the box 4. Groups 11, 12, 14, 15, $17-19$ are not seen at all. These groups consist of mostly D stars what points towards their possible origin through the outer Lindblad resonance (OLR) (for further explanations see Sect.~\ref{sec:sec:individual}).

In Fig.~\ref{_kde} we plot individually for each kinematic structure its $\rm [Mg/Fe]-[Fe/H]$ diagram and its metallicity distribution function which is a [Fe/H] versus a probability density, computed with the kernel density estimation (KDE) method. In each contour plot the positions of individual stars are shown as dots. The numbers in each panel indicate the numbers of the structures as listed in the legend to Fig.~\ref{_j3}. The horizontal red line at each density plot corresponds to the $\rm [Mg/Fe]=0.2$ showing the approximate separation between the thin and thick disks. Black dashed lines at each histogram show the probability density distribution for the full sample of 47\,849 stars with $S/N > 40$. The solid violet histogram at the top of each panel shows a probability density distribution for stars inside the current group. Each structure we will discussed in detail in Sect.~\ref{sec:sec:individual}.


\subsection{Individual structures}
\label{sec:sec:individual}

Individual structures will here be discussed in detail. Each case contains an overview of what is known about each group from the literature and how it compares with the results from the present study. The number in parentheses at the beginning of each paragraph indicates the number of the structure as listed in Table~\ref{_groups} and shown in Figs.~\ref{_j3} and \ref{_samples}.

\paragraph{{\bfseries Sirius} (1-3):}
\citet{_eggen92} studied the properties of the Sirius super-cluster, which is considered a part of the larger Ursa Major stream. They found that its stars fall into two distinct age groups, 6.3\,Gyr and 0.2\,Gyr, and that its chemical composition differs from the Hyades and Pleiades open clusters, showing heavy element abundances close to solar values. \citet{_famaey08} tried to reveal the origin of kinematic features including the Sirius stream by probing ages of stars that belong to the Sirius group and the evaporating Ursa Major star cluster. It was shown that only 30\,\% of the stars associated with the stellar stream fall on the same isochrone (300 Myr) as the open cluster, and, as was concluded in \citet{_famaey08}, not all stars of Sirius stream have an origin of being a remnant of an open cluster and favour a dynamical (resonant) origin for the Sirius stream.  
Later, through modelling of the dynamics of the Milky Way, \citet{_minchev10} showed that the low-velocity features including Sirius stream could be reproduced with the OLR of the Galactic bar.

\citet{_bovy10} studied the ages and metallicities of kinematic over-densities of nearby stars from Hipparcos \citet{esa1997} to investigate whether stellar streams consist of stars that belong to the same population, that could indicate that they originated from dissolved open clusters. Their main result was negative for the stellar streams they analysed, including the Sirius stream, and that it should not be associated with the Ursa Major open cluster. To test possible dynamical origins for the stellar streams (such as the OLR of the bar, or the inner Lindblad resonance (ILR) of the spiral structure) \citet{_bovy10} used the the Geneva-Copenhagen survey \citep{_holmberg09} to compare the metallicities of stellar stream stars to the background population of thin disk stars. \citet{_bovy10} assume that depending on the type of the resonance, orbits of stellar groups are located most of the time inside or outside the solar circle and consequently, these stars shows up the properties of different parts of the Galaxy. Metallicity is one of the main parameters that vary for kinematic groups that come from different parts of the Milky Way due to the existence of a metallicity gradient in the Galaxy. They found `weak evidence' that Sirius stream stars have lower metallicities than the thin disk population and could therefore be related to the ILR of the spiral arms.

We associate Sirius stream with structures 1-3 (see Fig.~\ref{_samples}). Sirius is elongated in both the $U$ and $V$ directions and is detected in all maps, although its shape and location vary from sample to sample. Structure~2 is the most significant sub-stream with more than 4800 stars located inside the `detection box', and 154\,\% of MC repeats. As the detection percentage exceeds 100\,\% the structure may consist of a few smaller groups like those detected at the scale $J=2$ (see Fig.~\ref{_uv_all}) that overlap with each other at the scale $J=3$. Below we provide a table of positions of the Sirius stream from this work and from the literature and a blue cross in Fig.~\ref{_j3} corresponds to the Sirius group from \citet{_antoja12}.

\begin{table}[h]
\begin{center}
\caption{Sirius positions}
\begin{tabular}{ll} 
\hline
\hline
\noalign{\smallskip}
 $(U,V)$  & Reference \\
 $[\kms]$ & \\
\noalign{\smallskip}
\hline
 $(30,-3)$    & group 1  \\ 
 $(0,8)$      & group 2  \\ 
 $(-11,9)$    & group 3  \\ 
 $(9,3)$      & \citet{_dehnen98} \\
 $(15,1)$     & \citet{_eggen92}  \\
 $(10,3)$     & \citet{_eggen96}  \\
 $(7,4)$      & \citet{_famaey05} \\
 $(5,2)$      & \citet{_famaey08} \\
 $(10,3)$     & \citet{_zhao09}   \\
 $(9,4)$      & \citet{_bovy10}   \\
 $(4,4)$      & \citet{_antoja12} \\
\hline
\end{tabular}
\end{center}
\end{table}

Our central peak 2 agrees with all the studies listed. Group 1 has a higher $U$ velocity and group 3 a lower $U$ velocity compared to the central peak but all have approximately the same $V$ velocity, so they could be members of one larger stream. Sirius is a nearby structure, while only stars from group 2 also appear in the distant BSN sample. Most of the stars appear to have chemical compositions comparable to what is observed for the Galactic thin disk stars, but group 2 is still strong in the thick disk sample. So Sirius could be a mixture of stars from both disk populations.

Plots 1-3 in Fig.~\ref{_kde} show the $\rm [Mg/Fe]-[Fe/H]$ diagrams for stars from groups 1-3 that we associate with the Sirius stream and at the top of each panel the metallicity distribution for each individual group is shown (solid violet distribution). The Sirius stream stars appear to have properties similar to the total sample (black dashed histogram) and do not show any particular metallicity trend inherent to the thick or thin disk populations.

Figure~\ref{_samples} also indicates that the Sirius stream is a large-scale structure that is observed in both SN and BSN samples and appears to be a mixture of both disk populations. Since we observe Sirius in both disks, we favour its dynamical origin possibly from the ILR of the spiral arms, but note that our thin/thick disk separation is uncertain due to the rather large errors in the RAVE chemical abundance ratios.

\paragraph{{\bfseries Coma Berenices} (4-6):}
\citet{_odenkirchen98} carried out an astrometric and photometric analysis of the region of the sky where the Coma Berenices open star cluster is located and found that the luminosity function of the core of the cluster decreases, while it increases towards fainter magnitudes in the edges of the cluster. \citet{_odenkirchen98} assume that there could be a lot of faint, low-mass members of the moving group that were not observed. The proximity of the moving group and the open cluster pointed \citet{_odenkirchen98} towards the idea that Coma Berenices moving group was formed due to a dissolution of the cluster. Conversely, \citep{_minchev10} through modelling of the dynamics of the Milky Way, reproduced a few main stellar streams including Coma Berenices assuming the OLR of the bar and thus, favour resonant mechanism of formation of also this kinematic over-density.

The table below lists the detection of the Coma Berenices kinematic over-density in the $U-V$ plane that is available in the literature. In our study Coma Berenices is associated with the structures 4-6 (see Fig.~\ref{_samples}) and the table below shows that the positions we detect are in agreement with results from the other studies.

\begin{table}[h]
\begin{center}
\caption{Coma Berenices positions}
\begin{tabular}{ll} 
 \hline
 \hline
\noalign{\smallskip}
 $(U,V)$  & Reference \\
 $[\kms]$ & \\
\noalign{\smallskip}
 \hline
 $(9,-12)$     & group 4 \\ 
 $(-2,-11)$    & group 5 \\ 
 $(-15,-7)$    & group 6 \\ 
 $(-10,-5)$    & \citet{_dehnen98} \\
 $(-10,-10)$   & \citet{_famaey07} \\
 $(-11,-7)_d$  & \citet{_zhao09} dwarf sample  \\
 $(-13,-6)_g$  & \citet{_zhao09} giant sample  \\
 $(-7,-6)$     & \citet{_antoja12} \\
 $(-7,-6)$     & \citet{_bobylev16} \\
\hline
\end{tabular}
\end{center}
\end{table}

All three groups 4-6 share similar space velocities. Peak 6 has higher detection percentage in MC simulation (79\%) than peaks 4 and 5. While group 5 is the biggest and contains over 2700 stars inside the `detection box'. The blue cross inside box 5 in Fig.~\ref{_j3} corresponds to the detection of Coma Berenices from \citet{_antoja12} at ($-$7, $-$6) $\kms$.

Figure~\ref{_kde} plots 4-6 show the $\rm [Mg/Fe]-[Fe/H]$ diagrams and metallicity histograms (at the top of each plot) for groups 4-6, respectively. Coma Berenices stream stars show metallicity properties similar to the total sample, and does not show any particular metallicity trend to either the thick or the thin disks. Figure~\ref{_samples} shows a similar result: Coma Berenices unites stars that belong to both thin and thick disk samples with more stars in the thin disk sample. It is a large scale over-density because it is seen in both distance samples.

As Coma Berenices shares properties similar to the Sirius moving group, both combining stars of different populations, it could also originate from the ILR of the spiral arms, again with the remark that the thin/thick disk separation is uncertain due to the low precision of the RAVE abundances.

\paragraph{{\bfseries Hyades} (7):}
Being first discovered by \citet{_proctor1869}, the Hyades stream, or Hyades super-cluster, was for a long time considered to be a remnant of the eponymous Hyades open stellar cluster. However, recent studies have shown the opposite. For instance, \citet{_famaey08} found that only half of stars of the Hyades stream could originate from the Hyades open cluster as only about 50\,\% of stars fall onto the same isochrone as would have been expected for an open cluster. They favour the dynamical origin for the Hyades stream.

Later, \citet{_pompeia11} compared chemical abundances and metallicities of stars belonging to Hyades stream with stars that are members of the Hyades open cluster, that is believed to be chemically homogeneous. It was found that only 2 of the 21 selected Hyades stream stars have similar chemical properties with the open cluster. Furthermore, \citet{_pompeia11} showed that the Hyades stream stars have a metallicity excess of about $0.06-0.15$\,dex compared to thin disk stars, which is consistent with an origin caused by the ILR of the spiral arms. They also performed a particle simulation test that supported the same conclusion, showing that the Hyades stream could be reproduced with the 4:1 resonance of the spiral arms.

Another chemical tagging study of the Hyades stream was presented by \citet{_tabernero12} that further supported the idea of dynamical origin of the Hyades stream. They analysed stellar spectra of 61 Hyades stream stars and compared the results with a reference star vB~153 that is a verified member of the Hyades open cluster. Only 26 stars were found to have similar parameters with the Hyades open cluster. \citet{_tabernero12} conclude that the Hyades stream does not completely originate from the Hyades open cluster.

\citet{_mcmillan11} used a simple dynamical modelling of the Milky Way to study the origin of the Hyades stream and to check whether it could originate from a Lindblad resonance. The author conclude that Hyades stream has a resonant (dynamical) nature, but that it is not possible to say exactly which resonance due to selection effects associated with the dynamics. 

bf{We associate Hyades with group 7 (see Fig.~\ref{_samples}). This group contains 2344 stars inside the detection box and has a high MC detection of 90\,\%.} The blue cross in Fig.~\ref{_j3} marks the detection of Hyades from \citet{_antoja12}. Below we show a table of positions of Hyades from this work and from the literature.

\begin{table}[h]
\begin{center}
\caption{Hyades positions}
\begin{tabular}{ll} 
 \hline
 \hline
\noalign{\smallskip}
 $(U,V)$  & Reference \\
 $[\kms]$ & \\
\noalign{\smallskip}
 \hline
 $(-44,-18)$   & group 7            \\ 
 $(-40,-20)$   & \citet{_dehnen98}  \\
 $(-35,-18)$   & \citet{_famaey08}  \\
 $(-38,-18)_d$ & \citet{_zhao09} dwarf sample   \\
 $(-38,-17)_g$ & \citet{_zhao09} giant sample   \\
 $(-40,-20)$   & \citet{_bovy10}    \\
 $(-30,-13)$   & \citet{_antoja12}  \\
 $(-30,-15)$   & \citet{_bobylev16} \\
\hline
\end{tabular}
\end{center}
\end{table}

The $\rm [Mg/Fe]-[Fe/H]$ diagram the and [Fe/H] distribution for structure 7 is shown in  Fig.~\ref{_kde}. The Hyades stream shows properties that are similar to the full sample. From the analysis of SN/BSN and D/TD sub-samples (see Fig.~\ref{_samples}) it is seen that Hyades stream sample mostly consists of nearby stars. It is also more distinct in the thin disk subsample, although the structure is detected in the thick disk subsample too. So, it appears as if the Hyades stream is nearby structure which consists of mixture of disk populations. This does not support the hypothesis for Hyades to be a dissolved open cluster as this theory implies all stars to have a solid chemical composition and could have a dynamical origin, again with the remark on a low precision of abundances given in RAVE.

\paragraph{{\bfseries Pleiades} (8):} 
The Pleiades was the first ever discovered moving group. \citet{_madler1846} found it through observations of Pleiades open cluster that there was a large number of stars located a few degrees far from the cluster, that were moving in the same direction. It was the Pleiades moving group. Its origin has been investigated in several studies. For example, \citet{_famaey08} conclude that Pleiades moving group has a dynamical (resonant) origin, since only 46\,\% of the moving groups stars fall onto the 100\,Myr isochrone that is the assumed age for the Pleiades open cluster.

Through galactic dynamics modelling \citet{_minchev10} reproduced stellar streams being due to the OLR. However, to be consistent with the number of stars in the Pleiades they assumed that the Milky Way bar was formed 2\,Gyr ago. This paper stands against the idea that the Pleiades and the Hyades share a common dynamical origin. \citet{_bovy10} analysed age and metallicity properties the Pleiades moving group and found that it could not originate through a dissolved Pleiades open cluster as their stellar populations differ. \citet{_bovy10} also compared the metallicity of the Pleiades and Hyades moving groups with the metallicity of the thin disk population and found similar metallicities for the Pleiades and the thin disk stars, while the Hyades shows a higher metallicity than thin disk. Hence, also \citet{_bovy10} does not support the idea of common dynamical origin for the Pleiades and Hyades. 

We detect one large structure that we associate with Pleiades, group 8 in Fig.~\ref{_samples}. This group could consist of a few separate groups that overlap since the percentage of detection in MC simulations is 170\,\%. Interestingly, at the $J=2$ scale the Pleiades detection consists of three separate structures, number 12-14 (see Fig.~\ref{_uv_all}). Group 8 (in the $J=3$ scale) is one of the largest groups with about 4200 stars inside the detection box. The table below gives the positions for Pleiades stream from this work and from the literature. The blue cross corresponding to Pleiades in Fig.~\ref{_j3} refers to the detection by \citet{_antoja12}.

\begin{table}[h]
\begin{center}
\caption{Pleiades positions}
\begin{tabular}{ll} 
 \hline
 \hline
\noalign{\smallskip}
 $(U,V)$  & Reference \\
 $[\kms]$ & \\
\noalign{\smallskip}
 \hline
 $(-22,-23)$   & group 8  \\ 
 $(-12,-22)$   & \citet{_dehnen98} \\
 $(-12,-21)$   & \citet{_eggen96}  \\
 $(-16,23)$    & \citet{_famaey08} \\
 $(-12,-23)_d$ & \citet{_zhao09} dwarf sample  \\
 $(-15,-23)_g$ & \citet{_zhao09} giant sample  \\
 $(-15,-20)$   & \citet{_bovy10}   \\
 $(-16,-22)$   & \citet{_antoja12} \\
 $(-13,-24)$   & \citet{_bobylev16} \\
\hline
\end{tabular}
\end{center}
\end{table}

Plot 8 in Fig.~\ref{_kde} shows the  $\rm [Mg/Fe]-[Fe/H]$ diagram and the [Fe/H] histogram the Pleiades stars. The metallicity distribution is almost equal to the full sample. The structure that we associate with the Pleiades does not have any particular distance or abundance dependence  as it is observed in both SN/BSN, and both D/TD samples (see Fig.~\ref{_samples}). The structure has a higher detection percentage for the thin disk sample, but this could be because the thin disk sample contains three times as many stars than the thick disk sample. The position and the shape of Pleiades group do not vary much between the different sub-samples and leads to the conclusion that it is a large-scale structure, composed of a mixture of different populations of stars. Thus, as it appears to be chemically inhomogeneous, unlike open clusters and moving groups, it could originate from the ILR of the spiral arms and not from the Pleiades open cluster. Again, a better thin/thick disk separation could be achieved with more precise chemical abundances than what RAVE is providing.

\paragraph{{\bfseries Hercules} (10-11):}
Being the largest and the most elongated structure in the $U$ direction, the origin of the Hercules stream has been investigated by many authors. For example, \citet{_dehnen00} favour a hypothesis that Hercules stream is a dynamical feature caused by the Galactic bar resonances (the OLR). \citet{_chakrabarty07} showed that a combined dynamical effect of spiral arms and a Galactic bar can explain main kinematic structures including the Hercules. \citet{_bensby07} performed a detailed chemical characterisation of its stars. They favour a dynamical origin through the Galactic bar as the Hercules stream stars appeared to be a mixture of thick and thin disk stars. Also \cite{_bovy10} performed a hypothesis testing to check whether moving groups consist of homogeneous population of stars. The results was negative and \cite{_bovy10} further found indications that Hercules stream stars have a higher average metallicity than the local thin disk, hence, concluding that it could be a structure caused by the OLR of the Galactic bar. Later, \citet{_bensby14} re-examined the chemical composition of Hercules stream stars and found that it mainly consists of stars that chemically can be associated with both the thin and thick disks. \citet{_ramya16} on the other hand studied 58 Hercules stream red giants and found that they are mostly metal-rich stars from the thin disk. The somewhat discrepant results could be explained with different target selection methods used by the two studies. 

\citet{_perez17} carried out a dynamical modelling of the Hercules stream ``in the framework'' of a slow bar and compared obtained results with data from the RAVE and LAMOST catalogues. They found that Hercules is more prominent in the Galactic inner disk and should consist in average of more metal-rich and older stars comparing to the Solar negihbourhood.

Hercules is identified as structures 10 and 11 in this study (see Fig.~\ref{_samples}). This kinematic structure is the most elongated feature in the $U$ direction and has a detection percentage for group number 10 that exceeds 100\,\%. An explanation of this result is that it appears to consist of a few separate structures that overlap in the MC simulations (see the $J=2$ scale in Fig.~\ref{_uv_all}, where Hercules is detected as the two peaks number 19 and 20. The blue crosses in Fig.~\ref{_j3} (one is inside the Hercules box 10 and another is just outside on the left-hand side) mark the results from \citet{_antoja12}. The table below gives the positions of the Hercules stream from this work and from the literature.

\begin{table}[h]
\begin{center}
\caption{Hercules stream positions}
\begin{tabular}{ ll } 
 \hline
 \hline
\noalign{\smallskip}
 $(U,V)$  & Reference \\
 $[\kms]$ & \\
\noalign{\smallskip}
 \hline
 $(-38,-49)$   & group 10  \\ 
 $(-16,-48)$   & group 11  \\ 
 $(-30,-50)$   & \citet{_eggen96}  \\
 $(-42,-51)$   & \citet{_famaey05} \\
 $(-35,-51)$   & \citet{_famaey08} \\
 $(-32,-48)_d$ & \citet{_zhao09} dwarf sample  \\
 $(-35,-51)_g$ & \citet{_zhao09} giant sample  \\
 $(-20,-33)$     & \citet{_bovy10}   \\
 $(-57,-48)_I$   & \citet{_antoja12} \\
 $(-28,-50)_{II}$& \citet{_antoja12} \\
 $(-57,-48)_I$   & \citet{_bobylev16} \\
 $(-35,-50)_{II}$& \citet{_bobylev16} \\
\hline
\end{tabular}
\tablefoot{
\tablefoottext{$\dagger$}{
$I$ and ${II}$ mark sub-streams found in the structure}}
\end{center}
\end{table}

The $(U,V)$ velocities of groups 10 and 11 are in agreement with most of the previous studies except \citet{_bovy10}, whose position differs from others by about $10\kms$. \citet{_antoja12} and \citet{_bobylev16} defined two sub-streams in the Hercules, we have only one centred peak, but the size of the structure covers both of them.

The panels 10 and 11 in Fig.~\ref{_kde} corresponds to the Hercules stream and shows its metallicity distribution and $\rm [Mg/Fe]-[Fe/H]$ diagram. It appears to contain more metal-rich stars, and it is also clearly a thin disk structure located in the nearby sample as it is observed only in the SN sample  (see Fig.~\ref{_samples}). Our results support recent findings that the Hercules stream mainly belong to the thin disk population and could be due to the OLR of the Galactic bar.

\paragraph{{\bfseries Wolf 630} (9):}
Wolf 630 was first identified by \citet{_eggen65} and its origin is still unclear. \citet{_bubar10} analysed spectra of 34 stars of the Wolf 630 stream and 19 stars were found to be chemically homogeneous. This sub-sample of 19 stars was fitted with a 2.7\,Gyr isochrone and a metallicity of $\rm [Fe/H]=-0.01$. \citet{_bubar10} suggest that the sub-sample of 19 stars could be a remnant of an open cluster since its stars share similar features, but the rest of the sample is inhomogeneous, which makes the origin of Wolf 630 uncertain.

We identify Wolf 630 as the group 9 (see Fig.~\ref{_samples}). It has a 168\% MC detection rate and 1777 stars of our sample can be associated within the group. At the $J=2$ scale the same region of the $U-V$ plane consists of four individual groups. This could indicate that group 9 consists of at two structures that overlap: Wolf 630 and Dehnen98 (to be discussed below). The result from \citet{_antoja12} is marked by the blue cross inside the structure 9 (see Fig.~\ref{_j3}). The table below gives positions of Wolf 630 obtained in this work and from the literature.

\begin{table}[h]
\begin{center}
\caption{Wolf 630 positions}
\begin{tabular}{ll} 
 \hline
 \hline
\noalign{\smallskip}
 $(U,V)$  & Reference \\
 $[\kms]$ & \\
\noalign{\smallskip}
 \hline
 $(43,-22)$    & group 9  \\ 
 $(23,-33)$    & \citet{_eggen65} \\
 $(28,-21)$    & \citet{_antoja12} \\
 $(29,-21)$    & \citet{_bobylev16} \\
\hline
\end{tabular}
\end{center}
\end{table}

Our $U$-component differs from the other works by at least in $10 \kms$. This can be a consequence that box 9 corresponds to at least two independent groups and thus its position represent mean coordinates for both groups. Plot 9 in Fig.~\ref{_kde} corresponds to the Wolf~630 stream and shows its metallicity distribution and $\rm [Mg/Fe]-[Fe/H]$ diagram. It has a metallicity properties very similar to the full sample, but maybe with a few more metal-rich stars. Based on the analysis of SN/BSN and D/TD sub-samples in Fig.~\ref{_samples}, Wolf 630 appears to be a thin disk structure belonging to the nearby sample. It could have a resonant origin with the remark on the uncertainties of the RAVE abundances that makes the disk separation less reliable.

\paragraph{{\bfseries Dehnen98} (9):} 
This structure is detected inside box 9 in Fig.~\ref{_samples}. It is a small kinematic group that was first discovered by \citet{_dehnen98} and has later been confirmed by other studies \citep[e.g.][]{_antoja12, _bobylev16}. \citet{_antoja08} found a group with the same $(U,V)$ coordinates, but after the analysis of the branch structure using modified equations as was first proposed by \citet{_skuljan99} to fit four branches of groups based on its motion, they concluded that the group could belong to the Coma Berenices stream. \citet{_antoja12} sub-sequently detected a kinematic over-density which they associated with the Dehnen98 structure. This result is marked by a blue cross at the right-hand side of the box 9 in Fig.~\ref{_j3}. The table below gives positions of this structure found in this work and from the literature.

\begin{table}[h]
\begin{center}
\caption{Dehnen98 positions}
\begin{tabular}{ll} 
 \hline
 \hline
\noalign{\smallskip}
 $(U,V)$  & Reference \\
 $[\kms]$ & \\
\noalign{\smallskip}
 \hline
 $(43,-22)$    & group 9  \\ 
 $(50,-25)$    & \citet{_dehnen98}  \\
 $(48,-24)$    & \citet{_antoja12}  \\
 $(43,-24)$    & \citet{_bobylev16} \\
\hline
\end{tabular}
\end{center}
\end{table}

Our detection of the Dehnen98 structure is in agreement with previous works. Dehnen98 has a very high percentage of detection in the MC simulations compared to other groups that we have detected: $\sim$98\% MC catches and it contains 58 stars. The metallicity distribution and the $\rm [Mg/Fe]-[Fe/H]$ diagram of Dehnen98 is given in plot 9 in Fig.~\ref{_kde}. Dehnen98 has similar metallicity properties to the total sample, with more thin disk stars. From the analysis of Fig.~\ref{_samples} Dehnen98 contains stars of different populations and belongs to the nearby sample. Concerning the assumption stated in \citet{_antoja08} that Dehnen98 is a part of a Coma Berenices branch, we can say that this group has similar properties with Wolf~630 and Coma Berenices streams, and they all could form to one large-scale structure that has a dynamical origin. A detailed chemical tagging of stars that belong to these groups is required to properly speculate on its origin.

\paragraph{{\bfseries $\gamma$Leo} (12):}
This structure is shown as group 12 in Fig.~\ref{_samples}, and has a relatively low detection percentage of 27\% in the MC simulations. It is rather small with only 96 stars from our sample. Figure~\ref{_j3} shows two blue crosses for this group from \citet{_antoja12}. The table below gives velocity positions of our detection of $\gamma$Leo together with those from the literature. 

\begin{table}[h]
\begin{center}
\caption{$\gamma$Leo positions}
\begin{tabular}{ll} 
 \hline
 \hline
\noalign{\smallskip}
 $(U,V)$  & Reference \\
 $[\kms]$ & \\
\noalign{\smallskip}
 \hline
 $(52,0)$      & group 12  \\ 
 $(50,0)$      & \citet{_dehnen98}  \\ 
 $(56,2)_I$    & \citet{_antoja12}  \\
 $(68,1)_{II}$ & \citet{_antoja12}  \\
 $(65,1)$      & \citet{_bobylev16} \\
\hline
\end{tabular}
\tablefoot{
\tablefoottext{$\dagger$}{$I$ and ${II}$ mark sub-streams found in the structure}
}
\end{center}
\end{table}

Group 12 is consistent with \citet{_dehnen98} and \citet{_antoja12} peak $I$, while \citet{_bobylev16} is in agreement with the structure II from \citet{_antoja12}. All the groups have similar $V$-velocities. Plot 12 in Fig.~\ref{_kde} shows the $\rm [Mg/Fe]-[Fe/H]$ diagram and the [Fe/H] distribution for group 12. The $\gamma$Leo stream shows metallicity properties similar to the total sample and it appears to be a nearby thin disk structure (see Fig.~\ref{_samples}) with only a few stars in the TD sample. Thus, it could have formed due to dynamical reasons.

\paragraph{{\bfseries New} (13):}
Group 13 at $(37, 8)\,\kms$ in Fig.~\ref{_samples} has 201 stars and a high detection level of 74\% MC repeats. We cannot find any previous detections in the literature of a structure at these coordinates, and we therefore identify this as a New structure. It appears to be a nearby structure and is detected in both the thin and the thick disk sub-samples. It is, however, not detected in the more distant BSN sample, which could be due to smaller number of stars in the BSN sample compared to the SN sample. The metallicity distribution and $\rm [Mg/Fe]-[Fe/H]$ diagram for this new group 13 are shown in plot 13 in Fig.~\ref{_kde}. This group contains stars of both disk populations. It can be an elongation of larger nearby streams such as Sirius or $\gamma$Leo, as their properties are similar. A more precise detailed chemical analysis of stars associated with these groups is required to more precisely probe the origin of the new group 13.

\begin{table}[h]
\begin{center}
\caption{Position of new structure detected in this work}
\begin{tabular}{ll} 
 \hline
 \hline
\noalign{\smallskip}
 $(U,V)$  & Reference \\
 $[\kms]$ & \\
\noalign{\smallskip}
 \hline
 $(37,8)$    & group 13  \\ 
\hline
\end{tabular}
\end{center}
\end{table}

\paragraph{{\bfseries Antoja12(15)} (14):}
This structure was first reported in \citet{_antoja12} but was detected only at 2$\sigma$ confidence level and needed further confirmation. In Fig.~\ref{_j3} it is shown as a blue cross close to the box 14. We received a $3\sigma$-significant group 14 which is 10 $\kms$ higher in $U$, but can be associated with the one detected in \citep{_antoja12}. It has only 6\% of MC detection and accounts 8 stars. Below we show a list of positions we found in the literature for this structure and included our results.

\begin{table}[h]
\begin{center}
\caption{Antoja12(15) positions}
\begin{tabular}{ll} 
 \hline
 \hline
\noalign{\smallskip}
 $(U,V)$  & Reference \\
 $[\kms]$ & \\
\noalign{\smallskip}
 \hline
 $(48,-68)$ & group 14           \\ 
 $(60,-72)$ & \citet{_antoja12}  \\
 $(72,-64)$ & \citet{_bobylev16} \\
\hline
\end{tabular}
\end{center}
\end{table}

This group appears clearly in the nearby and the thin disk sub-samples (see Fig.~\ref{_kde}). Taking into account the low number of stars associated with this group it could not be observed in the BSN and TD samples as they consist of less stars than SN and D samples. The metallicity distribution and the $\rm [Mg/Fe]-[Fe/H]$ diagram for group 14 are shown in plot 14 in Fig.~\ref{_samples} and both point toward the thin disk population, which is coherent with the result from Fig.~\ref{_kde}. The {\it Gaia} DR2 data will provide astrometric data for more stars, thus, one could verify whether this group is observed in the BSN and TD samples too. With the current results a dynamical origin seems favoured.

\paragraph{{\bfseries Antoja12(12)} (15):} 
This group was stated as new in \citep{_antoja12} and is marked by a cross in Fig.~\ref{_j3} close to structure 15. In this study, as in \citep{_antoja12}, structure 15 was detected with a 3$\sigma$-significance. The table below gives the positions for this group obtained in this work and from the literature.

\begin{table}[h]
\begin{center}
\caption{Antoja12(12) positions}
\begin{tabular}{ll} 
 \hline
 \hline
\noalign{\smallskip}
 $(U,V)$  & Reference \\
 $[\kms]$ & \\
\noalign{\smallskip}
 \hline
 $(94,-13)$ & group 15           \\ 
 $(92,-23)$ & \citet{_antoja12}  \\
 $(91,-35)$ & \citet{_bobylev16} \\
\hline
\end{tabular}
\end{center}
\end{table}

Our group 15 shares the same $U$ velocity as in the other studies, but differs in $V$ direction by $-10\,\kms$ compared to \citep{_antoja12}. Interestingly, \citet{_bobylev16} obtained a structure which also differs by $10\,\kms$ in the $V$ direction, but in the other direction. It can be the same structure as it is located in the low-density region of the $U-V$ map, so it cannot be affected by other stronger streams. Antoja12(12) has a 38\,\% detection in the MC simulations and includes only 10 stars. Group 15 appears to be a thin disk structure mainly present in the nearby sample (see Fig.~\ref{_samples}). The metallicity distribution and $\rm [Mg/Fe]-[Fe/H]$ diagram for group 15 are shown in plot 15 in Fig.~\ref{_kde} and its properties are similar to the full sample. We suppose that this group is an independent one, but has to be confirmed in later studies that contain more stars. The {\it Gaia} DR2 data release may help to resolve this case.

\paragraph{{\bfseries Bobylev16} (16):} 
This group has 14 stars and 17\,\% of MC detection. It was first discovered in \citep{_bobylev16} and is shown with a blue cross on the left-hand side of structure 16 in Fig.~\ref{_samples}. We confirm this group and add that it belongs to all both nearby and distant, thin and thick disk samples, what allows to state that it is a mixture of different type stars. 

\begin{table}[h]
\begin{center}
\caption{Bobylev16 positions}
\begin{tabular}{ll} 
 \hline
 \hline
\noalign{\smallskip}
 $(U,V)$  & Reference \\
 $[\kms]$ & \\
\noalign{\smallskip}
 \hline
 $(-94,-5)$  & group 16           \\ 
 $(-96,-10)$ & \citet{_bobylev16} \\
\hline
\end{tabular}
\end{center}
\end{table}

The same as group 15, structure 16 is observed far aside from the majority of kinematic groups. This supports group's independence from other structures, but unlike group 15 is present all samples. The metallicity distribution and [Mg/Fe]-[Fe/H] diagram for group 16 are shown in pattern 16 (see Fig.~\ref{_kde}) and it appears to have more thin disk stars. We propose its dynamical origin similar to the Sirius group since the sample properties are alike.

\paragraph{{\bfseries $\epsilon$Ind} (17):}
The closest blue cross to the group 17 in Fig.~\ref{_j3} is the one previously found at $2\sigma$ confidence level by \citep{_antoja12} that is listed in the table below. Although the structure is detected at the $3\sigma$-significane level, it has a low percentage of detection, only 12\,\% and contains only 24 stars. This group appears to be detected only in the nearby sample, but this could be due to the fact that this group contains very few stars.

\begin{table}[h]
\begin{center}
\caption{$\epsilon$Ind positions}
\begin{tabular}{ll} 
 \hline
 \hline
\noalign{\smallskip}
 $(U,V)$  & Reference \\
 $[\kms]$ & \\
\noalign{\smallskip}
 \hline
 $(-88,-48)$ & group 17           \\
 $(-81,-42)$ & \citet{_antoja12}  \\
 $(-90,-49)$ & \citet{_bobylev16} \\
\hline
\end{tabular}
\end{center}
\end{table}

Group 17 is a small group and is thus easier to detect in the larger SN sample. However, it is not detected in the larger thin disk sample that has 5\,000 more stars than the SN sample. The metallicity distribution and $\rm [Mg/Fe]-[Fe/H]$ diagram for group 17 are shown in plot 17 in Fig.~\ref{_kde}, it appears to mainly be a thin disk structure. To speculate on the origin of this kinematic feature {\it Gaia}~DR2 data should be used to have larger stellar sample.

\paragraph{{\bfseries Two tentatively new structures} (18-19), $J=2$:} These two groups have a low structure count in MC simulations and contains 12 and 70 stars respectively. Group 19 can be associated with HR1614 peak detected at (15, $-60$) $\kms$ by \citep{_dehnen98} (marked by a blue cross in Fig.~\ref{_j3}), but none of the groups have similar velocities. Tentatively we define them as new structures, but require further confirmation with larger data samples.

\begin{table}[h]
\begin{center}
\caption{Position of two tentatively new structures}
\begin{tabular}{ll} 
 \hline
 \hline
\noalign{\smallskip}
 $(U,V)$  & Reference \\
 $[\kms]$ & \\
\noalign{\smallskip}
 \hline
 $(-88,-76)$ & group 18           \\
 $(-18,-67)$ & group 19           \\
\hline
\end{tabular}
\end{center}
\end{table}

The metallicity distributions and $\rm [Mg/Fe]-[Fe/H]$ diagrams for groups 18-19 are shown in plots 18-19 in Fig.~\ref{_kde}. Structure 19 appears to be more a thin disk structure, and structures 18 is seen only in SN sample, which could be a consequence of the groups small sizes. {\it Gaia}~DR2 will help us to further investigate the existence and origin of these two structures.

\section{Summary and discussion}
\label{sec:discussion}

We have analysed the velocity distribution of 55\,831 {\it Gaia}~DR1/TGAS stars in the Solar neighbourhood and sample properties relative to distance and metallicity using wavelet analysis. 19 kinematic structures were detected at scales 8-16 $\kms$, 32 groups found between 4-8 $\kms$ and 4 structures at 16-32 $\kms$ in the $U-V$ plane. Our analysis has several advantages comparing to previous works. As it is the first ever analysis of {\it Gaia}~DR1 data in such a kinematical context, and the most important benefit is the precision of astrometry provided by TGAS itself. High precision of the input data allow us to apply the analysis for a larger sample of stars than in previous works, and even after cutting the sample based on $\sigma_U$ and $\sigma_V<4 \kms$ we still have a competitive number of stars. This limit on velocity uncertainties is important to obtain robust measurements of positions of kinematic structures. In previous works velocity uncertainties were either not accounted at all, either were established too high to retain more stars in the sample, that could led to uncertain results in both cases.

A set of $3\sigma$-significant (99.8 \%) wavelet coefficients that indicate kinematic structures were received after applying the wavelet analysis and filtering the data. Although the output data was already smoothed with the auto-convolution histogram method, the question whether obtained structures are real remained due to the availability of velocity uncertainties. Then we run Monte Carlo simulations and apply the same analysis to them as for the real sample. This step is beneficial for the procedure in general as it allows to calculate the percentage of detection which indicates if the structures are real.

To investigate properties of obtained structures with respect to distance and chemical composition four sub-samples were defined: a Solar neighbourhood sample with stars closer than 300\,pc (SN), a sample with more distant stars (BSN), and based on [Mg/Fe] enhancement (from RAVE abundances) a thick disk sample (TD) and a thin disk sample (D). As shown is Sect.~\ref{sec:sec:sn}, \ref{sec:disks}, some structures are SN/BSN and/or D/TD structures. For example, group 10 (Hercules) is obviously SN/D structure, while group 4 (Coma B part) is a BSN/TD structure. Most of the moving groups are observed at close distances $d < 300$\,pc and at higher metallicities. This can be a repercussion of the selection effect since SN and D samples contain more stars compared to the BSN and TD samples. Some groups change their positions and shapes when considering distance and metallicity (e.g. group 7 (Hyades), and group 2 (Sirius)). These variations could be a consequence of how the sample is split, where the SN and D samples contain more stars than BSN and TD samples, but possibly can prove the dynamical origin of these groups since shifts in the velocity plane were also found in \citet{_antoja12}, when analysing nearby and distant samples of stars. They found that the observed shifts were consistent with the dynamical models of spiral arm effects discussed in \citet{_antoja11}. {\it Gaia}~DR2 data will cover more stars and can possibly resolve the question of shifted positions.

With a high probability we observe major peaks like Sirius, Coma B, Hyades, Pleiades, Hercules, Wolf 630. We confirm group 9 (Dehnen98), which was recently discovered in \citet{_dehnen98} and discuss the possibility to be a part of the Coma Berenices stream together with Wolf~630, since these groups share similar metallicity properties (see Fig.\ref{_kde}, Fig.\ref{_samples}). Groups 14 and 15 (Antoja12(15) and Antoja12(12)) were first reported in \citet{_antoja12} at 2 and $3\sigma$ confidence level. We confirm both of them at the $3\sigma$ level. Structure 16, which was first discovered in \citet{_bobylev16}, is also confirmed.

We report on a new group (number 13) which has not been discussed in the literature before. It appeared in 74\% of the MC runs and contains around 201 stars. This group belongs to the nearby sample and unites stars of both disks. Group 13 is located in the proximity to Sirius and $\gamma$Leo streams. The latest one, group 12, has rather low percentage of detection, but shares similar properties as the group 13. This new group could be an independent structure, but could also be an elongation of the Sirius or $\gamma$Leo streams, because the metallicity properties  are similar for all three groups (see Fig.\ref{_kde}, Fig.\ref{_samples}). To claim if this structure is independent, this case should be further investigated, possibly through a detailed chemical analysis of stars that belong to the structures. The $\epsilon$Ind and another two tentatively new structures have weak detection percentages in the MC simulations (less than about $25\,\%$). Hence, the tentatively new structures 18 and 19 need further confirmation.   

We discuss a possible origin of stellar streams 1-19 based on our results and previous findings form the literature. If found groups showed metallicity homogeneity it would point towards an origin through being remnants of open clusters. Most of the structures do not show any particular properties inherent to thin or thick disk populations and thus we consider them to be a mixture of different type stars caused through dynamical resonances. Those groups that are more likely thick or thin disk structures are either large-scale structures (e.g. Hercules), or are small-scale groups located far from the most dense regions in the $U-V$ plane, and thus, should be independent structures possibly also caused by resonances too. Our conclusions on the origin of kinematic structures are consistent with previous works, but should be verified with a better data which would include more stars with high-precision abundances and astrometry. 

We also want to discuss a few groups which are not observed in our work, but they were in the centre of discussions in a couples of recent works. Among them is a debatable structure at (35, $-$20) $\kms$ which was first reported in \citep{_antoja08}. Taking into account its proximity to Wolf~630, Dehnen98 and such bigger streams like Sirius or Coma Berenices, authors of the same paper claim that the structure at (35, $-$20) $\kms$ can be en elongation of these bigger groups. At the same time \citep{_zhao09} detected a distinct structure at (38, $-$20) $\kms$ with probability 98\% ($\sim 3\sigma$) and suggested that it is an independent group. However, in our analysis we detected all discussed above streams except the one at (35, $-$20) $\kms$, while Wolf~630 and Dehnen98 share similar metallicity properties to the Coma Berenices stream. Groups NGC~1901 \& IC~2391 were detected by \citet{_dehnen98} and \citet{_eggen96} at ($-$25, $-$10) and ($-$20.8, $-$15.9) $\kms$ respectively. Interestingly, later works with bigger stellar samples like \citet{_antoja12}, did not detect these structures. \citet{_antoja08} make an assumption that these groups are weak compared to super-streams like Sirius, Coma Berenices, Hyades and Pleiades as they did not detect them. We do not observe these groups too. We note that the $J=2$ scale (see Fig.~\ref{_uv_all}) which is almost two times as rich with kinematic structure detections than the scale $J=3$, all these smaller-scale $J=2$ structures could be associated with some of the $J=3$ streams (see Table~\ref{_groups2}). The question remains for groups 21 (part of $\gamma$Leo?), and groups 25-30 (parts of Antoja(12) and Antoja(15)?) detected on the $J=2$ scale. These structures could be also independent and new, the answer may be given later when {\it Gaia} DR2 data is available.

The next step should be a deeper investigation of the origin of these moving groups through a better detailed analysis of chemical composition and ages of stars associated with each group to better understand the Milky Way formation. This can be done on small scales for individual structures, but ongoing and upcoming large spectroscopic surveys such as for example the $\it Gaia$-ESO \citep{gilmore2012}, WEAVE \citep{dalton2014}, and 4MOST \citep{dejong2016} surveys will provide precise elemental abundances for millions of stars, that together with astrometry from $\it Gaia$ will allow us to probe the kinematic structures at greater detail throughout the Galactic disk.


\begin{acknowledgements}
T.B. was funded by the ``The New Milky Way'' project grant from the Knut and Alice Wallenberg Foundation. We thank Prof. F.~Murtagh for making available for us the MR software packages and for valuable and helpful comments.
\end{acknowledgements}

\bibliographystyle{aa}
\bibliography{bibliography_file}
\end{document}